\documentclass[12pt]{article}
\usepackage[titletoc, title]{appendix}
\usepackage{graphicx}

\usepackage[tbtags]{amsmath}
\usepackage{amsthm}
\usepackage{amssymb}
\usepackage{amsfonts}
\usepackage{textcomp}
\usepackage{gensymb}
\usepackage{mathtools}
\usepackage{bm}
\usepackage[english]{babel}

\usepackage{etoolbox}
\usepackage{tabularx}
\usepackage{caption}
\usepackage{csquotes}
\usepackage{enumerate}
\usepackage{enumitem}
\usepackage{multirow}
\usepackage{multicol}
\usepackage{tabu}
\usepackage{float}
\usepackage[linesnumbered,ruled,vlined]{algorithm2e}

\usepackage{url} 
\usepackage[colorlinks=true,linkcolor=blue,citecolor=blue,urlcolor=blue]{hyperref}
\usepackage[authoryear]{natbib}
\newcommand{\blind}{1}

\graphicspath{ {./images/} }

\newtheorem{theorem}{Theorem}

\newtheorem{definition}{Definition}

\newtheorem{assumption}{Assumption}
\newtheorem{remark}{Remark}

\SetKwInput{KwInput}{Input}
\SetKwInput{KwOutput}{Output}
\SetKwInput{KwInit}{Initialization}

\DeclareMathOperator*{\argmin}{arg\,min}

\begin{document}

\def\spacingset#1{\renewcommand{\baselinestretch}%
{#1}\small\normalsize} \spacingset{1}


\if0\blind
{
  \bigskip
  \bigskip
  \bigskip
  \begin{center}
    {\LARGE\bf High-Dimensional Multivariate VAR Processes Estimation for Environmental Data}
  \end{center}
  \medskip
} \fi

\if1\blind
{
  \title{\bf High-Dimensional Multivariate VAR Estimation with Spatio-Temporal Structure}
  \author{Peiliang Bai \thanks{ByteDance Email: \texttt{baiplstat@gmail}}
  }
  \maketitle
} \fi

\bigskip
\begin{abstract}
High-dimensional vector autoregressive (VAR) models provide a flexible framework for
characterizing dynamic dependence in multivariate spatio-temporal systems, but their
unrestricted estimation becomes infeasible when multiple variables are observed over many
spatial locations. This paper develops a structured estimation procedure for high-dimensional
multivariate VAR processes that explicitly incorporates spatial information. We decompose
each block transition matrix into a cross-variable dependence coefficient and a spatial
transition matrix, and constrain the spatial transition matrices through a pre-specified
spatial graph. The resulting estimator is formulated as a weighted $\ell_1$-regularized
least-squares problem, where the weights encode spatial proximity or topological similarity
and induce stronger shrinkage on spatially implausible interactions. Since the objective
function is bi-convex, we estimate the cross-variable dependence matrix and the spatial
transition matrices through an alternating convex-search algorithm implemented with ADMM.
Under stability and restricted-eigenvalue-type conditions for high-dimensional VAR
processes, we establish convergence to a blockwise stationary point in the
subgradient sense and derive high-probability estimation error bounds for both
components of the model. Simulation studies demonstrate
that the proposed estimator accurately recovers sparse transition structures and improves
over existing two-step $\ell_1$-regularized methods in support recovery and estimation
accuracy. An application to North American climate data illustrates that the method
recovers interpretable variable-dependence networks and spatial interaction patterns across
different climate regions.
\end{abstract}


\newpage
\spacingset{1.4} 

\section{Introduction}\label{sec:intro}
Vector autoregressive (VAR) models are widely used to characterize temporal dependence in multivariate dynamical systems, with applications in macroeconomics \citep{sims1980macroeconomics}, finance \citep{frisen2008financial}, neuroscience \citep{seth2015granger}, air-pollution monitoring, and ecological studies \citep{nobre1994monitoring, hampton2013quantifying, ensor2013case, chen2015traffic, schweinberger2017high}. In modern applications, the number of component time series can be comparable to, or even larger than, the number of observations. Unrestricted VAR estimation is then statistically unstable and computationally expensive, and additional structure must be imposed to reduce the effective dimensionality of the transition matrices. A common approach is to assume sparsity or structured sparsity in the transition matrices and to estimate them through regularized least-squares or likelihood-based criteria, using penalties such as the $\ell_1$ penalty, sparse group lasso, or ordered weighted $\ell_1$ penalty \citep{basu2015regularized,davis2016sparse,melnyk2016estimating,lin2017regularized}. Another related line of work combines sparsity with low-rank structure in the VAR transition matrices \citep{basu2019low}.

Spatio-temporal systems contain additional information that is not fully exploited by generic high-dimensional VAR estimators. In this paper, we consider $m$ variables observed over $p$ spatial locations, so that the full VAR process has dimension $mp$. A standard sparse VAR estimator treats all $mp$ component series symmetrically and does not distinguish between cross-variable dependence and spatial dependence. This can lead to inefficient estimation and unstable support recovery, especially when the number of spatial locations is large. Spatial information provides a natural source of structural regularization: nearby locations, or locations with similar neighborhood structure, are typically more likely to interact than distant or topologically unrelated locations.

Several existing methods incorporate spatial structure into VAR estimation. \citet{schweinberger2017high} considered a single spatially distributed variable and imposed an $\ell_1$ penalty only on location pairs within a pre-specified distance threshold. \citet{wang2020regularized} proposed a weighted $\ell_1$ regularized estimator that incorporates spatial information and temporal lags. These methods demonstrate the usefulness of spatial regularization, but they do not explicitly separate variable-level dependence from spatial transition structure in a multivariate spatio-temporal VAR system. This separation is important when multiple climate, pollution, ecological, or neuroimaging variables are observed on the same spatial domain and may interact through different spatial patterns.

This paper proposes a structured estimation framework for high-dimensional multivariate VAR processes with spatio-temporal dependence. For the block transition matrix from variable $j$ to variable $i$, we use the decomposition
\[
B_{ij}^\star = \gamma_{ij}^\star \Theta_{ij}^\star,
\]
where $\gamma_{ij}^\star$ represents the overall lag-one dependence from variable $j$ to variable $i$, and $\Theta_{ij}^\star$ captures the corresponding spatial transition pattern. The spatial transition matrices are constrained by a pre-specified spatial graph and are estimated using a weighted $\ell_1$ penalty. The weights can be constructed from geographic distance or topological similarity, allowing the estimator to penalize spatially implausible interactions more strongly while preserving plausible local dependence.

The main contributions of this paper are as follows. First, we introduce a weighted $\ell_1$-regularized estimator for multivariate high-dimensional VAR models that jointly captures cross-variable and spatial dependence. Second, we develop an alternating convex-search algorithm, implemented through ADMM updates, for estimating the sparse cross-variable dependence matrix and the spatial transition matrices. Third, we establish high-probability estimation error bounds under stability and restricted-eigenvalue-type conditions for high-dimensional VAR processes. Simulation studies show that the proposed method accurately recovers sparse transition structures and improves over existing two-step regularized estimators. We further apply the method to a North American climate dataset and obtain interpretable variable-dependence networks and spatial interaction patterns.

\textit{Outline of this paper.} The remainder of this paper is organized as follows. Section~\ref{sec:model} introduces the model formulation, structural assumptions, estimation procedure, tuning-parameter selection, and construction of the spatial weight matrix. Section~\ref{sec:theory} establishes the convergence and estimation-error properties of the proposed estimator under high-dimensional scaling. Section~\ref{sec:simu} reports simulation studies and comparisons with competing methods. Section~\ref{sec:application} applies the proposed method to climate data from North America. Section~\ref{sec:discuss} concludes the paper. Technical arguments and supporting lemmas are provided in the Appendix.

\bigskip
\noindent
\textit{Notations.} Throughout this paper, we use a superscript ``$\star$" to denote the true value of the model parameters. We denote matrices by capital letter $A, B, \dots$, and blocked matrices by boldface capital letters $\mathbf{A}, \mathbf{B}, \dots$, respectively. For a matrix $A$, we use $\|A\|_F$, $\|A\|_1$, $\|A\|_\infty$, $A^\prime$, and $A^H$ to denote its Frobenius norm, vectorized $\ell_1$ norm, max norm, transpose, and conjugate transpose, respectively. For any $p\times p$ sparse matrix $S$, we use $s \overset{\text{def}}{=} \|S\|_0$ to denote the sparsity level of $S$, i.e. the number of non-zero entries in $S$, and we use $\mathcal{J}(S)$ to represent the sparse pattern of $S$, i.e. $\mathcal{J}(S) = \{S_{ij} \neq 0: 1\leq i,j \leq p\}$. As for any set $J$, we denote the cardinality of $J$ as $|J|$, and use $J^c$ to denote its complementary set. For a symmetric or Hermitian matrix $A$, $\Lambda_{\min}(A)$ and $\Lambda_{\max}(A)$ represent the smallest and largest eigenvalue respectively, and we use $A \otimes B$ and $A \circ B$ to denote Kronecker product and element-wise Hadamard product. For simplicity, we use $\mathbf{1}_n$ and $\mathbf{0}_n$ to denote a vector of 1 and 0 with length $n$, respectively. As for a graph $\mathcal{G}=(V,E)$, we use $\textsf{Adj}(\mathcal{G})$ to denote its adjacent matrix.

\section{Model Formulation}\label{sec:model}
To properly establish the model and the essential technical results, we start by considering a linear dynamical system comprising of $m$ stationary $p$-dimensional VAR(1) processes $\{X_t^j\}$ for $j=1,2,\dots,m$ observed at $T$ time points, whose structure is given by:
\begin{equation}
    \label{eq:1}
    \begin{cases}
        \begin{aligned}
        &X_t^1 = B_{11}^\star X_{t-1}^1 + B_{12}^\star X_{t-1}^2 + \cdots + B_{1m}^\star X_{t-1}^m + \epsilon_t^1, \\
        &X_t^2 = B_{21}^\star X_{t-1}^1 + B_{22}^\star X_{t-1}^2 + \cdots +
        B_{2m}^\star X_{t-1}^m + \epsilon_t^2, \\
        &\ \ \vdots \quad \quad \quad \vdots \quad \quad \quad \quad \quad \vdots \quad \quad \quad \quad \quad \quad \quad \ \ \vdots \\
        &X_t^m = B_{m1}^\star X_{t-1}^1 + B_{m2}^\star X_{t-1}^2 + \cdots + B_{mm}^\star X_{t-1}^m + \epsilon_t^m,
    \end{aligned}
    \end{cases}
\end{equation}
where $t=1,2,\dots,T$, $B_{ij}^\star$ is a $p\times p$ transition matrix which captures the dependencies between the $i$-th and $j$-th processes $X_t^i$ and $X_{t-1}^j$, and the $j$-th noise process $\{\epsilon_t^j\}$ is assumed to follow zero mean Gaussian distributions with covariance matrix $\Sigma_\epsilon^j = \sigma^2\mathbf{I}_p$, for $j=1,2,\dots,m$, i.e. $\epsilon_t^j \sim \mathcal{N}(0,\sigma^2\mathbf{I}_p)$. Based on the model setup above, the model parameters of interest are transition matrices $B^\star_{ij} \in \mathbb{R}^{p\times p}$.

In high-dimensional settings, different structural assumptions can be imposed on those transition matrices to enable the estimation procedure from the given time series data. In the posited linear dynamical system \eqref{eq:1}, any variable $X_t^i$ is assumed to connect to other variables as well as itself. Hence, we express the model \eqref{eq:1} as the following multivariate linear regression form:
\begin{equation}
    \label{eq:2}
    \begin{bmatrix}
        X_t^1 \\
        X_t^2 \\
        \vdots \\
        X_t^m
    \end{bmatrix} =
    \begin{bmatrix}
        B_{11}^\star & B_{12}^\star & \cdots & B_{1m}^\star \\
        B_{21}^\star & B_{22}^\star & \cdots & B_{2m}^\star \\
        \vdots & \vdots & \ddots & \vdots \\
        B_{m1}^\star & B_{m2}^\star & \cdots & B_{mm}^\star
    \end{bmatrix}
    \begin{bmatrix}
        X_{t-1}^1 \\
        X_{t-1}^2 \\
        \vdots \\
        X_{t-1}^m
    \end{bmatrix}
    +
    \begin{bmatrix}
        \epsilon_t^1 \\
        \epsilon_t^2 \\
        \vdots \\
        \epsilon_t^m
    \end{bmatrix},
\end{equation}
where each block $B_{ij}^\star \in \mathbb{R}^{p\times p}$ is assumed to be sparse, and we use $B_{ij}^\star(s,s^\prime)$ to denote the $(s,s^\prime)$-th entry of $B_{ij}^\star$, hence, $B_{ij}^\star(s,s^\prime)$ represents the 1-lagged influence of location $s$ on location $s^\prime$ between variable $i$ and variable $j$. Furthermore, we can also write \eqref{eq:2} in a compact matrix form by stacking all variables with respect to the time from $1$ to $T$:
\begin{equation}
    \label{eq:3}
    \underbrace{
    \begin{bmatrix}
        X_1^{1^\prime} & \cdots & X_1^{m^\prime} \\
        X_2^{1^\prime} & \cdots & X_2^{m^\prime} \\
        \vdots & \vdots & \vdots \\
        X_T^{1^\prime} & \cdots & X_T^{m^\prime}
    \end{bmatrix}}_{\mathbf{Y}} =
    \underbrace{
    \begin{bmatrix}
        X_0^{1^\prime} & \cdots & X_0^{m^\prime} \\
        X_1^{1^\prime} & \cdots & X_1^{m^\prime} \\
        \vdots & \vdots & \vdots \\
        X_{T-1}^{1^\prime} & \cdots & X_{T-1}^{m^\prime}
    \end{bmatrix}
    }_{\mathbf{X}}
    \underbrace{
    \begin{bmatrix}
        B_{11}^{\star^\prime} & \cdots & B_{m1}^{\star^\prime} \\
        \vdots & \ddots & \vdots \\
        B_{1m}^{\star^\prime} & \cdots & B_{mm}^{\star^\prime}
    \end{bmatrix}
    }_{\mathbf{B}^{\star^\prime}} +
    \underbrace{
    \begin{bmatrix}
        \epsilon_1^{1^\prime} & \cdots & \epsilon_1^{m^\prime} \\
        \vdots & \vdots & \vdots \\
        \epsilon_T^{1^\prime} & \cdots & \epsilon_T^{m^\prime}
    \end{bmatrix}
    }_{\mathbf{E}},
\end{equation}
Thus, the model can be written compactly as
\begin{equation}
    \label{eq:4}
    \mathbf{Y} = \mathbf{X}\mathbf{B}^{\star^\prime} + \mathbf{E},
\end{equation}
where $\mathbf{Y}, \mathbf{X}, \mathbf{E} \in \mathbb{R}^{T \times mp}$, and the block transition matrix $\mathbf{B}^\star \in \mathbb{R}^{mp \times mp}$.

\subsection{Structure Setup and Model Assumptions}\label{sec:2.1}
In this section, we introduce the structural assumptions used to regularize the multivariate regression form in \eqref{eq:2}. We assume that each block matrix $B_{ij}^\star$ can be represented by two components: a scalar coefficient $\gamma_{ij}^\star$, which measures the overall lagged dependence from variable $j$ to variable $i$, and a spatial transition matrix $\Theta_{ij}^\star$, which describes how this dependence is distributed across spatial locations:
\begin{equation}
    \label{eq:5}
    B_{ij}^\star = \gamma_{ij}^\star\Theta_{ij}^\star,\ \text{for}\ 1 \leq i,j \leq m.
\end{equation}

The decomposition in \eqref{eq:5} is used to separate variable-level dependence from spatial transition structure. Since a product decomposition is otherwise invariant to rescaling, we impose the following normalization to make $\Gamma^\star$ and $\bm{\Theta}^\star$ separately identifiable.
\begin{assumption}[Identifiability normalization]
\label{assump:identifiability}
For each active variable pair $(i,j)$ with $\gamma_{ij}^\star\neq 0$, the spatial transition component is normalized as
\begin{equation*}
    \|\Theta_{ij}^\star\|_F = 1,
\end{equation*}
and the sign of $\gamma_{ij}^\star$ is chosen so that the first nonzero entry of $\Theta_{ij}^\star$ is positive. For inactive pairs, we set $\gamma_{ij}^\star=0$ and $\Theta_{ij}^\star=\mathbf{0}_{p\times p}$.
\end{assumption}
Under Assumption~\ref{assump:identifiability}, the error bounds for $\widehat{\Gamma}$ and $\widehat{\bm{\Theta}}$ in Section~\ref{sec:theory} are interpreted for this normalized parameterization. Without such a normalization, only the product blocks $B_{ij}^\star=\gamma_{ij}^\star\Theta_{ij}^\star$ are identifiable.

\begin{figure*}[!ht]
    \centering
        \includegraphics[scale=.32]{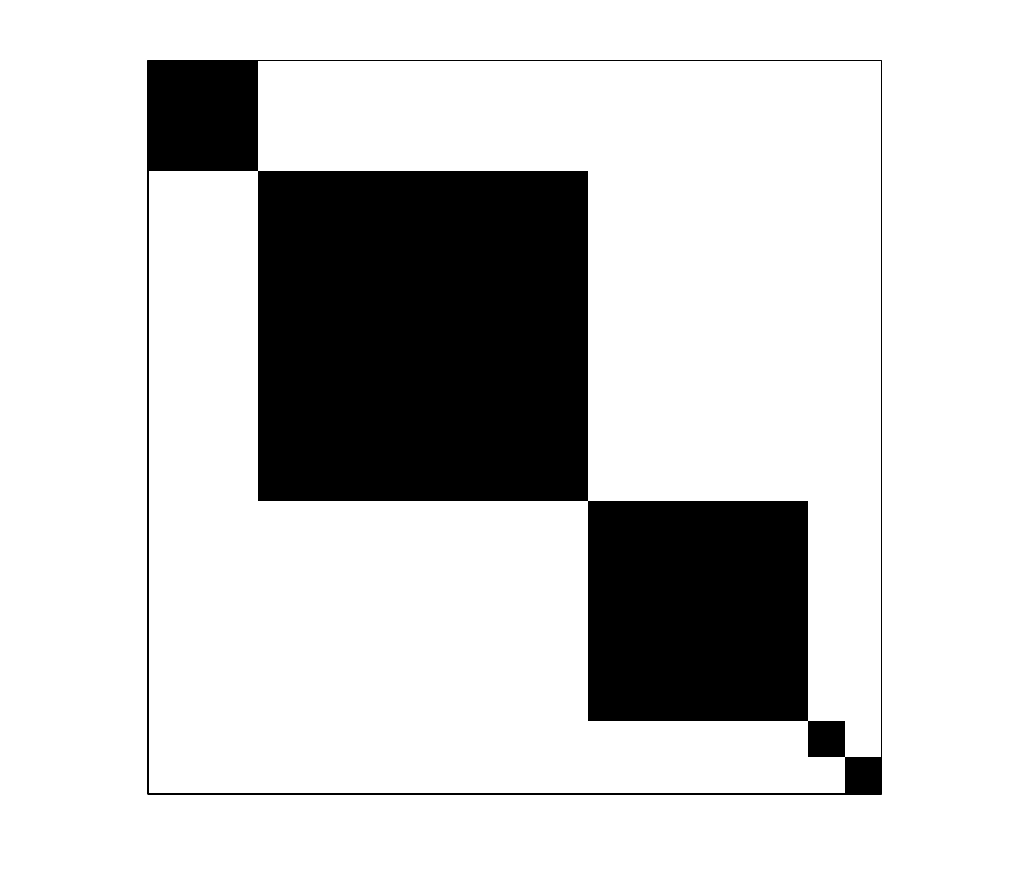}%
        \includegraphics[scale=.32]{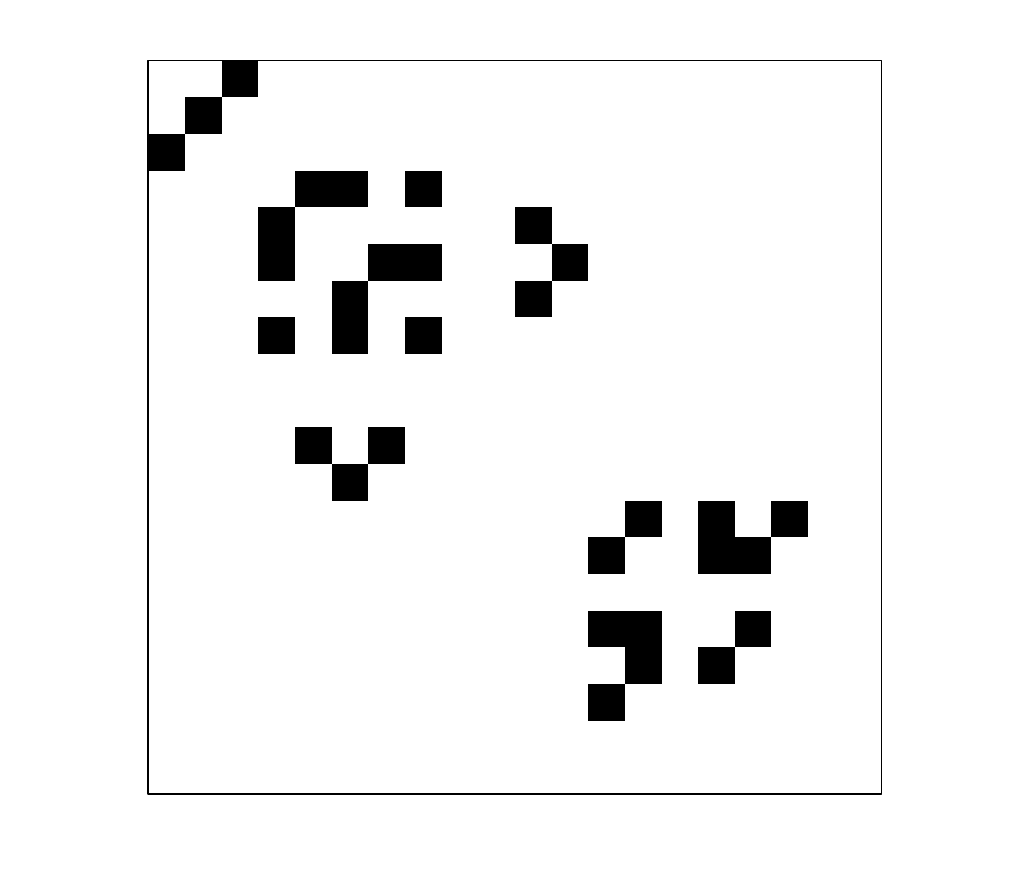}%
        \includegraphics[scale=.32]{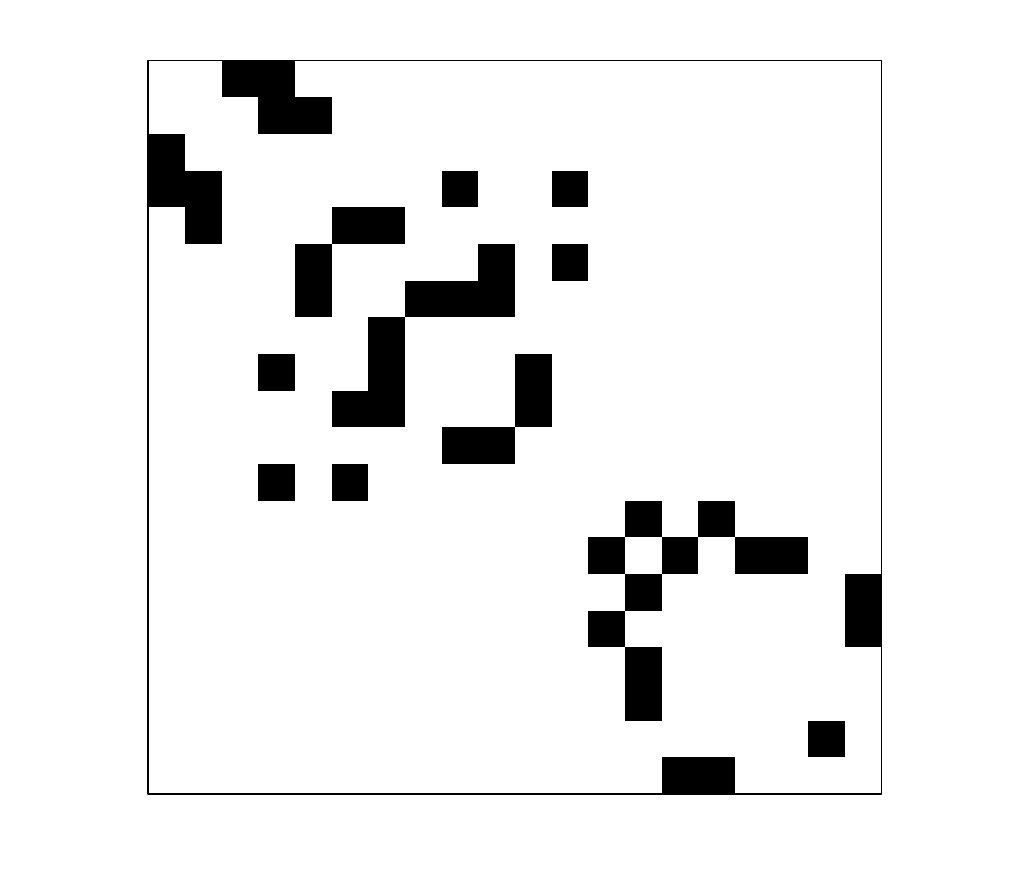}%
    \caption{Plots of the spatial structure of transition matrices. Left: the pre-determined spatial structure $\mathcal{J}_0$; Middle: the spatial structure of an \emph{acceptable} estimated transition matrix $\widehat{\Theta}$ (satisfying $\mathcal{J}(\widehat{\Theta}) \subseteq \mathcal{J}_0$); Right: the spatial structure of an \textit{invalid} estimated transition matrix $\widetilde{\Theta}$ (satisfying $\mathcal{J}(\widetilde{\Theta}) \nsubseteq \mathcal{J}_0$).}
    \label{fig:spatial-structure}
\end{figure*}

Furthermore, for any spatial structure matrix $\Theta_{ij}^\star \in \mathbb{R}^{p \times p}$, we assume that the sparsity pattern $\mathcal{J}(\Theta_{ij}^\star) \overset{\text{def}}{=} \mathcal{J}_{ij}^\star$ is a subset of a pre-determined spatial structure $\mathcal{J}_0$: $\mathcal{J}_{ij}^\star \subseteq \mathcal{J}_0$. To be specific, we provide Figure~\ref{fig:spatial-structure} to exhibit the sparsity structure between the pre-determined spatial structure $\mathcal{J}_0$ as well as the \emph{acceptable} estimated spatial structure $\widehat{\mathcal{J}}_{ij}$ with respect to the $i$ and $j$-th variables: the left panel presents the pre-determined spatial structure $\mathcal{J}_0$ where the black blocks indicate the non-zero entries; the middle plot indicates an acceptable spatial structure of estimated transition matrix $\widehat{\Theta}$, since the sparsity pattern of $\widehat{\Theta}$ is the subset of $\mathcal{J}_0$; the right plot presents an example of unacceptable estimated one $\widetilde{\Theta}$, as one can easily observe that the sparsity pattern of $\mathcal{J}(\widetilde{\Theta}) \nsubseteq \mathcal{J}_0$.

According to the spatial structure decomposition provided in \eqref{eq:5} and the spatial structure illustrated above, we are able to reformulate the model and introduce a series of assumptions on the spatial structure. We first introduce an indicator matrix $\Omega$ to indicate the spatial pattern of $\mathcal{J}_0$:
\begin{equation}
    \label{eq:6}
    \Omega_{ss^\prime} =
    \begin{cases}
        1,\quad \text{if locations pair } (s, s^\prime) \in \mathcal{J}_0, \\
        0,\quad \text{otherwise}.
    \end{cases}
\end{equation}
Based on the indicator matrix $\Omega$ in \eqref{eq:6}, we impose the following assumptions on the spatial structure and the cross-variable dependence matrix:
\begin{assumption}
\label{assump:1}
Let the pre-determined spatial structure denote as $\mathcal{J}_0$, then for any block transition matrix $\Theta_{ij}^\star$, $1\leq i,j\leq m$, denote the corresponding spatial structure (sparsity pattern) $\mathcal{J}_{ij}^\star$ satisfying: $\mathcal{J}_{ij}^\star \subseteq \mathcal{J}_0$. Moreover, the sparsity level of corresponding indicator matrix $\Omega$ of spatial structure $\mathcal{J}_0$ is given by: $s_\Omega^\star = |\mathcal{J}_0| = \|\Omega\|_0$, and $s_\Omega^\star \ll p^2$.
\end{assumption}
This assumption indicates that the spatial model parameter $\Theta_{ij}^\star$ for the variable pair $(i,j)$ is constrained in the pre-determined spatial structure $\mathcal{J}_0$. Hence, the sparsity level for $\Theta^\star_{ij}$ satisfies $s_{ij}^\star \leq s^\star_{\Omega}$. Moreover, $\mathcal{J}_0$ is assumed sparse, which is essential to establish the consistency theorem in Section \ref{sec:theory}.

For the variable-level dependence parameters $\gamma_{ij}^\star$, we define an $m \times m$ matrix $\Gamma^\star = (\gamma_{ij}^\star)_{m\times m}$, where $\gamma_{ij}^\star$ measures the overall lag-one influence from variable $j$ to variable $i$. We impose the following sparsity assumption on $\Gamma^\star$.
\begin{assumption}
\label{assump:2}
The cross-variable dependence matrix $\Gamma^\star$ is sparse. Its sparsity level is denoted by $s_\Gamma^\star \overset{\text{def}}{=} \|\Gamma^\star\|_0$, and satisfies $s_\Gamma^\star \ll m^2$.
\end{assumption}
Similar to Assumption \ref{assump:1}, this assumption states that the variable dependence model parameter $\Gamma^\star$ should also be sparse. Furthermore, by using the sparsity levels of $\Gamma^\star$ and $\Theta_{ij}^\star$'s, one can derive that the sparsity level of model parameter $\mathbf{B}^\star$ satisfies $s^\star_{\mathbf{B}} \leq s_{\Omega}^\star s^\star_{\Gamma}$.

Therefore, by employing the specific spatial structure of $B_{ij}$ in \eqref{eq:3}, and spatial structure indicator matrix $\Omega$ introduced in \eqref{eq:6} into the model \eqref{eq:2}, we obtain that:
\begin{equation*}
    \mathbf{B}^\star =
    \begin{pmatrix}
        \gamma_{11}^\star\Theta^\star_{11} & \gamma_{12}^\star\Theta^\star_{12} & \cdots & \gamma_{1m}^\star\Theta^\star_{1m} \\
        \gamma_{21}^\star\Theta^\star_{21} & \gamma_{22}^\star\Theta^\star_{22} & \cdots & \gamma_{2m}^\star\Theta^\star_{2m} \\
        \vdots & \vdots & \ddots & \vdots \\
        \gamma_{m1}^\star\Theta^\star_{m1} & \gamma_{m2}^\star\Theta^\star_{m2} & \cdots & \gamma_{mm}^\star\Theta^\star_{mm}
    \end{pmatrix} =
    (\Gamma^\star \otimes \Omega) \circ \bm{\Theta}^\star,
\end{equation*}
where $\Omega$ is the spatial structure indicator matrix defined in \eqref{eq:6}, and $\bm{\Theta}^\star \in \mathbb{R}^{mp \times mp}$ is blocked matrix consisted of all $\Theta_{ij}^\star$'s, i.e.
\begin{equation*}
    \bm{\Theta}^\star \overset{\text{def}}{=} \begin{pmatrix}
        \Theta_{11}^\star & \Theta_{12}^\star & \cdots & \Theta_{1m}^\star \\
        \Theta_{21}^\star & \Theta_{22}^\star & \cdots & \Theta_{2m}^\star \\
        \vdots & \vdots & \ddots & \vdots \\
        \Theta_{m1}^\star & \Theta_{m2}^\star & \cdots & \Theta_{mm}^\star
    \end{pmatrix}.
\end{equation*}
Therefore, by applying the specific structure proposed above, the model \eqref{eq:4} can be written in the following form:
\begin{equation}
    \label{eq:7}
    \mathbf{Y} = \mathbf{X}((\Gamma^\star \otimes \Omega) \circ \mathbf{\Theta}^\star)^\prime + \mathbf{E}.
\end{equation}

Therefore, according to \eqref{eq:7}, we proposed the following weighted $\ell_1$ regularized least squares, which penalizes both variable dependency matrix $\Gamma^\star$ as well as the spatial structures $\Theta_{ij}^\star$ for any $1\leq i,j\leq m$:
\begin{equation}
    \label{eq:8}
    (\widehat{\Gamma}, \widehat{\mathbf{\Theta}}) \overset{\text{def}}{=}
    \argmin_{\Gamma, \mathcal{J}(\mathbf{\Theta}) \subseteq \mathcal{J}_0} \left\{ \frac{1}{T}\left\| \mathbf{Y} - \mathbf{X}\left( (\Gamma \otimes \Omega) \circ \mathbf{\Theta} \right)^\prime \right\|_F^2 + \lambda_T\|\Gamma\|_1 + \mu_T\sum_{1\leq i,j \leq m}\|W\circ\Theta_{ij}\|_1\right\},
\end{equation}
where $W = (w_{ss^\prime}) \in \mathbb{R}^{p\times p}$ with $w_{ss^\prime} \geq 0$ is the spatial penalty weight, $\circ$ denotes the Hadamard product, and $\lambda_T>0$ and $\mu_T>0$ are tuning parameters for $\Gamma$ and $\bm{\Theta}$, respectively. Throughout the paper, the weighted penalty is interpreted entrywise: equivalently, if $D_W=\operatorname{diag}\{\operatorname{vec}(W)\}$, then $\|W\circ\Theta_{ij}\|_1=\|D_W\operatorname{vec}(\Theta_{ij})\|_1$. This entrywise form is used in the support-decomposition arguments for the weighted Lasso penalty. The entries of $\Theta_{ij}$ encode dependence across spatial locations. Since dependence between distant locations is typically weaker, $w_{ss^\prime}$ is chosen to increase with spatial separation or, more generally, to penalize spatially implausible pairs more heavily. Thus, the estimator uses both the sparsity of the cross-variable dependence matrix $\Gamma$ and the spatially weighted sparsity of the matrices $\Theta_{ij}$. The next assumption imposes a mild positivity condition on the spatial weights:
\begin{assumption}[Positive and bounded spatial weights]
    \label{assump:3}
    The spatial weights are deterministic and entrywise positive on the admissible spatial graph. After an arbitrary global rescaling absorbed into $\mu_T$, we assume
    \begin{equation*}
        1 \leq w_{ss^\prime} \leq r_w < \infty,\qquad (s,s^\prime)\in \mathcal{J}_0.
    \end{equation*}
    Entries outside $\mathcal{J}_0$ are constrained to be zero through the feasible set $\mathcal{J}(\Theta)\subseteq\mathcal{J}_0$; if they are retained in a vectorized representation, their weights are also taken to be at least one.
\end{assumption}
This assumption ensures that the weighted penalty is decomposable with respect to the true support and that no potentially active coordinate is left unpenalized. The constant $r_w$ measures the largest relative penalty assigned to an admissible spatial edge and appears in the estimation error bounds. Since multiplying all weights by a common positive constant is equivalent to rescaling $\mu_T$, the lower bound one is imposed without loss of generality.

Next, we introduce the assumption for stability of the linear dynamical system \eqref{eq:1} with respect to the compact matrix form presented in \eqref{eq:4}. Consider a $mp$-dimensional VAR process $\{\mathcal{W}_t\}$ indicates the VAR process with stacking all variables together, i.e. $\mathcal{W}_t \overset{\text{def}}{=} [X_t^{1^\prime}, X_t^{2^\prime}, \cdots, X_t^{m^\prime}]^\prime \in \mathbb{R}^{mp}$ for $t=0, 1, \dots, T$. Then, we introduce the autocovariance function $\Gamma_\mathcal{W}(h) = \text{Cov}(\mathcal{W}_t, \mathcal{W}_{t+h})$. We make the following assumption:
\begin{assumption}
    \label{assump:4}
    Define the spectral density function of the joint process $\{\mathcal{W}_t\}$ as follows:
    \begin{equation*}
        f_\mathcal{W}(\theta) \overset{\text{def}}{=} \frac{1}{2\pi}\sum_{l = -\infty}^{\infty}\Gamma_\mathcal{W}(l)\exp(-il\theta), \quad \theta \in [-\pi, \pi]
    \end{equation*}
    and its maximum eigenvalue is upper bounded and its minimum eigenvalue are lower bounded away from zero almost surely on $[-\pi, \pi]$, i.e.
    \begin{equation*}
        \mathcal{M}(f_\mathcal{W}) \overset{\text{def}}{=} \sup_{\theta \in [-\pi, \pi]}\Lambda_{\max}(f_\mathcal{W}(\theta)) < +\infty, \quad
        \mathfrak{m}(f_\mathcal{W}) \overset{\text{def}}{=} \inf_{\theta \in [-\pi, \pi]}\Lambda_{\min}(f_{\mathcal{W}}(\theta)) > 0
    \end{equation*}
    Moreover, we define the matrix-valued characteristic polynomial function as follows:
    \begin{equation*}
        \mathcal{B}(z) \overset{\text{def}}{=} \mathbf{I}_{mp} - \mathbf{B}^\star z,
    \end{equation*}
    then it satisfies that if $z \in \{z \in \mathbb{C}: |z|=1 \}$, then $|\mathcal{B}(z)| \neq 0$.
\end{assumption}

This is a basic assumption in high-dimensional time series analysis, and this requirement is sufficient and necessary condition for the existence of a stationary system \citep{lutkepohl2013introduction, basu2015regularized}, since it guarantees the spectral density of the joint process $\{\mathcal{W}_t\}$ exists, which is sufficient in establishing the theories.

Under this assumption, \cite{basu2015regularized} applied spectral density to establish measure of dependence and proved the estimator of model parameters satisfy two important conditions. Specifically, Proposition 4.2 and 4.3 in \cite{basu2015regularized} indicate that there exist universal constants $c_i$, such that for any $N \geq \max\{1, \omega^2\}(2\log p + 2\log m)$, the restricted eigenvalue (RE) condition and deviation bound condition hold with probability at least $1 - c_1\exp(-c_2 N) - c_3\exp(-c_4(\log p + \log m))$. We provide detailed discussion and corresponding definitions in Section \ref{sec:theory}.


\subsection{Model Parameters Estimation Algorithm}\label{sec:estimate}
In this section, we outline the algorithm for obtaining the regularized maximum likelihood estimates of the transition matrices components $\Gamma^\star$ and $\Theta_{ij}^\star$'s from the observed data.

Recall that for the given $m$ different VAR(1) processes from time 0 to $T$ together with the pre-determined spatial structure $\mathcal{J}_0$ and its corresponding indicator matrix $\Omega$. By using the same notations as in the model \eqref{eq:2} and \eqref{eq:4}, we subsequently define the objective function with respect to the model parameters $\Gamma$ and $\mathbf{\Theta}$ as follows:
\begin{equation}
    \label{eq:9}
    f(\Gamma, \bm{\Theta}) \overset{\text{def}}{=} \frac{1}{T}\left\| \mathbf{Y} - \mathbf{X}((\Gamma \otimes \Omega) \circ \bm{\Theta})^\prime\right\|_F^2 + \lambda_T\|\Gamma\|_1 + \mu_T\sum_{1\leq i,j\leq m}\|W\circ\Theta_{ij}\|_1.
\end{equation}

It can be seen that minimizing the objective function $f$ is not a \emph{jointly convex} optimization problem, but only convex in $\Gamma$ with fixed $\mathbf{\Theta}$ and convex in $\mathbf{\Theta}$ with fixed $\Gamma$. This feature indicates that the objective function $f$ is a bi-convex function, which implies the proposed algorithm might fail to converge to the global optimum. Therefore, good initial parameters are essential for fast convergence of the algorithm as well as validating the convergence of the algorithm. The detailed computational procedures are discussed as follows.

\textit{Initialization.} To initialize the spatial variable $\mathbf{\Theta}$ and the cross-variable dependence matrix $\Gamma$ in a proper procedure, we consider the following initialization method. According to the assumption of the model, for any two variables $X_t^i$ and $X_t^j$, the cross-variable dependence is the common property for all dimensions, specifically, if the $m$-th dimension of $X_t^i$ connects to $X_t^j$, then for any other dimension of $X_t^i$, they should connect to $X_t^j$. Therefore, one natural initialization method for $\Gamma$ is given by: establishing the mean process $\{x_t^j\}$ which indicates the average of $\{X_t^j\}$ across all the dimensions, i.e. $x_t^j = \frac{1}{p}\mathbf{1}^\prime_p X_t^j \in \mathbb{R}$. Therefore, we define $\widetilde{X}_t \overset{\text{def}}{=} (x_t^1, x_t^2, \cdots, x_t^m )^\prime \in \mathbb{R}^m$, then we initialize $\Gamma$ by solving the lasso problem:
\begin{equation}\label{eq:10}
    \widehat{\Gamma}^{(0)} = \argmin_\Gamma\left\{ \frac{1}{T}\|\widetilde{\mathbf{Y}} - \widetilde{\mathbf{X}}\Gamma^\prime\|_F^2 + \lambda_T^0\|\Gamma\|_1 \right\},
\end{equation}
where we define $\widetilde{\mathbf{Y}} \overset{\text{def}}{=} (\widetilde{X}_1^\prime, \widetilde{X}_2^\prime, \cdots, \widetilde{X}_T^\prime)^\prime \in \mathbb{R}^{T\times m}$, and $\widetilde{\mathbf{X}} \overset{\text{def}}{=} (\widetilde{X}_0^\prime, \widetilde{X}_1^\prime \cdots, \widetilde{X}_{T-1}^\prime)^\prime \in \mathbb{R}^{T\times m}$.

Next, by using the defined "response" and "design" matrices $\mathbf{Y}$ and $\mathbf{X}$ above, an initial estimate for $\widehat{\bm{\Theta}}^{(0)}$ is obtained by solving the following sparse fused lasso problem:
\begin{equation}\label{eq:11}
    \widehat{\bm{\Theta}}^{(0)} = \argmin_{\mathcal{J}(\mathbf{\Theta}) \subseteq \mathcal{J}_0} \left\{ \frac{1}{T}\|\mathbf{Y} - \mathbf{X}((\widehat{\Gamma}^{(0)} \otimes \Omega) \circ \bm{\Theta})^\prime\|_F^2 + \mu_T^0\sum_{1\leq i,j\leq m}\|W\circ\Theta_{ij}\|_1 \right\},
\end{equation}
where $\lambda^0$ and $\mu^0$ represent the initial tuning parameters.

\textit{Alternating Convex Search.} After we determine the initializers, in this step, we are going to utilize \textit{Alternate Convex Search} (ACS) method \citep{hastie2015statistical} to estimate $\Gamma$ and $\bm{\Theta}$ using the bi-convex objective function $f$ defined in \eqref{eq:9} for iterations $k=0,1,2,\dots$. Specifically, we consider the following two penalized optimization problems:
    \begin{equation}
    \label{eq:12}
    \begin{aligned}
        \widehat{\Gamma}^{(k+1)} &= \argmin_{\Gamma} \left\{ \frac{1}{T}\|\mathbf{Y} - \mathbf{X}((\Gamma \otimes \Omega) \circ \widehat{\bm{\Theta}}^{(k)})^\prime\|_F^2  + \lambda_T^{k+1}\|\Gamma\|_1 \right\}, \\
        \widehat{\bm{\Theta}}^{(k+1)} &= \argmin_{\mathcal{J}(\mathbf{\Theta}) \subseteq \mathcal{J}_0} \left\{ \frac{1}{T}\|\mathbf{Y} - \mathbf{X}((\widehat{\Gamma}^{(k+1)} \otimes \Omega) \circ \bm{\Theta})^\prime\|_F^2  + \mu_T^{k+1}\sum_{1\leq i,j\leq m}\|W\circ\Theta_{ij}\|_1 \right\},
    \end{aligned}
\end{equation}
where $\lambda_T^{k+1}$ and $\mu_T^{k+1}$ are the properly selected regularization parameters for the $(k+1)$-th iteration.

To solve the updating equations \eqref{eq:12}, we use ADMM subroutines that match the entrywise weighted penalty in \eqref{eq:8}. The notation below is written in vectorized form to avoid ambiguity. For a matrix $A$, let $\operatorname{vec}(A)$ stack its columns, and let $K_{p,p}$ denote the commutation matrix satisfying $\operatorname{vec}(A^\prime)=K_{p,p}\operatorname{vec}(A)$ for any $p\times p$ matrix $A$. Define
\[
    D_W=\operatorname{diag}\{\operatorname{vec}(W)\}.
\]
Before presenting the updates, define $\mathcal{Y}_i=[X_1^{i\prime},\ldots,X_T^{i\prime}]^\prime\in\mathbb{R}^{T\times p}$ and $\mathcal{X}_j=[X_0^{j\prime},\ldots,X_{T-1}^{j\prime}]^\prime\in\mathbb{R}^{T\times p}$. The following updates are applied within each outer ACS iteration.
\begin{itemize}
    \item {\it Estimation of $\Gamma$}: Fix $\bm{\Theta}=\widehat{\bm{\Theta}}^{(k)}$. For each response variable $i$, define
    \begin{equation*}
        y_i=\operatorname{vec}(\mathcal{Y}_i),\qquad
        A_i^{(k)}=\left[\operatorname{vec}(\mathcal{X}_1\widehat{\Theta}_{i1}^{(k)\prime}),\ldots,
        \operatorname{vec}(\mathcal{X}_m\widehat{\Theta}_{im}^{(k)\prime})\right]\in\mathbb{R}^{Tp\times m}.
    \end{equation*}
    The $i$-th row $\gamma_i=(\gamma_{i1},\ldots,\gamma_{im})^\prime$ of $\Gamma$ is updated by solving
    \begin{equation*}
        \min_{\gamma_i\in\mathbb{R}^m}\left\{ {1\over T}\|y_i-A_i^{(k)}\gamma_i\|_2^2+\lambda_T^{k+1}\|\gamma_i\|_1\right\}.
    \end{equation*}
    A standard scaled-form ADMM implementation introduces an auxiliary variable $z_i$ and scaled dual variable $u_i$, and iterates
    \begin{equation}
        \label{eq:13}
        \begin{aligned}
            b_i^{(r+1)}&=\left( {2\over T}A_i^{(k)\prime}A_i^{(k)}+\rho I_m\right)^{-1}
            \left( {2\over T}A_i^{(k)\prime}y_i+\rho(z_i^{(r)}-u_i^{(r)})\right),\\
            z_i^{(r+1)}&=S_{\lambda_T^{k+1}/\rho}\left(b_i^{(r+1)}+u_i^{(r)}\right),\\
            u_i^{(r+1)}&=u_i^{(r)}+b_i^{(r+1)}-z_i^{(r+1)},
        \end{aligned}
    \end{equation}
    where $S_a(x)=\operatorname{sign}(x)(|x|-a)_+$ is applied entrywise. At convergence, we set $\widehat{\gamma}_i^{(k+1)}=z_i$.

    \item {\it Estimation of $\bm{\Theta}$}: Fix $\Gamma=\widehat{\Gamma}^{(k+1)}$. The update for a spatial block $\Theta_{ij}$ is written as a weighted Lasso over the admissible support $\mathcal{J}_0$. Let
    \begin{equation*}
        \theta_{ij}=\operatorname{vec}(\Theta_{ij}),\qquad
        a_{ij}=\operatorname{vec}(W),
    \end{equation*}
    and define the design matrix
    \begin{equation*}
        A_{ij}^{(k+1)}=\widehat{\gamma}_{ij}^{(k+1)}(I_p\otimes \mathcal{X}_j)K_{p,p},
    \end{equation*}
    so that $A_{ij}^{(k+1)}\theta_{ij}=\operatorname{vec}(\widehat{\gamma}_{ij}^{(k+1)}\mathcal{X}_j\Theta_{ij}^{\prime})$. In a block-coordinate ADMM step, the partial residual excluding block $(i,j)$ is
    \begin{equation*}
        r_{ij}=\operatorname{vec}\left(\mathcal{Y}_i-
        \sum_{\ell<j}\widehat{\gamma}_{i\ell}^{(k+1)}\mathcal{X}_{\ell}\widehat{\Theta}_{i\ell}^{(k+1)\prime}
        -\sum_{\ell>j}\widehat{\gamma}_{i\ell}^{(k+1)}\mathcal{X}_{\ell}\widehat{\Theta}_{i\ell}^{(k)\prime}\right).
    \end{equation*}
    The corresponding subproblem is
    \begin{equation*}
        \min_{\theta_{ij}:\,\operatorname{supp}(\theta_{ij})\subseteq\mathcal{J}_0}
        \left\{{1\over T}\|r_{ij}-A_{ij}^{(k+1)}\theta_{ij}\|_2^2+
        \mu_T^{k+1}\|D_W\theta_{ij}\|_1\right\}.
    \end{equation*}
    With auxiliary variable $z_{ij}$ and scaled dual variable $u_{ij}$, the diagonal-weight ADMM updates are
    \begin{equation}
        \label{eq:14}
        \begin{aligned}
            h_{ij}^{(r+1)}&=\left( {2\over T}A_{ij}^{(k+1)\prime}A_{ij}^{(k+1)}+\rho I_{p^2}\right)^{-1}
            \left( {2\over T}A_{ij}^{(k+1)\prime}r_{ij}+\rho(z_{ij}^{(r)}-u_{ij}^{(r)})\right),\\
            z_{ij}^{(r+1)}&=\mathcal{T}_{\mathcal{J}_0}\left\{S_{\mu_T^{k+1}a_{ij}/\rho}\left(h_{ij}^{(r+1)}+u_{ij}^{(r)}\right)\right\},\\
            u_{ij}^{(r+1)}&=u_{ij}^{(r)}+h_{ij}^{(r+1)}-z_{ij}^{(r+1)}.
        \end{aligned}
    \end{equation}
    Here $S_{\mu_T^{k+1}a_{ij}/\rho}$ denotes entrywise soft-thresholding with coordinate-specific thresholds:
    \[
        \left[S_{\mu_T^{k+1}a_{ij}/\rho}(x)\right]_\ell
        =\operatorname{sign}(x_\ell)\left(|x_\ell|-{\mu_T^{k+1}a_{ij,\ell}\over \rho}\right)_+,
    \]
    and $\mathcal{T}_{\mathcal{J}_0}$ sets all coordinates outside the spatial graph $\mathcal{J}_0$ to zero. At convergence, $\widehat{\Theta}_{ij}^{(k+1)}$ is obtained by reshaping $z_{ij}$ into a $p\times p$ matrix.

    \item {\it Inner and outer stopping criteria}: The ADMM iterations in \eqref{eq:13} and \eqref{eq:14} are stopped using the usual primal and dual residuals. For the $\Gamma$-update,
    \begin{equation*}
        r_{\Gamma,i}^{(r)}=b_i^{(r)}-z_i^{(r)},\qquad
        s_{\Gamma,i}^{(r)}=\rho\{z_i^{(r)}-z_i^{(r-1)}\},
    \end{equation*}
    and for the $\bm{\Theta}$-update,
    \begin{equation}
        \label{eq:15}
        r_{\Theta,ij}^{(r)}=h_{ij}^{(r)}-z_{ij}^{(r)},\qquad
        s_{\Theta,ij}^{(r)}=\rho\{z_{ij}^{(r)}-z_{ij}^{(r-1)}\}.
    \end{equation}
    Each inner ADMM subproblem is iterated until the corresponding primal and dual residual norms are below $\epsilon^{\mathrm{pri}}$ and $\epsilon^{\mathrm{dual}}$. The outer ACS iteration is stopped when
    \begin{equation*}
        \Delta^{(k+1)}=
        \|\widehat{\Gamma}^{(k+1)}-\widehat{\Gamma}^{(k)}\|_F+
        \|\widehat{\bm{\Theta}}^{(k+1)}-\widehat{\bm{\Theta}}^{(k)}\|_F
        \leq \epsilon^{\mathrm{ACS}}.
    \end{equation*}

    \item {\it Dynamical selection of step-size $\rho$.} It is shown in \cite{boyd2011distributed} that dynamic selection of $\rho$ improves ADMM convergence. For each inner ADMM subproblem, $\rho$ is updated according to
    \begin{equation}
        \label{eq:16}
        \rho^{(r+1)} =
        \begin{cases}
            2\rho^{(r)}, \quad &\text{if} \quad \|r^{(r)}\|_2 > \alpha\|s^{(r)}\|_2, \\
            \rho^{(r)} / 2, \quad &\text{if} \quad \|s^{(r)}\|_2 > \alpha\|r^{(r)}\|_2, \\
            \rho^{(r)}, \quad &\text{otherwise},
        \end{cases}
    \end{equation}
    where $r^{(r)}$ and $s^{(r)}$ denote the relevant inner primal and dual residuals for the current subproblem; we use $\alpha=10$.
\end{itemize}

The following Algorithm \ref{algo:1} summarizes the essential steps in the proposed alternating updating procedure.

\begin{algorithm}[!ht]
    \DontPrintSemicolon
    \KwInput{Multivariate time series data $\{X_t^i\}_{t=0}^T$, where $i=1,2,\dots,m$, the spatial weight matrix $W$, tuning parameters $\lambda$ and $\mu$, inner ADMM tolerances $\epsilon^{\mathrm{pri}}$ and $\epsilon^{\mathrm{dual}}$, and outer ACS tolerance $\epsilon^{\mathrm{ACS}}$.}
    \KwInit{Initialize $\widehat{\Gamma}^{(0)}$ by solving the Lasso problem \eqref{eq:10}; initialize $\widehat{\bm{\Theta}}^{(0)}$ by solving \eqref{eq:11}; set $\Delta^{(0)}=+\infty$ and $k=0$.}
    \While{$\Delta^{(k)} > \epsilon^{\mathrm{ACS}}$}
    {
        Update $\widehat{\Gamma}^{(k+1)}$ row by row with fixed $\bm{\Theta}=\widehat{\bm{\Theta}}^{(k)}$ using the ADMM rules in \eqref{eq:13}\; 
        Update $\widehat{\Theta}_{ij}^{(k+1)}$ block by block with fixed $\Gamma=\widehat{\Gamma}^{(k+1)}$ using the diagonal-weight ADMM rules in \eqref{eq:14}, and apply $\mathcal{T}_{\mathcal{J}_0}$ to enforce $\mathcal{J}(\widehat{\Theta}^{(k+1)}_{ij})\subseteq\mathcal{J}_0$\;
        Update the ADMM step-size $\rho$ within each subproblem using \eqref{eq:16}\;
        Compute $\Delta^{(k+1)}=\|\widehat{\Gamma}^{(k+1)}-\widehat{\Gamma}^{(k)}\|_F+\|\widehat{\bm{\Theta}}^{(k+1)}-\widehat{\bm{\Theta}}^{(k)}\|_F$ and set $k\leftarrow k+1$\;
    }
    \KwOutput{The estimated sparse cross-variable dependence matrix $\widehat{\Gamma}$ and spatial dependency matrices $\widehat{\Theta}_{ij}$.}
    \caption{Estimating $\bm{\Theta}$ and $\Gamma$ by Alternating Convex Search (ACS).}
    \label{algo:1}
\end{algorithm}

\subsection{Tuning Parameter Selection}
To select suitable tuning parameters, we employ the Bayesian Information Criterion (BIC), which is the summation of the log-likelihood and a penalty term on the effective parameters. To be specific, the BIC function with respect to any pair of $(\lambda, \mu)$ is given by:
\begin{equation*}
    \mathrm{BIC}(\lambda, \mu)
    =
    Tmp \log\left(
    \frac{1}{Tmp}\| \mathbf{Y} - \mathbf{X}\widehat{\mathbf{B}}(\lambda, \mu)^\prime \|_F^2
    \right)
    + \widehat{s}(\lambda,\mu)\log T,
\end{equation*}
where $\widehat{s}(\lambda,\mu) = \|\widehat{\mathbf{B}}(\lambda, \mu)\|_0$ is the number of non-zero entries in $\widehat{\mathbf{B}}(\lambda, \mu)$. We select $(\lambda^\star, \mu^\star)$ over a two-dimensional grid of candidate values by minimizing the BIC evaluated at the converged estimator $(\widehat{\Gamma}^\infty, \widehat{\bm{\Theta}}^\infty)$.

\subsection{Construction of Spatial Weight Matrix}\label{sec:2.4}
As discussed in Section \ref{sec:2.1}, the spatial weight matrix $W$ depends on the distance between any two of the spatial locations. There are several ways to construct the weight matrix $W$, for example, we can directly use the distance $d_{ss^\prime}$ between any two of the spatial locations $s$ and $s^\prime$, then we have:
\begin{equation}
    \label{eq:17}
    w_{ss^\prime} = 1+c_1\frac{d_{ss^\prime}}{d_{\max}},
\end{equation}
where $d_{\max}$ is the maximum of $d_{ss^\prime}$.

We can also consider the topological structure of the spatial locations. Assuming that the neighborhood set of a vertex $v_i$ in the spatial structure is $N(v_i)$, then we use the Jaccard index as the weight:
\begin{equation}
    \label{eq:18}
    w_{ss^\prime} = 1+c_2\left(1-\frac{|N(s) \cap N(s^\prime)|}{|N(s) \cup N(s^\prime)|}\right),
\end{equation}
where $c_1, c_2>0$ are universal constants to be determined by cross validation. The performance of different weight functions are evaluated in the simulation studies Section \ref{sec:simu}.


\section{Theoretical Properties}\label{sec:theory}
In this section, we establish theoretical properties of the proposed estimation procedure in Algorithm~\ref{algo:1}. The main difficulty is that the objective function in \eqref{eq:9} is bi-convex rather than jointly convex in $(\Gamma,\bm{\Theta})$. Moreover, because the objective contains $\ell_1$ penalties, stationarity must be understood in the first-order subgradient sense. Following the general theory of alternating minimization for bi-convex problems \citep{gorski2007biconvex, hastie2015statistical}, we first show that any limit point of the alternating updates is a blockwise optimum, and hence a first-order stationary point, provided that the iterates remain in a compact neighborhood of the true parameters.

A second difficulty is that each block update uses estimated quantities from the previous iteration rather than the original design alone. Therefore, the estimation error must be controlled uniformly over the iterations. The proof proceeds in two steps. Section~\ref{sec:converge-alteralgo} establishes the deterministic convergence of the alternating algorithm for fixed realizations. Section~\ref{sec:consistent-rates} then shows that, for the random VAR process, the iterates remain in a controlled neighborhood of $(\Gamma^\star,\bm{\Theta}^\star)$ with high probability.

Throughout this section, we denote the limit point (if there exists) of $\{\widehat{\Gamma}^{(k)}\}$ and $\{\widehat{\bm{\Theta}}^{(k)}\}$ by $\Gamma^{\infty} = \lim_{k\to \infty}\widehat{\Gamma}^{(k)}$ and $\bm{\Theta}^\infty = \lim_{k\to \infty}\widehat{\bm{\Theta}}^{(k)}$, respectively. Denote the domain of the objective function $f$ by $\operatorname{dom}(f)$, where $f$ is defined in \eqref{eq:9}.

\subsection{Convergence of the Alternating Algorithm}\label{sec:converge-alteralgo}
In this subsection, we first establish a deterministic convergence statement for the alternating convex-search algorithm. The stochastic condition required by this statement, namely that the iterates remain in a compact neighborhood of the true parameters, is verified with high probability in Section~\ref{sec:consistent-rates}.

Recall that the objective function in \eqref{eq:9} consists of a smooth least-squares loss and nonsmooth $\ell_1$ penalties. We use the following stationarity notion.

\begin{definition}[First-order stationary point]
    Let $f$ be the objective function in \eqref{eq:9}, and let $z=(\Gamma,\bm{\Theta})\in\operatorname{dom}(f)$. We call $z$ a first-order stationary point if
    \[
        0 \in \partial f(z),
    \]
    where $\partial f(z)$ denotes the subdifferential of $f$ at $z$.
\end{definition}

\begin{definition}[Blockwise optimum]
    Let $f$ be the objective function in \eqref{eq:9}. A point $(\Gamma^\infty,\bm{\Theta}^\infty)$ is called a blockwise optimum if
    \begin{equation*}
        \begin{aligned}
            f(\Gamma^\infty,\bm{\Theta}) &\geq f(\Gamma^\infty,\bm{\Theta}^\infty),
            \quad \forall\ \bm{\Theta}\ \text{such that}\ \mathcal{J}(\bm{\Theta})\subseteq \mathcal{J}_0,\\
            f(\Gamma,\bm{\Theta}^\infty) &\geq f(\Gamma^\infty,\bm{\Theta}^\infty),
            \quad \forall\ \Gamma \in \mathbb{R}^{m\times m}.
        \end{aligned}
    \end{equation*}
\end{definition}

Because each block subproblem is convex, a blockwise optimum satisfies the corresponding blockwise KKT conditions. Under standard regularity conditions for bi-convex alternating minimization \citep{gorski2007biconvex}, any accumulation point of exact block minimization is therefore a first-order stationary point.

\begin{theorem}[Convergence for fixed design]
    \label{thm:1}
    Suppose that, for fixed realizations $\{X_t\}$, the sequence $\{(\widehat{\Gamma}^{(k)},\widehat{\bm{\Theta}}^{(k)})\}_{k\geq 1}$ generated by Algorithm~\ref{algo:1} is contained in a compact neighborhood of the true parameters; that is, for some finite $U(p,T)>0$,
    \begin{equation*}
        \|(\widehat{\Gamma}^{(k)},\widehat{\bm{\Theta}}^{(k)})-(\Gamma^\star,\bm{\Theta}^\star)\|_F
        \leq U(p,T),\quad \forall k\geq 1.
    \end{equation*}
    Then every limit point $(\Gamma^\infty,\bm{\Theta}^\infty)$ of the sequence generated by Algorithm~\ref{algo:1} is a blockwise optimum and hence a first-order stationary point of $f$.
\end{theorem}

\begin{remark}\label{remark:1}
    Theorem~\ref{thm:1} is a deterministic convergence statement. It does not by itself imply statistical consistency; rather, it shows that if the iterates remain in a compact neighborhood of the truth, then any accumulation point is a first-order stationary point of the penalized objective. The high-probability error bounds in Section~\ref{sec:consistent-rates} provide the corresponding statistical control of this neighborhood.
\end{remark}

\subsection{Estimation Consistency Rate}\label{sec:consistent-rates}
In this subsection, we prove that given a random process of $\mathbf{X}$, $\mathbf{Y}$ and $\mathbf{E}$ defined in \eqref{eq:3}, the sequence of estimations $\left\{ (\widehat{\Gamma}^{(k)}, \widehat{\mathbf{\Theta}}^{(k)}) \right\}^\infty_{k=1}$ lies in a uniformly controlled ball around the true values $(\Gamma^\star, \mathbf{\Theta}^\star)$ with high probability, which implies that the requirement in Theorem \ref{thm:1} is satisfied.

Recall that the alternate updating rules are based on the two penalized optimization problems presented in \eqref{eq:12}. Here, we introduce the following linear regression models with design matrix incorporated with $\widehat{\Gamma}^{(k)}$ and $\widehat{\mathbf{\Theta}}^{(k)}$, respectively.

To be specific, we construct the design matrices as follows: for the $k$-th iteration, given estimated $\widehat{\Gamma}^{(k)}$, we are able to obtain $\widehat{\Theta}^{(k)}$ by considering the following linear regression form. Here the spatial model parameters $\Theta_{ij}$ is thresholded according to the spatial indicator matrix $\Omega$ (i.e. $\mathcal{J}(\Theta_{ij}) \subseteq \mathcal{J}_0$), therefore, we obtain that
\begin{equation*}
    \underbrace{
    \begin{pmatrix}
        \mathcal{Y}_1 \\
        \mathcal{Y}_2 \\
        \vdots \\
        \mathcal{Y}_m
    \end{pmatrix}}_{\mathbf{y}}
    =
    \underbrace{
    \begin{pmatrix}
        [\widehat{\gamma}_{11}^{(k)}\mathcal{X}_1, \cdots, \widehat{\gamma}_{1m}^{(k)}\mathcal{X}_m] & \cdots & \mathbf{0} \\
        \vdots & \ddots & \vdots \\
        \mathbf{0} & \cdots & [\widehat{\gamma}_{m1}^{(k)}\mathcal{X}_1, \cdots, \widehat{\gamma}_{mm}^{(k)}\mathcal{X}_m]
    \end{pmatrix}}_{\mathbf{X}_{\Gamma}^{(k)}}
    \underbrace{
    \begin{pmatrix}
        {\Theta}_{11} \\
        \vdots \\
        {\Theta}_{1m} \\
        \vdots \\
        {\Theta}_{m1} \\
        \vdots \\
        {\Theta}_{mm}
    \end{pmatrix}}_{\widetilde{\mathbf{\Theta}}}
    +
    \underbrace{
    \begin{pmatrix}
        \mathcal{E}_1 \\
        \vdots \\
        \mathcal{E}_m
    \end{pmatrix}}_{\mathbf{e}},
\end{equation*}
where $\mathbf{y} \in \mathbb{R}^{mT\times p}$, $\mathbf{X}_{\Gamma}^{(k)}$ is a $mT \times m^2p$ diagonal blocked matrix, $\mathbf{\widetilde{\Theta}}$ is a $m^2p \times p$ matrix constructed by stacking columns of $\mathbf{\Theta}$, and the error term $\mathbf{e}$ is a $mT \times p$ matrix and $\mathcal{E}_j \overset{\text{def}}{=} [\epsilon_1^{j^\prime}, \dots, \epsilon_T^{j^\prime}]^\prime \in \mathbb{R}^{T\times p}$. Then, the $k$-th iteration of $\bm{\Theta}$ is estimated by solving the following regularized linear regression problem:
\begin{equation}
    \label{eq:19}
    \widehat{\bm{\Theta}}^{(k)} = \argmin_{\bm{\Theta}}\left\{ \frac{1}{T}\|\mathbf{y} - \mathbf{X}_{\Gamma}^{(k)}\mathbf{\Theta}\|_F^2 + \mu_T\sum_{1\leq i,j\leq m}\|W\circ\Theta_{ij}\|_1 \right\},
\end{equation}
where the weighted penalty is applied entrywise to each spatial block. In vectorized notation this is equivalent to applying a block-diagonal diagonal-weight matrix with repeated blocks $D_W$ to $\beta_\Theta=\operatorname{vec}(\bm{\Theta})$. Similarly, for the iteration $k$ and given $\bm{\Theta}^{(k)}$, we analogously update $\Gamma$ by using the following linear regression:
\begin{equation*}
    \underbrace{
    \begin{pmatrix}
        \mathcal{Y}_1 \\
        \vdots \\
        \mathcal{Y}_m
    \end{pmatrix}}_{\mathbf{y}} =
    \underbrace{
    \begin{pmatrix}
        [\mathcal{X}_1\widehat{\Theta}_{11}^{(k)}, \cdots, \mathcal{X}_m\widehat{\Theta}_{1m}^{(k)}] & \cdots & \mathbf{0} \\
        \vdots & \ddots & \vdots \\
        \mathbf{0} & \cdots & [\mathcal{X}_1\widehat{\Theta}_{m1}^{(k)}, \cdots, \mathcal{X}_m\widehat{\Theta}_{mm}^{(k)}]
    \end{pmatrix}}_{\mathbf{X}_{\Theta}^{(k)}}
    \underbrace{
    \begin{pmatrix}
        \gamma_{11}\mathbf{I}_p \\
        \vdots \\
        \gamma_{1m}\mathbf{I}_p \\
        \vdots \\
        \gamma_{m1}\mathbf{I}_p \\
        \vdots \\
        \gamma_{mm}\mathbf{I}_p
    \end{pmatrix}}_{\widetilde{\Gamma}}
    +
    \underbrace{
    \begin{pmatrix}
        \mathcal{E}_1 \\
        \vdots \\
        \mathcal{E}_m
    \end{pmatrix}}_{\mathbf{e}},
\end{equation*}
where $\mathbf{X}_{\Theta}^{(k)}$ is a $mT \times m^2p$, which is similar to $\mathbf{X}_\Gamma^{(k)}$. Therefore, the updating rules for the $(k+1)$-th iteration of $\mathbf{\Gamma}$ is derived by solving the following regularized linear regression problem:
\begin{equation}\label{eq:20}
    \widehat{{\Gamma}}^{(k+1)} = \argmin_{{\Gamma}}\left\{ \frac{1}{T}\|\mathbf{y} - \mathbf{X}^{(k)}_\Theta{\Gamma}\|_F^2 + \lambda_T\|{\Gamma}\|_1 \right\}.
\end{equation}

Note that, for both \eqref{eq:19} and \eqref{eq:20}, we let $\beta_\Theta \overset{\text{def}}{=} \text{vec}(\mathbf{\Theta})$ or $\beta_\Gamma \overset{\text{def}}{=} \text{vec}(\Gamma)$ be the vectorized model parameters, respectively. Let $D_{W,m}$ denote the block-diagonal diagonal-weight matrix that applies $D_W=\operatorname{diag}\{\operatorname{vec}(W)\}$ to each of the $m^2$ spatial blocks, so that $\|D_{W,m}\beta_\Theta\|_1=\sum_{i,j}\|W\circ\Theta_{ij}\|_1$. Consequently, we denote $\widehat{\beta}_\Gamma$ and $\widehat{\beta}_\Theta$ to represent the estimated model parameters at $k$-th iteration. Therefore, we remove the superscripts and use $\widehat{\beta}_\Gamma$ and $\widehat{\beta}_\Theta$ to represent the generic estimators given by
\begin{equation}
    \label{eq:21}
    \widehat{\beta}_\Gamma = \argmin_{\beta_\Gamma \in \mathbb{R}^{m^2}} \left\{ -2\beta_\Gamma^\prime \widehat{\gamma}_\Theta + \beta^\prime_\Gamma\widehat{\mathfrak{X}}_\Theta\beta_\Gamma + \lambda_T\|\beta_\Gamma\|_1 \right\},
\end{equation}
\begin{equation}
    \label{eq:22}
    \widehat{\beta}_\Theta = \argmin_{\beta_\Theta \in \mathbb{R}^{m^2p^2}, \mathcal{J}(\beta_\Theta) \subseteq \mathcal{J}_0} \left\{ -2\beta^\prime_\Theta \widehat{\gamma}_\Gamma + \beta^\prime_\Theta \widehat{\mathfrak{X}}_\Gamma\beta_\Theta + \mu_T\|D_{W,m}\beta_\Theta\|_1 \right\},
\end{equation}
where
\begin{equation*}
    \begin{aligned}
        &\widehat{\gamma}_\Theta = \left( \mathbf{I}_p \otimes \mathbf{X}_\Theta^\prime \right)\text{vec}(\mathbf{y})/T,\quad \widehat{\gamma}_\Gamma = \left( \mathbf{I}_p \otimes \mathbf{X}_\Gamma^\prime \right)\text{vec}(\mathbf{y})/T, \\
        &\widehat{\mathfrak{X}}_\Theta = \left(\mathbf{I}_p \otimes \frac{\mathbf{X}^\prime_\Theta \mathbf{X}_\Theta}{T}\right),\quad \widehat{\mathfrak{X}}_\Gamma = \left( \mathbf{I}_p \otimes \frac{\mathbf{X}^\prime_\Gamma\mathbf{X}_\Gamma}{T} \right).
    \end{aligned}
\end{equation*}

Next, we introduce some additional definitions and assumptions which are essential to establish the theory.
\begin{definition}[Restricted eigenvalue (RE) condition \citep{loh2012high}]
    The empirical Gram matrix $\widehat{\mathfrak{X}}_\Theta$ used in the $\Gamma$-update satisfies $RE(\alpha_\Gamma,\phi_\Gamma)$ if
    \begin{equation*}
        u^\prime\widehat{\mathfrak{X}}_\Theta u
        \geq \alpha_\Gamma\|u\|_2^2-\phi_\Gamma\|u\|_1^2,
        \qquad \forall u\in\mathbb{R}^{m^2}.
    \end{equation*}
    The empirical Gram matrix $\widehat{\mathfrak{X}}_\Gamma$ used in the $\bm{\Theta}$-update satisfies $RE(\alpha_\Theta,\phi_\Theta)$ if
    \begin{equation*}
        v^\prime\widehat{\mathfrak{X}}_\Gamma v
        \geq \alpha_\Theta\|v\|_2^2-\phi_\Theta\|v\|_1^2,
        \qquad \forall v\in\mathbb{R}^{m^2p^2}.
    \end{equation*}
\end{definition}

\begin{definition}[Deviation bound]
    For any given stationary time series $\{X_t\}$, noise series $\{\epsilon_t\}$ and corresponding $\widehat{\gamma}$ and $\widehat{\mathfrak{X}}$, there exists a deterministic function $\mathbb{Q}$, which depends on the true model parameters $\Gamma^\star$, $\mathbf{\Theta}^\star$ and covariance matrix $\Sigma_\epsilon^j$'s of the error term, such that
    \begin{equation*}
        \left\| \widehat{\gamma}_\Theta - \widehat{\mathfrak{X}}_\Theta\beta^\star_\Gamma \right\|_\infty \leq \mathbb{Q}(\mathbf{\Theta}^\star)\sqrt{\frac{\log m}{T}},\quad \left\| \widehat{\gamma}_\Gamma - \widehat{\mathfrak{X}}_\Gamma \beta^\star_\Theta \right\|_\infty \leq \mathbb{Q}(\Gamma^\star)\sqrt{\frac{\log p + \log m}{T}}.
    \end{equation*}
\end{definition}
Here the universal constants $\alpha_\Gamma, \alpha_\Theta, \phi_\Gamma, \phi_\Theta$ as well as the deterministic function $\mathbb{Q}$ are determined by the true model parameters.

By using the same procedure of Proposition 4.2 and 4.3 in \cite{basu2015regularized}, we are able to show that there exist universal constants $c_i$, such that the quantities $\widehat{\gamma}_\Gamma$, $\widehat{\gamma}_\Theta$, $\widehat{\mathfrak{X}}_\Gamma$ and $\widehat{\mathfrak{X}}_\Theta$ satisfy the RE condition and the Deviation bound with probability at least
\begin{equation*}
    1 - c_1\exp(-c_2T\min\{\omega^{-2}, 1\}) - c_3\exp(-c_4(\log p + \log m)).
\end{equation*}
We provide a brief proof in the appendix and refer the readers to \cite{basu2015regularized} for technical details.

Besides the essential ingredients RE condition and Deviation bound, the following assumptions on the true model parameters $\Gamma^\star$, $\mathbf{\Theta}^\star$ are also important to establish the theoretical results.
\begin{assumption}
    \label{assump:5}
    Recall that the true model parameters are denoted as $\Gamma^\star$ and $\Theta_{ij}^\star$ for $1\leq i,j \leq m$. Then, we assume that there exist some large enough constants $C_\Theta>0$ and $C_\Gamma > 0$ such that $\|\Theta_{ij}^\star\|_\infty \leq C_\Theta < +\infty$, and $\|\Gamma^\star\|_\infty \leq C_\Gamma < +\infty$.
\end{assumption}

\begin{remark}\label{remark:2}
Assumption \ref{assump:5} provides the constraints of the magnitude of the entries in model parameters $\Gamma^\star$ and $\Theta_{ij}^\star$. It also relates to the sparsity level of the model, which is mentioned in Assumption \ref{assump:1} and \ref{assump:2} as well.
\end{remark}

\begin{assumption}[Local initializer]
\label{assump:init}
The initializer $(\widehat{\Gamma}^{(0)},\widehat{\bm{\Theta}}^{(0)})$ lies in a local neighborhood of the truth with high probability. Specifically, for constants $C_\Gamma^{(0)},C_\Theta^{(0)}>0$,
\begin{equation*}
    \|\widehat{\beta}_\Gamma^{(0)}-\beta_\Gamma^\star\|_\infty
    \leq C_\Gamma^{(0)}\sqrt{\frac{\log m}{T}},\qquad
    \|\widehat{\beta}_\Theta^{(0)}-\beta_\Theta^\star\|_\infty
    \leq C_\Theta^{(0)}\sqrt{\frac{\log m+\log p}{T}}.
\end{equation*}
The mean-process Lasso initializer in Section~\ref{sec:estimate} is used as a practical warm start. A sufficient condition for it to satisfy this assumption is that the spatial transition blocks are normalized so that the averaged processes preserve the cross-variable transition coefficients; for example, $p^{-1}\mathbf{1}_p^\prime\Theta_{ij}^\star\mathbf{1}_p$ is bounded away from zero and infinity uniformly over active pairs $(i,j)$.
\end{assumption}

With these definitions and assumptions in place, we are in a position to prove that with \emph{any deterministic $\widehat{\beta}_\Theta$} within a small ball around $\beta_\Theta^\star$, the estimation error for $\widehat{\beta}_\Gamma$ is upper bounded. Moreover, with \emph{any deterministic $\widehat{\beta}_\Gamma$} lying in a small ball around $\beta_\Gamma^\star$, the estimation error for $\widehat{\beta}_\Theta$ is upper bounded as well. The following two results are stated on the event where the corresponding empirical Gram matrix satisfies the restricted eigenvalue condition and the stochastic gradient satisfies the deviation bound; these events hold with high probability under Assumption~\ref{assump:4} and the stated local-neighborhood conditions. The randomness of $\widehat{\beta}_\Gamma$ and $\widehat{\beta}_\Theta$ in the proposed alternating algorithm \ref{algo:1} will be addressed in Theorem~\ref{thm:4}.

\begin{theorem}[Error bound for $\widehat{\beta}_\Gamma$ with $\mathbf{X}_\Theta$ and $\mathbf{E}$]
    \label{thm:2}
    For any observations $\mathbf{X}_\Theta$ defined based on $\widehat{\beta}_\Theta$, consider $\widehat{\beta}_\Theta$ obtained by \eqref{eq:22} satisfying:
    \begin{equation*}
        \|\widehat{\beta}_\Theta - \beta_\Theta^\star\|_\infty \leq \nu_\Theta = \eta_\Theta\sqrt{\frac{\log p + \log m}{T}},
    \end{equation*}
    where $\eta_\Theta>0$ is some sufficiently large constant. Then, for sample size $T \succsim \log m + \log p$, and for any tuning parameter $\lambda_T \geq 4\mathbb{Q}(\nu_\Theta)\sqrt{\frac{\log m}{T}}$, there exist universal constants $c_i>0$, $i=1,2,3,4$ such that with probability at least
    \begin{equation*}
        1 - c_1\exp(-c_2(\log m + \log p)) - c_3T^{-c_4},
    \end{equation*}
    the following estimation error bound holds
    \begin{equation*}
        \left\| \widehat{\beta}_\Gamma - \beta_\Gamma^\star \right\|_1 \leq \frac{48s_\Gamma^\star}{\alpha_\Gamma}\lambda_T,
    \end{equation*}
\end{theorem}

\begin{theorem}[Error bound for $\widehat{\beta}_\Theta$ with $\mathbf{X}_\Gamma$ and $\mathbf{E}$]
    \label{thm:3}
    For any observations $\mathbf{X}_\Gamma$ defined based on $\widehat{\beta}_\Gamma$, and consider $\widehat{\beta}_\Gamma$ obtained by \eqref{eq:21} satisfying:
    \begin{equation*}
        \|\widehat{\beta}_\Gamma - \beta_\Gamma^\star\|_\infty \leq \nu_\Gamma = \eta_\Gamma\sqrt{\frac{\log m}{T}},
    \end{equation*}
    where $\eta_\Gamma>0$ is some large enough constant. Then, for sample size $T \succsim \log m + \log p$, and for regularization parameter $\mu_T \geq 4\mathbb{Q}(\nu_\Gamma)\sqrt{\frac{\log m + \log p}{T}}$, there exist universal constants $c_i^\prime > 0$, $i=1,2,3,4$ such that with probability at least
    \begin{equation*}
        1 - c_1^\prime \exp(-c_2^\prime(\log m + \log p)) - c_3^\prime T^{-c_4^\prime},
    \end{equation*}
    the error bound holds:
    \begin{equation*}
        \left\| \widehat{\beta}_\Theta - \beta_\Theta^\star \right\|_1 \leq \frac{2 + 6r_ws_\Gamma^\star + 4r_w^2(s_\Gamma^\star)^2}{\alpha_\Theta}s_\Theta^\star\mu_T,
    \end{equation*}
\end{theorem}

\begin{remark}
\label{remark:3}
Theorems~\ref{thm:2} and \ref{thm:3} show that, when one block of parameters is sufficiently close to the truth, the other block can be estimated with a controlled $\ell_1$ error.
\end{remark}

After providing the error bounds for $\widehat{\beta}_\Gamma$ and $\widehat{\beta}_\Theta$ separately, we next state a uniform version along the computed alternating updates. Since the design matrices in later iterations depend on previous estimates, the uniform statement is made over a fixed finite number of iterations, or equivalently up to a finite stopping time determined by Algorithm~\ref{algo:1}.

\begin{theorem}[Uniform error bounds along the computed iterates]
    \label{thm:4}
    Let $K$ denote a fixed maximum number of ACS iterations, or a finite stopping time determined by the stopping criterion in Algorithm~\ref{algo:1}. Suppose Assumption~\ref{assump:init} holds and that, for every iterate $k=0,1,\ldots,K-1$, the empirical Gram matrices and stochastic gradients associated with the two block updates satisfy the restricted eigenvalue and deviation-bound conditions in Definitions~3--4. Suppose further that
    \begin{equation*}
        T \succsim s_\Gamma^\star s_\Theta^\star(\log m+\log p),
    \end{equation*}
    and the tuning parameters satisfy
    \begin{equation*}
        \lambda_T^k \asymp \sqrt{\frac{\log m}{T}},\qquad
        \mu_T^k \asymp \sqrt{\frac{\log m+\log p}{T}},\qquad k=0,1,\ldots,K.
    \end{equation*}
    Then there exist constants $C_\Gamma,C_\Theta,C_1,C_2,C_3,C_4>0$ such that, with probability at least
    \begin{equation*}
        1-C_1K\exp\{-C_2(\log m+\log p)\}-C_3KT^{-C_4},
    \end{equation*}
    the following bounds hold simultaneously for all computed iterates $k=1,\ldots,K$:
    \begin{equation}
        \label{eq:uniform-error-bounds}
        \left\| \widehat{\beta}_\Gamma^{(k)} - \beta_\Gamma^\star \right\|_1
        \leq C_\Gamma s_\Gamma^\star\sqrt{\frac{\log m}{T}},\qquad
        \left\| \widehat{\beta}_\Theta^{(k)} - \beta_\Theta^\star \right\|_1
        \leq C_\Theta s_\Theta^\star s_\Gamma^\star\sqrt{\frac{\log m + \log p}{T}}.
    \end{equation}
    The constants $C_\Gamma$ and $C_\Theta$ depend on the stability and curvature constants, the upper bounds on the true model parameters, and the spatial-weight constant $r_w$, but not on $T$, $p$, or $m$.
\end{theorem}

\section{Performance Evaluation}\label{sec:simu}
In this section, we present the results from several numerical experiments to evaluate the performance of the proposed alternatingly convex search algorithm in a multivariate VAR model. The VAR processes $\{X_t^j\}$ for $j=1,2,\dots,m$ are generated from the model posited in \eqref{eq:1} with the pre-determined true spatial structure pattern $\mathcal{J}_0$ for all the spatial dependency matrices $\Theta_{ij}^\star$. The details are listed as follows:
\begin{itemize}
    \item {\bf Data generation:} For all settings, the data are generated from the model \eqref{eq:1}. Specifically, for the true cross-variable dependence component $\Gamma^\star = (\gamma_{ij}^\star)_{m\times m}$, each entry in $\Gamma^\star$ is nonzero with probability $2/m$, and the nonzero entries are generated from \textsf{Unif}(0.5, 1) and all diagonal entries are forced to be nonzero. The spatial structure $\mathcal{J}_0$ is constructed in accordance to the spatial positions generated from a two-dimensional uniform distribution $\textsf{Unif}([-3,3] \times [-3,3])$.

    More precisely, we first generate $c$ points as centroids on the 2-dimensional square $[-3,3] \times [-3,3]$, then we generate $\lfloor p/c \rfloor$ points around each centroid by using bivariate Gaussian distribution for the longitude and latitude, respectively. Hence, we obtain the spatial structure pattern $\mathcal{J}_0$ with dimension $p$, which are approximately clustered into $c$ cliques.

    For the true spatial dependency matrices $\Theta^\star_{ij}$, the sparsity pattern of each $\Theta_{ij}^\star$ is restricted by the spatial structure $\mathcal{J}_0$, i.e. $\mathcal{J}(\Theta_{ij}^\star) \subseteq \mathcal{J}_0$. The nonzero entries of $\Theta^\star_{ij}$ are generated from $\textsf{Unif}(0.1,0.5)$ and then normalized according to Assumption~\ref{assump:identifiability}. After constructing the unscaled transition matrix $\widetilde{\mathbf{B}}$, we rescale it as
    \begin{equation*}
        \mathbf{B}^\star = \rho\,{\widetilde{\mathbf{B}}\over \rho_{\mathrm{sp}}(\widetilde{\mathbf{B}})},
    \end{equation*}
    where $\rho_{\mathrm{sp}}(\widetilde{\mathbf{B}})$ denotes the spectral radius of $\widetilde{\mathbf{B}}$. This step ensures that the generated VAR process is stable with the target spectral radius $\rho$.

    \item {\bf Spatial weight matrix construction:} We recall that the entries in the spatial weight matrix $W = (w_{ss^\prime})_{p\times p}$ is determined by the following two different methods in Section \ref{sec:2.4}: the method is according to the distance between spatial locations $s$ and $s^\prime$ provided in \eqref{eq:17}, and the second method is using the topological structure and Jaccard index illustrated in \eqref{eq:18}.

    \item {\bf Model evaluation methods:} We evaluate the performance of the estimation results by using the sensitivity (SEN), specificity (SPC), accuracy (ACC), $\text{F}_1$-score, Matthews dependence coefficient (MCC) and relative error in Frobenius norm (Rel. Err) as criteria to evaluate the performance of the estimates of transition matrices $B_{ij}$. Specifically, the formulas of those measurements are provided as follows:
    \begin{equation*}
        \begin{aligned}
            &\text{SEN} = \frac{\text{TP}}{\text{TP}+\text{FN}},\quad \text{SPC} = \frac{\text{TN}}{\text{TN}+\text{FP}},\quad \text{ACC} = \frac{\text{TP}+\text{TN}}{\text{TP}+\text{TN}+\text{FP}+\text{FN}}, \\
            &\text{F}_1\text{-score} = \frac{2\text{TP}}{2\text{TP}+\text{FP}+\text{FN}},\  \text{MCC} = \frac{\text{TP} \times \text{TN} - \text{FP} \times \text{FN}}{\sqrt{(\text{TP}+\text{FP})(\text{TP}+\text{FN})(\text{TN}+\text{FP})(\text{TN}+\text{FN})}}, \\
            &\text{Rel. Err} = \frac{\| \text{Est.} - \text{Truth.}\|_F}{\|\text{Truth.}\|_F}.
        \end{aligned}
    \end{equation*}
\end{itemize}

\subsection{Simulation Results}
Table~\ref{tab:1} summarizes all the model parameters under different model settings. Specifically, we consider three major settings in which the size of the model, the number of observations, and the stability level of the linear system \eqref{eq:1}. The spatial weight matrix $W$ is obtained by using the first method unless otherwise specified.
\begin{table}[!ht]
    \spacingset{1}
    \centering
    \caption{Parameters under different model settings}
    \label{tab:1}
    \begin{tabular}{c|c|c|c|c|c}
        \hline\hline
         &  & $p$ & $m$ & $T$ & $\rho$ \\
        \hline
        \multirow{6}*{Model dimension} & A.1 & 20 & 5 & 500 & 0.8 \\
                                       & A.2 & 40 & 5 & 500 & 0.8 \\
                                       & A.3 & 60 & 5 & 500 & 0.8 \\
                                       & A.4 & 20 & 20 & 500 & 0.8 \\
                                       & A.5 & 40 & 20 & 500 & 0.8 \\
                                       & A.6 & 60 & 20 & 500 & 0.8 \\
        \hline
        \multirow{4}*{Sample size} & B.1 & 20 & 20 & 300 & 0.8 \\
                                   & B.2 & 40 & 20 & 300 & 0.8 \\
                                   & B.3 & 20 & 20 & 800 & 0.8 \\
                                   & B.4 & 40 & 20 & 800 & 0.8 \\
        \hline
        \multirow{2}*{Spectral radius} & C.1 & 20 & 20 & 500 & 0.5 \\
                                       & C.2 & 40 & 20 & 500 & 0.5 \\
        \hline\hline
    \end{tabular}
\end{table}

All numerical simulations are run in R 3.6.0 on a PC equipped with an Intel i7-7700 12-core 3.60 GHz CPU and 16 GB memory. The code and scripts for simulation examples are available on GitHub \footnote{\url{https://github.com/peiliangbai92/ts_estimation}}.

Table~\ref{tab:2} reports the performance for estimating the transition matrix $\mathbf{B}^\star$ under different simulation settings presented in Table~\ref{tab:1}. The results are based on an average of 50 replications and their standard deviations are given in parentheses. Overall, the results show that the transition matrix $\mathbf{B}^\star$ is estimated with a high degree of accuracy for identifying the sparsity patterns. For most cases in setting A, the algorithm provides near perfect results. In scenario B, we examine the small sample size cases in B.1 and B.2. As expected, the performance deteriorates when the sample size is smaller. In scenario C, we investigate performance under a smaller spectral radius and observe different estimation behavior compared with settings A.1 and A.2.
\begin{table}[!ht]
    \spacingset{1}
    \centering
    \caption{Performance evaluation results under different settings}
    \label{tab:2}
    \begin{tabular}{c|c|c|c|c|c|c}
        \hline\hline
         & SEN & SPE & ACC & $\text{F}_1$-score & MCC & Rel.Err \\
        \hline
        A.1 & $1.000_{(0.000)}$ & $0.944_{(0.017)}$ & $0.947_{(0.016)}$ & $0.689_{(0.067)}$ & $0.706_{(0.059)}$ & $0.299_{(0.016)}$ \\
        A.2 & $1.000_{(0.000)}$ & $0.933_{(0.015)}$ & $0.937_{(0.014)}$ & $0.657_{(0.051)}$ & $0.676_{(0.045)}$ & $0.342_{(0.032)}$ \\
        A.3 & $0.999_{(0.001)}$ & $0.969_{(0.005)}$ & $0.970_{(0.005)}$ & $0.766_{(0.028)}$ & $0.776_{(0.024)}$ & $0.618_{(0.007)}$ \\
        A.4 & $1.000_{(0.001)}$ & $0.996_{(0.000)}$ & $0.996_{(0.001)}$ & $0.919_{(0.003)}$ & $0.920_{(0.004)}$ & $0.551_{(0.031)}$ \\
        A.5 & $0.999_{(0.001)}$ & $0.995_{(0.001)}$ & $0.995_{(0.001)}$ & $0.898_{(0.021)}$ & $0.901_{(0.020)}$ & $0.428_{(0.023)}$ \\
        A.6 & $0.996_{(0.002)}$ & $0.968_{(0.006)}$ & $0.969_{(0.006)}$ & $0.612_{(0.045)}$ & $0.654_{(0.037)}$ & $0.796_{(0.022)}$ \\
        \hline
        B.1 & $0.970_{(0.029)}$ & $0.996_{(0.000)}$ & $0.995_{(0.001)}$ & $0.905_{(0.015)}$ & $0.905_{(0.015)}$ & $0.535_{(0.040)}$ \\
        B.2 & $0.995_{(0.012)}$ & $0.993_{(0.003)}$ & $0.993_{(0.003)}$ & $0.867_{(0.050)}$ & $0.872_{(0.045)}$ & $0.724_{(0.030)}$ \\
        B.3 & $1.000_{(0.000)}$ & $0.997_{(0.000)}$ & $0.997_{(0.001)}$ & $0.921_{(0.000)}$ & $0.922_{(0.001)}$ & $0.521_{(0.036)}$ \\
        B.4 & $1.000_{(0.000)}$ & $0.994_{(0.001)}$ & $0.994_{(0.001)}$ & $0.878_{(0.026)}$ & $0.882_{(0.024)}$ & $0.721_{(0.020)}$ \\
        \hline
        C.1 & $1.000_{(0.000)}$ & $0.994_{(0.001)}$ & $0.994_{(0.001)}$ & $0.875_{(0.021)}$ & $0.879_{(0.020)}$ & $0.469_{(0.013)}$ \\
        C.2 & $1.000_{(0.000)}$ & $0.994_{(0.002)}$ & $0.994_{(0.002)}$ & $0.878_{(0.039)}$ & $0.882_{(0.035)}$ & $0.729_{(0.026)}$ \\
        \hline\hline
    \end{tabular}
\end{table}

\subsection{A Comparison with the Two-step Regularized Model}
Next, we present baseline results for the two-step $\ell_1$-regularized maximum likelihood estimator with a known spatial structure graph $\mathcal{J}_0$ \citep{schweinberger2017high}. This baseline imposes a short-distance restriction on the candidate transition entries. Table~\ref{tab:3} reports the results for estimating $\mathbf{B}^\star$ using the baseline two-step $\ell_1$ model. Compared with the proposed estimator in Table~\ref{tab:2}, the baseline has substantially lower sensitivity in higher-dimensional settings, indicating that explicitly separating cross-variable dependence from spatial transition structure improves support recovery.
\begin{table}[!ht]
    \spacingset{1}
    \centering
    \caption{Performance evaluation results for two-step $\ell_1$-penalized model}
    \label{tab:3}
    \begin{tabular}{c|c|c|c|c|c|c}
        \hline\hline
         & SEN & SPE & ACC & $\text{F}_1$-score & MCC & Rel.Err \\
        \hline
        A.1 & $0.690_{(0.033)}$ & $0.916_{(0.013)}$ & $0.903_{(0.011)}$ & $0.448_{(0.011)}$ & $0.435_{(0.018)}$ & $0.784_{(0.011)}$ \\
        A.2 & $0.175_{(0.032)}$ & $0.979_{(0.006)}$ & $0.932_{(0.004)}$ & $0.231_{(0.026)}$ & $0.215_{(0.016)}$ & $0.976_{(0.005)}$ \\
        A.3 & $0.061_{(0.006)}$ & $0.998_{(0.000)}$ & $0.953_{(0.000)}$ & $0.112_{(0.011)}$ & $0.192_{(0.010)}$ & $0.986_{(0.002)}$ \\
        A.4 & $0.068_{(0.020)}$ & $0.996_{(0.002)}$ & $0.976_{(0.001)}$ & $0.107_{(0.024)}$ & $0.126_{(0.013)}$ & $0.994_{(0.002)}$ \\
        A.5 & $0.027_{(0.007)}$ & $0.998_{(0.001)}$ & $0.978_{(0.000)}$ & $0.049_{(0.011)}$ & $0.082_{(0.006)}$ & $1.000_{(0.001)}$ \\
        A.6 & $0.007_{(0.006)}$ & $0.998_{(0.001)}$ & $0.978_{(0.001)}$ & $0.031_{(0.009)}$ & $0.039_{(0.002)}$ & $1.001_{(0.002)}$ \\
        \hline
        B.1 & $0.010_{(0.008)}$ & $0.999_{(0.001)}$ & $0.977_{(0.000)}$ & $0.017_{(0.016)}$ & $0.010_{(0.000)}$ & $1.000_{(0.001)}$ \\
        B.2 & $0.007_{(0.004)}$ & $0.999_{(0.000)}$ & $0.979_{(0.000)}$ & $0.014_{(0.007)}$ & $0.038_{(0.007)}$ & $1.002_{(0.001)}$ \\
        B.3 & $0.229_{(0.029)}$ & $0.989_{(0.003)}$ & $0.974_{(0.002)}$ & $0.254_{(0.010)}$ & $0.244_{(0.008)}$ & $0.963_{(0.004)}$ \\
        B.4 & $0.096_{(0.007)}$ & $0.995_{(0.001)}$ & $0.976_{(0.000)}$ & $0.145_{(0.008)}$ & $0.159_{(0.006)}$ & $0.990_{(0.002)}$ \\
        \hline
        C.1 & $0.001_{(0.000)}$ & $0.999_{(0.000)}$ & $0.979_{(0.000)}$ & $0.003_{(0.003)}$ & $0.000_{(0.000)}$ & $1.000_{(0.000)}$ \\
        C.2 & $0.000_{(0.000)}$ & $1.000_{(0.000)}$ & $0.979_{(0.001)}$ & $0.000_{(0.001)}$ & $0.000_{(0.000)}$ & $1.001_{(0.001)}$ \\
        \hline\hline
    \end{tabular}
\end{table}

\subsection{Comparison with the Different Spatial Weight Matrix}
In Table~\ref{tab:4}, we perform some additional experiments to compare different methods by which we construct the spatial weight matrix $W$ mentioned in Section \ref{sec:2.4}. Here we present the results for A.1, B.1, B.4 and C.1 using the spatial weight matrix $W$ generated based on the second method using Jaccard index \eqref{eq:18}.
\begin{table}[!ht]
    \spacingset{1}
    \centering
    \caption{Performance evaluation results for $W$ generated by Jaccard index.}
    \label{tab:4}
    \begin{tabular}{c|c|c|c|c|c|c}
        \hline\hline
         & SEN & SPE & ACC & $\text{F}_1$-score & MCC & Rel.Err \\
        \hline
        A.1 & $1.000_{(0.000)}$ & $0.946_{(0.017)}$ & $0.950_{(0.016)}$ & $0.700_{(0.068)}$ & $0.713_{(0.060)}$ & $0.298_{(0.014)}$ \\
        A.4 & $1.000_{(0.000)}$ & $0.994_{(0.001)}$ & $0.994_{(0.001)}$ & $0.887_{(0.013)}$ & $0.890_{(0.012)}$ & $0.482_{(0.020)}$ \\
        B.1 & $0.991_{(0.012)}$ & $0.997_{(0.000)}$ & $0.997_{(0.001)}$ & $0.916_{(0.006)}$ & $0.917_{(0.007)}$ & $0.556_{(0.031)}$ \\
        B.4 & $1.000_{(0.000)}$ & $0.994_{(0.001)}$ & $0.992_{(0.001)}$ & $0.858_{(0.026)}$ & $0.872_{(0.024)}$ & $0.732_{(0.020)}$ \\
        C.1 & $0.995_{(0.008)}$ & $0.996_{(0.000)}$ & $0.996_{(0.000)}$ & $0.916_{(0.006)}$ & $0.917_{(0.006)}$ & $0.564_{(0.033)}$ \\
        \hline\hline
    \end{tabular}
\end{table}

According to Table~\ref{tab:4}, we conclude that the results of the spatial weight matrix $W$ constructed by the first method is slightly better than the second method which considers the topological spatial structure only.

\subsection{Additional Experiments on VAR(2)}
Finally, we consider the performance for the VAR model with lag-2 under model parameter settings A.1, B.1, B.3 and C.1, respectively. Recall that any VAR($d$) models can be rewritten as a VAR(1) model \citep{lutkepohl2013introduction}, the performance of estimation is also evaluated by the measurements presented previously.
\begin{table}[!ht]
    \spacingset{1}
    \centering
    \caption{Performance evaluation results for VAR(2) model.}
    \label{tab:5}
    \begin{tabular}{c|c|c|c|c|c|c}
        \hline\hline
          & SEN & SPE & ACC & $\text{F}_1$-score & MCC & Rel.Err \\
        \hline
        A.1 & $0.990_{(0.015)}$ & $0.987_{(0.006)}$ & $0.986_{(0.006)}$ & $0.885_{(0.047)}$ & $0.884_{(0.045)}$ & $0.655_{(0.082)}$ \\
        A.4 & $1.000_{(0.000)}$ & $0.994_{(0.001)}$ & $0.994_{(0.001)}$ & $0.887_{(0.013)}$ & $0.890_{(0.012)}$ & $0.482_{(0.020)}$ \\
        B.1 & $0.991_{(0.012)}$ & $0.997_{(0.000)}$ & $0.997_{(0.001)}$ & $0.916_{(0.006)}$ & $0.917_{(0.007)}$ & $0.556_{(0.031)}$ \\
        B.3 & $1.000_{(0.000)}$ & $0.994_{(0.001)}$ & $0.992_{(0.001)}$ & $0.858_{(0.026)}$ & $0.872_{(0.024)}$ & $0.732_{(0.020)}$ \\
        C.1 & $0.995_{(0.008)}$ & $0.996_{(0.000)}$ & $0.996_{(0.000)}$ & $0.916_{(0.006)}$ & $0.917_{(0.006)}$ & $0.564_{(0.033)}$ \\
        \hline\hline
    \end{tabular}
\end{table}

Comparing to the results presented in Table~\ref{tab:2}, Table~\ref{tab:5} presents a close result to Table~\ref{tab:2}, though the relative error exhibits a slight exacerbation. These results promise that our proposed estimation algorithm can be extended to a generic time lag $d$ and maintain the favorable performance.

\section{Application to Climate Modeling}\label{sec:application}
In this section, we apply the proposed estimation procedure to an environmental dataset to study the relationships among climate-related variables collected at multiple spatially dependent locations in North America.

\subsection{Data Description}
The data consist of monthly measurements from January 2001 to December 2010 on $m=15$ different variables including mean temperature (TMP), diurnal temperature range (DTR), maximum and minimum temperature (TMX, TMN), precipitation (PRE), vapor pressure (VAP), cloud cover (CLD), rain days (WET), potential evapotranspiration (PET), frost days (FRS), greenhouse gases (carbon dioxide ($\text{CO}_2$), methane ($\text{CH}_4$), and nitrous oxide ($\text{N}_2\text{O}$)) and aerosols (AER) from CRU\footnote{\url{http://www.cru.uea.ac.uk/data/}}, NOAA \footnote{\url{https://www.esrl.noaa.gov/gmd/dv/ftpdata.html}} and NASA \footnote{\url{https://disc-beta.gsfc.nasa.gov/}}.

\begin{figure}[!ht]
    \centering
    \includegraphics[scale=.35]{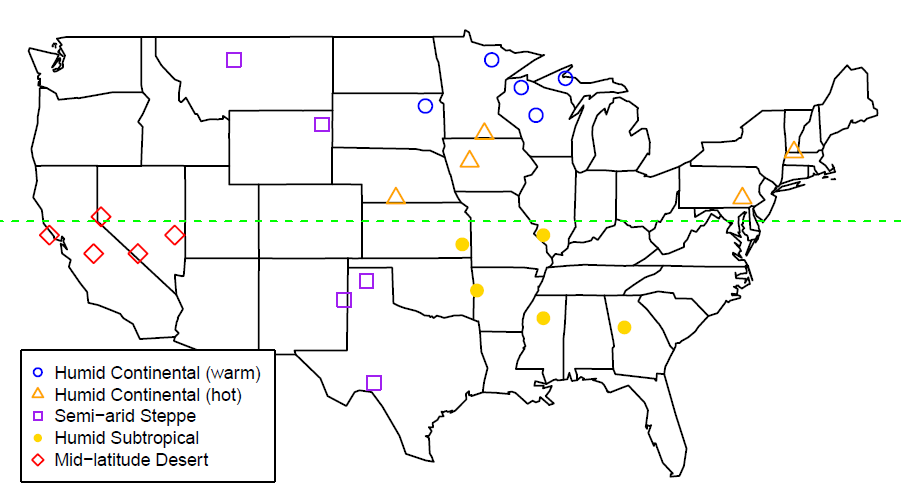}
    \caption{The selected 25 locations based on climate classification. The dashed line separates the south and north of North America and corresponds to $39^{\circ}$ N.}
    \label{fig:1}
\end{figure}

The dataset is organized on a $2.5^\circ$ latitude by $2.5^\circ$ longitude grid across North America. We aggregate the monthly series into three-month intervals and take first differences of the quarterly data to reduce seasonality and autocorrelation. This preprocessing yields $T=40$ time points, and the resulting series are normalized. We then divide North America into five climate zones: Humid Continental (warm summer), Humid Continental (hot summer), Semi-arid Steppe, Humid Subtropical, and Mid-latitude Desert. From these zones, we randomly select $p=25$ locations; the selected locations are plotted in Figure~\ref{fig:1}.

After preprocessing and removing variables with incomplete coverage, the final dataset contains $T=40$ observations, $p=25$ locations, and $m=14$ climate-related variables. Additional details on preprocessing and location selection are provided in the Appendix.

\begin{figure}[!ht]
    \centering
    \includegraphics[scale=.425]{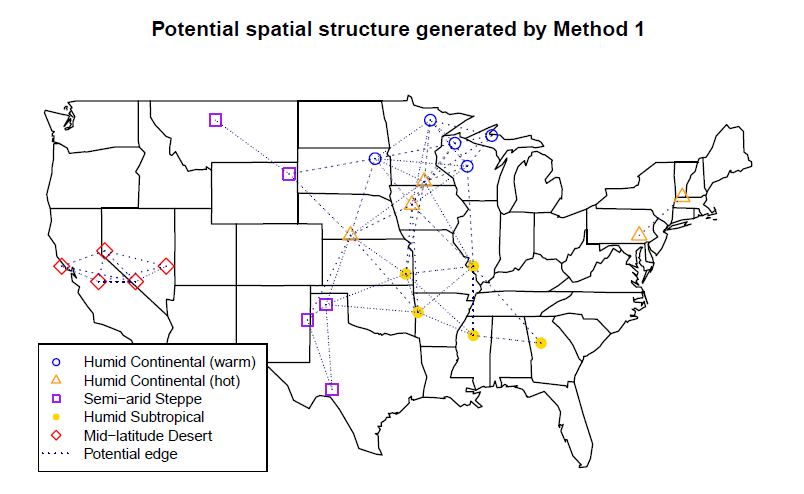}
    \caption{Spatial structure $\mathcal{J}_0$ generated by truncating long distances.}
    \label{fig:graph-structure}
\end{figure}

The spatial structure $\mathcal{J}_0$ is constructed by following the short-distance criterion \citep{rao1997space}: for any two spatial locations $s$ and $s^\prime$, we denote the maximum distance as $d_{\max}$, we construct the spatial structure $\mathcal{J}_0$ by truncating the distances longer than $0.2d_{\max}$. The spatial structure $\mathcal{J}_0$ for our 25 selected locations is presented in Figure~\ref{fig:graph-structure}.

\subsection{Estimation Results for VAR(1) Model}
\begin{figure*}[!ht]
    \centering
        \begin{multicols}{3}
            \includegraphics[width=\linewidth]{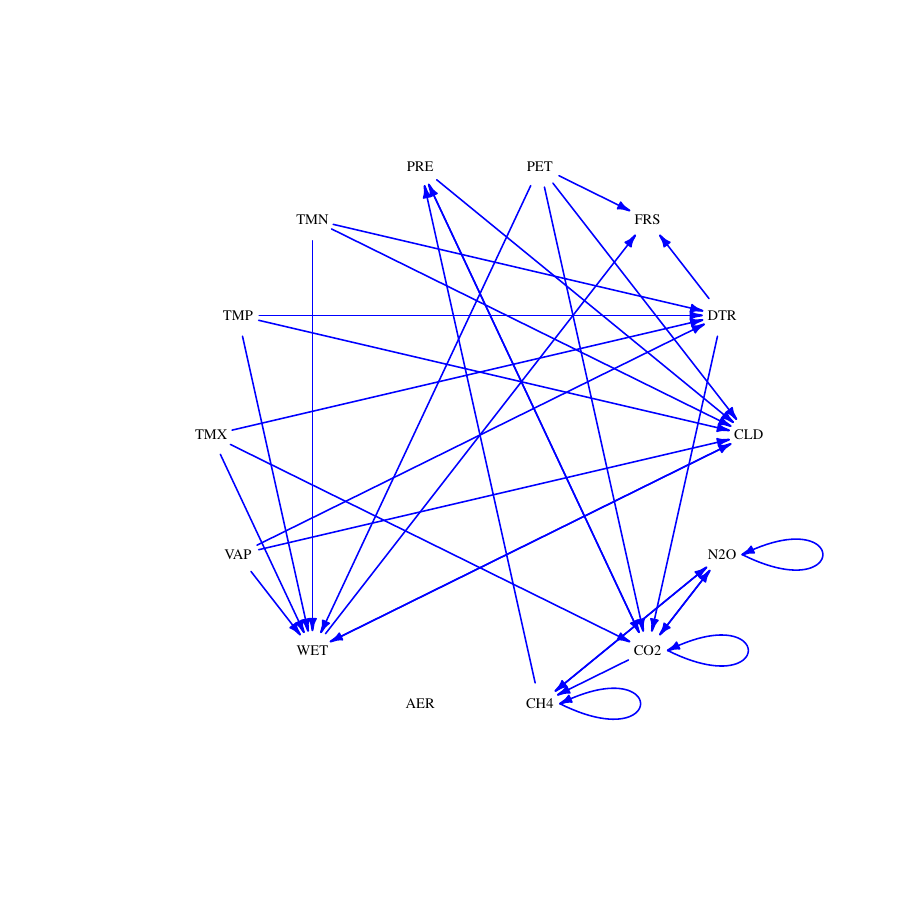}%
            \includegraphics[width=\linewidth]{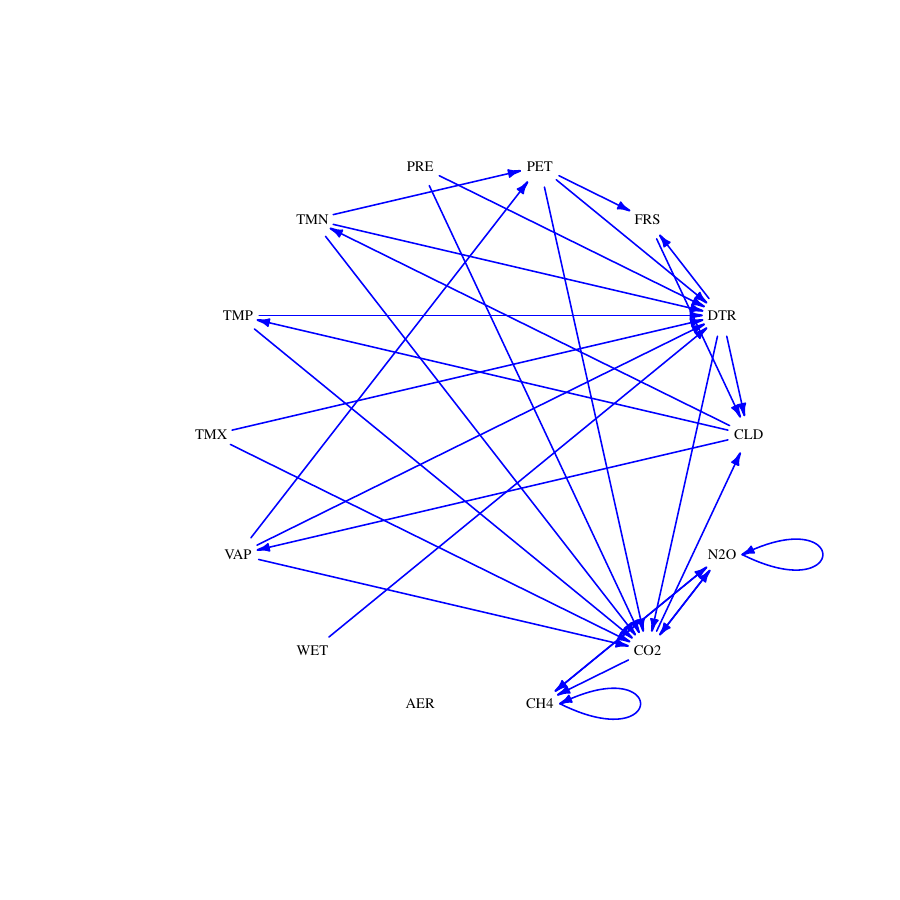}%
            \includegraphics[width=\linewidth]{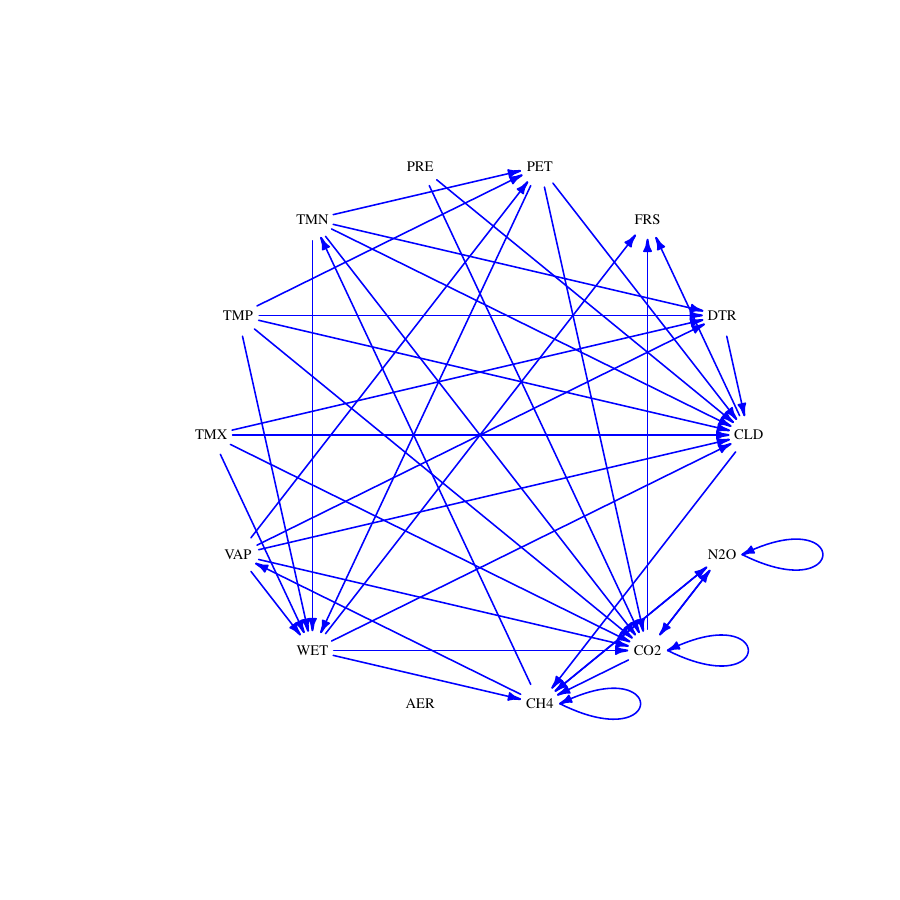}%
        \end{multicols}
        \begin{multicols}{3}
            \includegraphics[width=\linewidth]{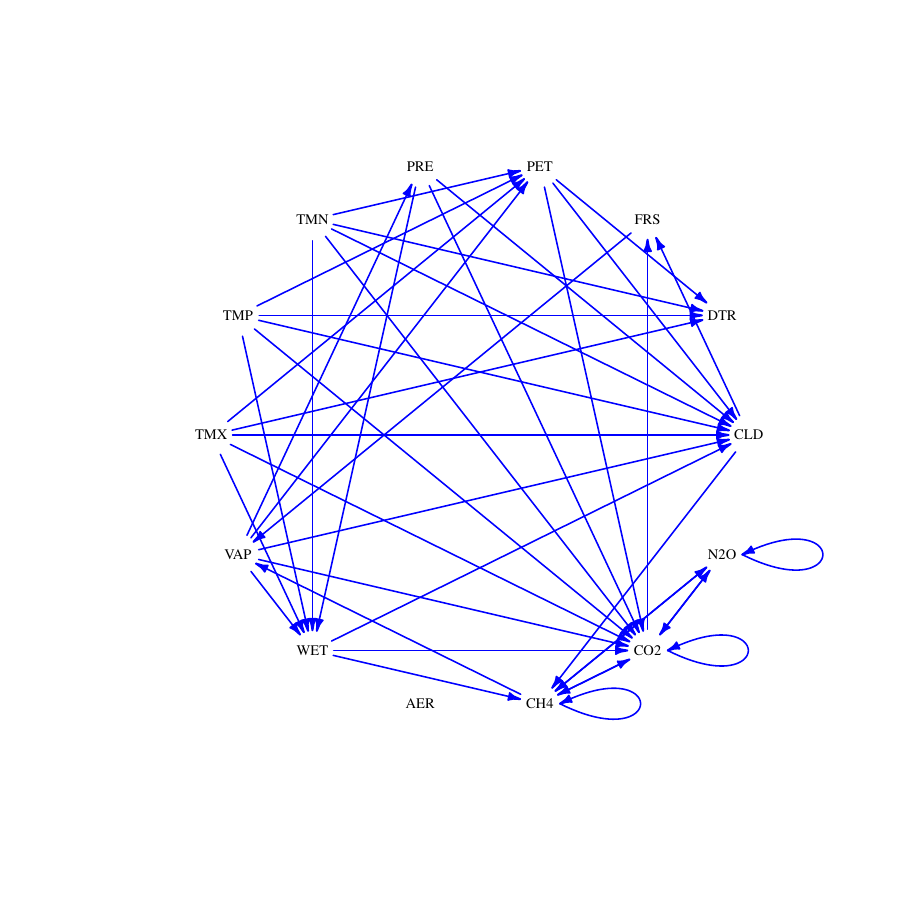}%
            \includegraphics[width=\linewidth]{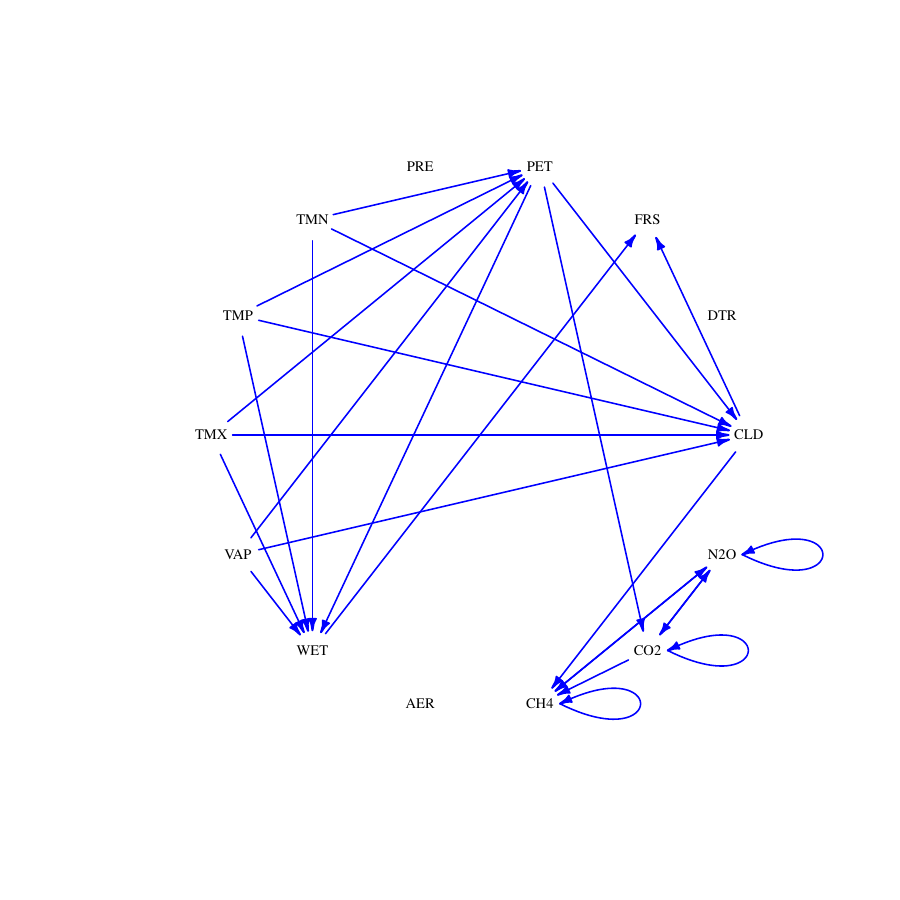}%
            \includegraphics[width=\linewidth]{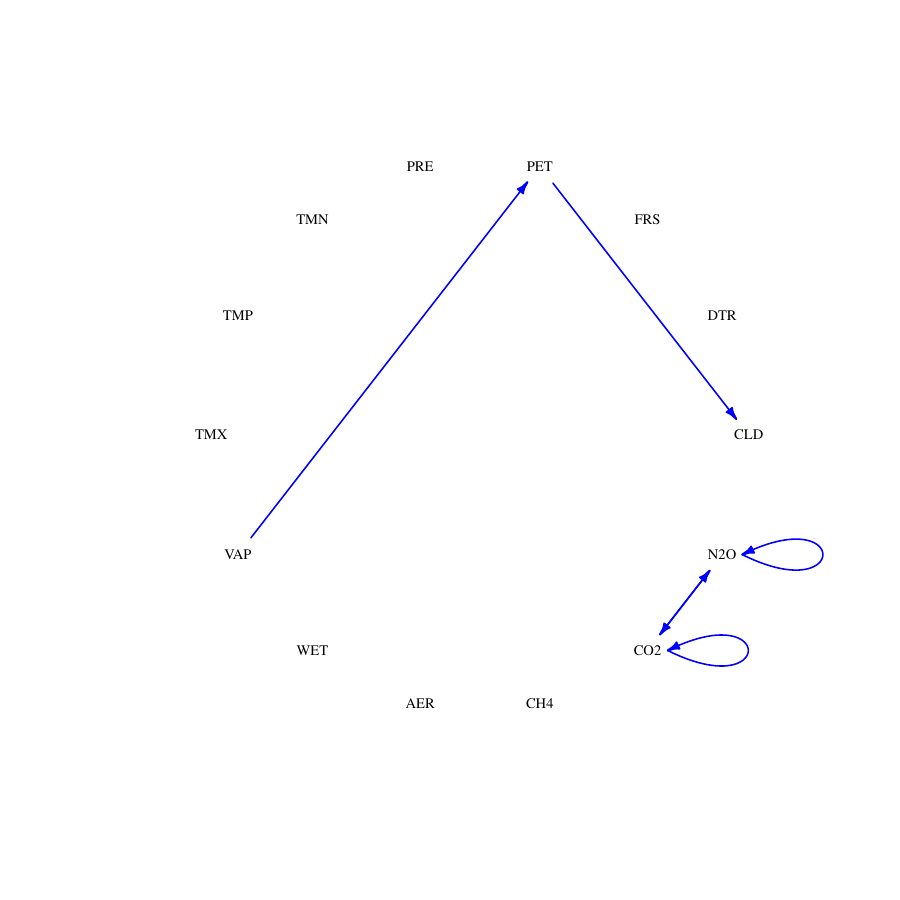}%
        \end{multicols}
    \vspace{-25pt}
    \caption{Estimated variable dependence networks at all locations and five distinct climate zones. Top left to bottom right: All locations; Humid continental (warm summer); Humid continental (hot summer); Semi-arid steppe; Humid subtropical; Mid-latitude desert.}
    \label{fig:networks}
\end{figure*}

For conciseness, Figure~\ref{fig:networks} presents the estimated dependence networks among climate-related variables for all locations and for the five climate zones using the proposed VAR(1) estimation algorithm. The corresponding heatmaps are shown in Figure~\ref{fig:heatmaps}, and selected spatial structures for specific variable pairs are shown below.
\begin{figure*}[!ht]
    \centering
        \begin{multicols}{3}
            \includegraphics[width=\linewidth]{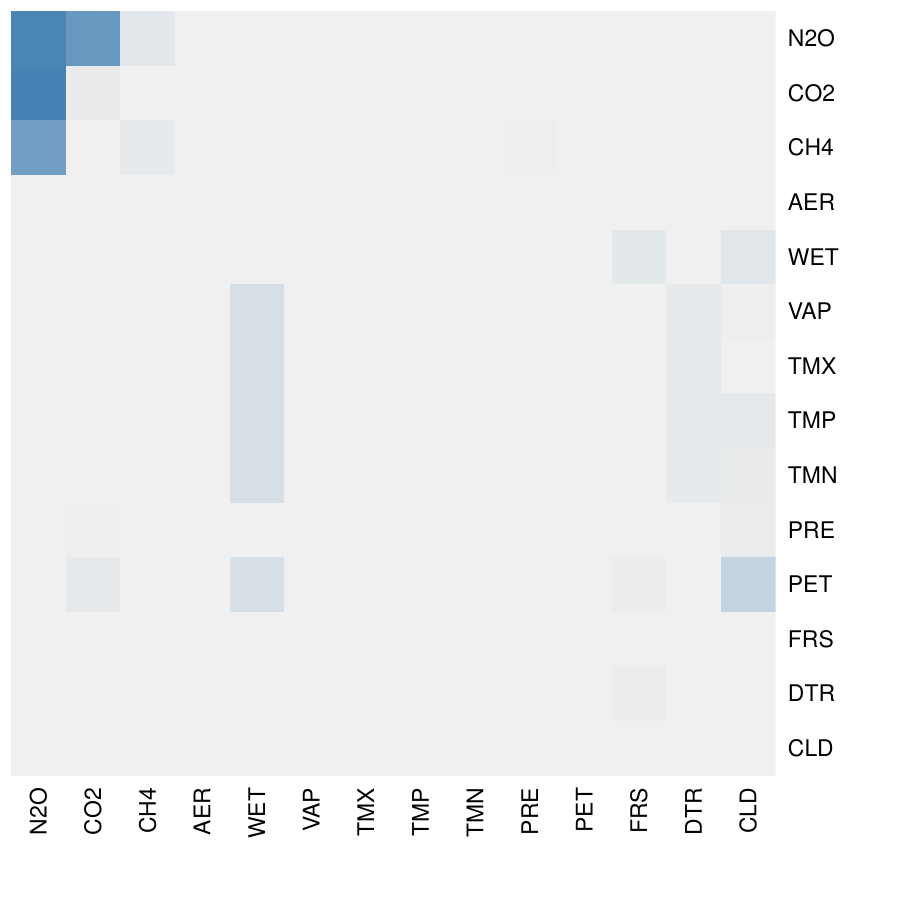}%
            \includegraphics[width=\linewidth]{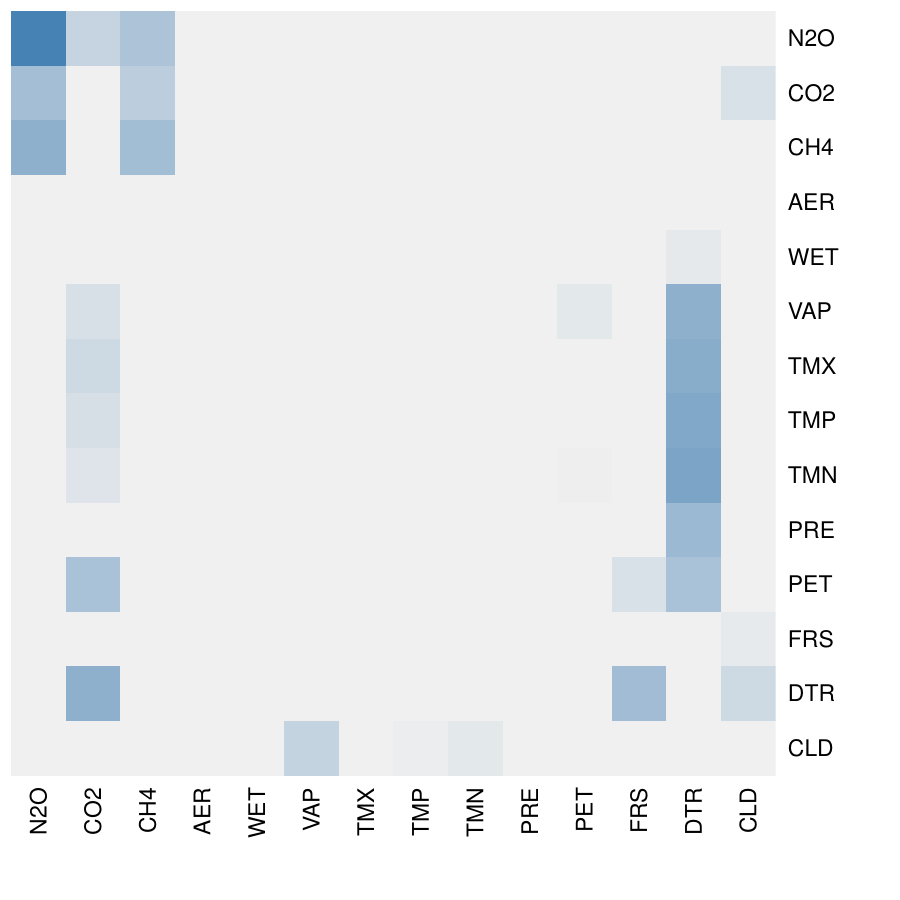}%
            \includegraphics[width=\linewidth]{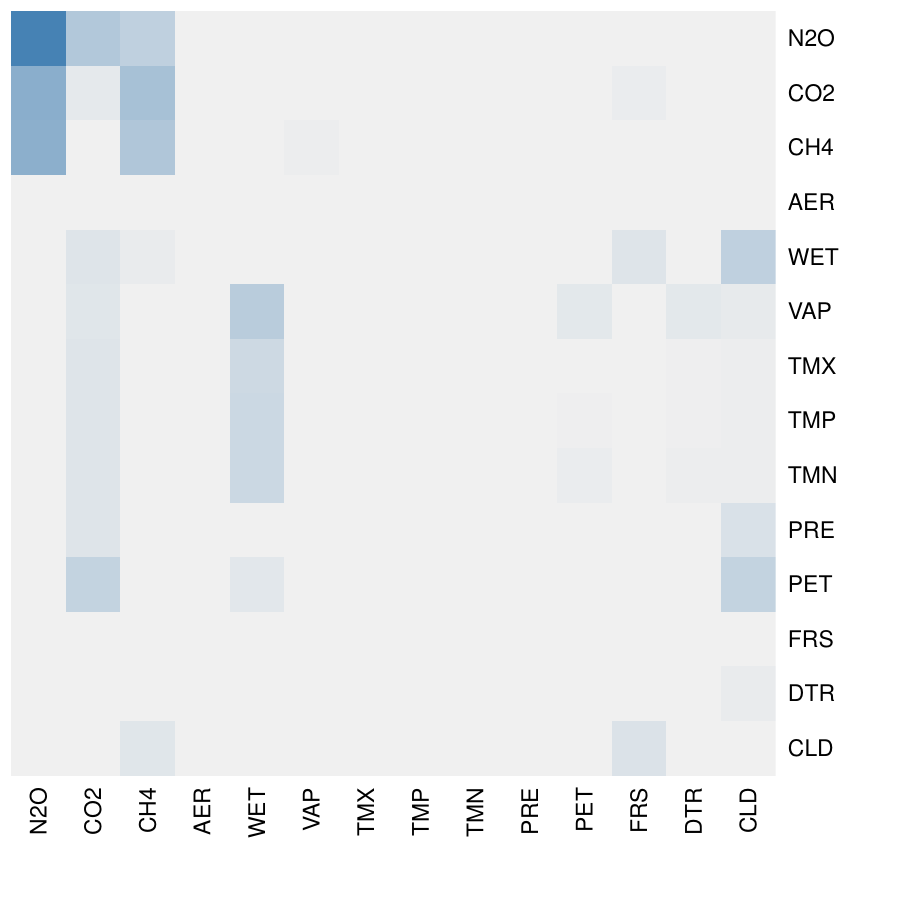}%
        \end{multicols}
        \begin{multicols}{3}
            \includegraphics[width=\linewidth]{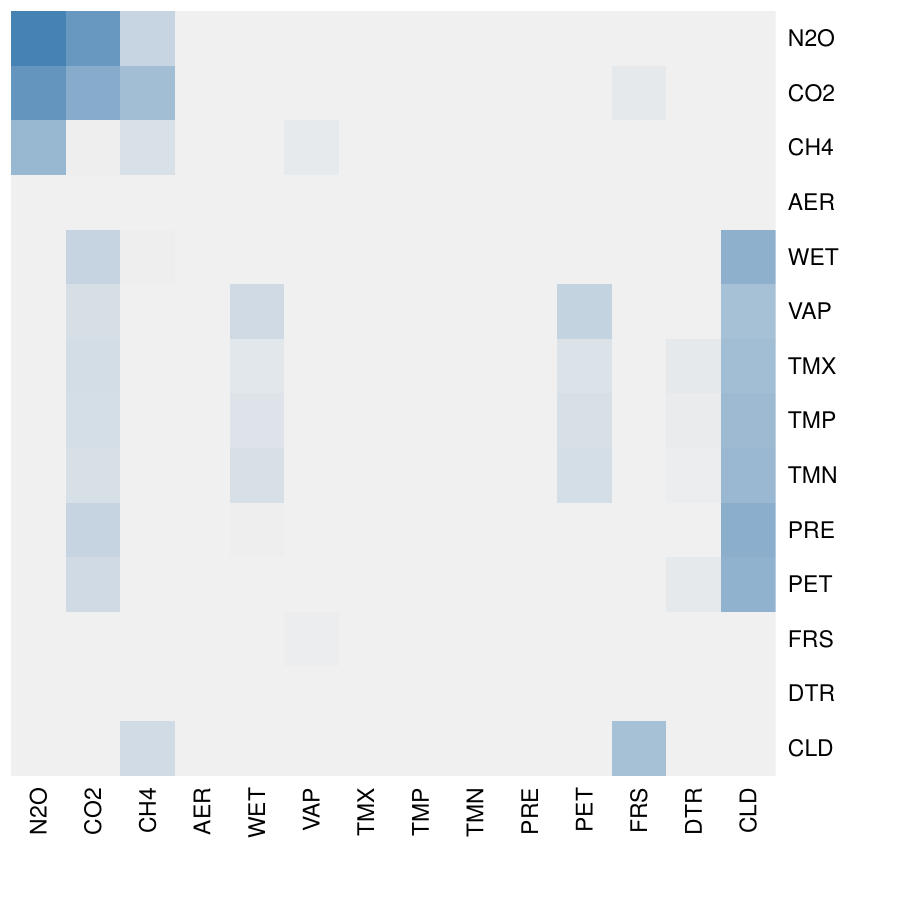}%
            \includegraphics[width=\linewidth]{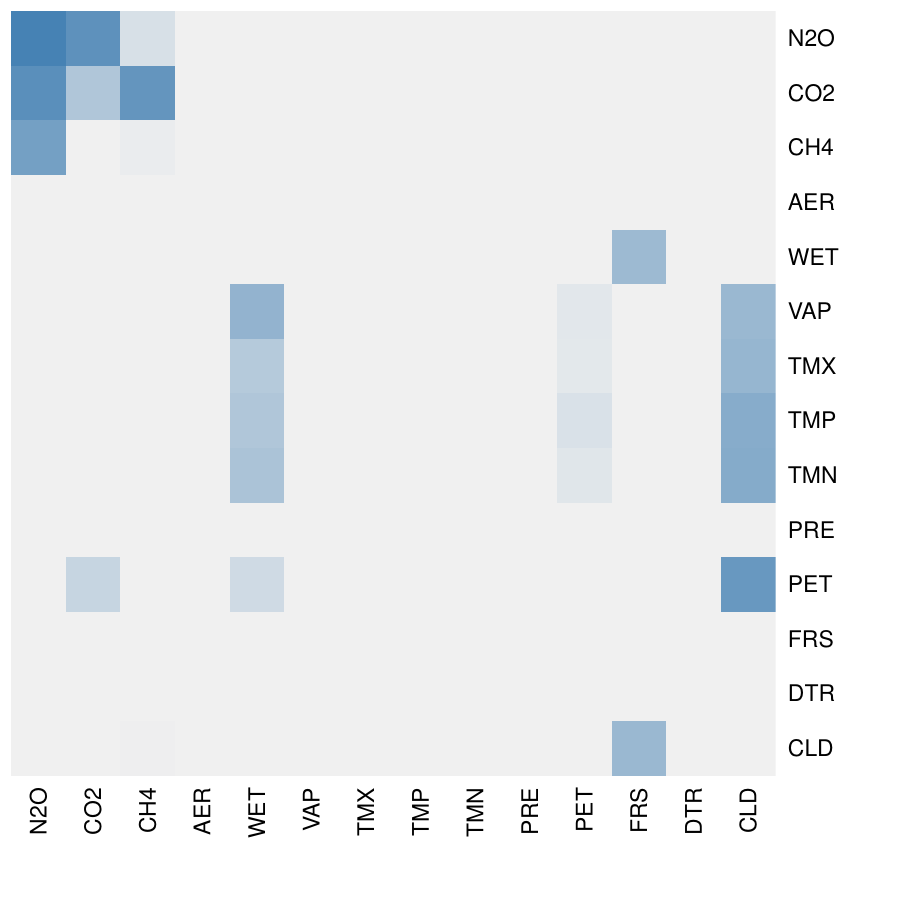}%
            \includegraphics[width=\linewidth]{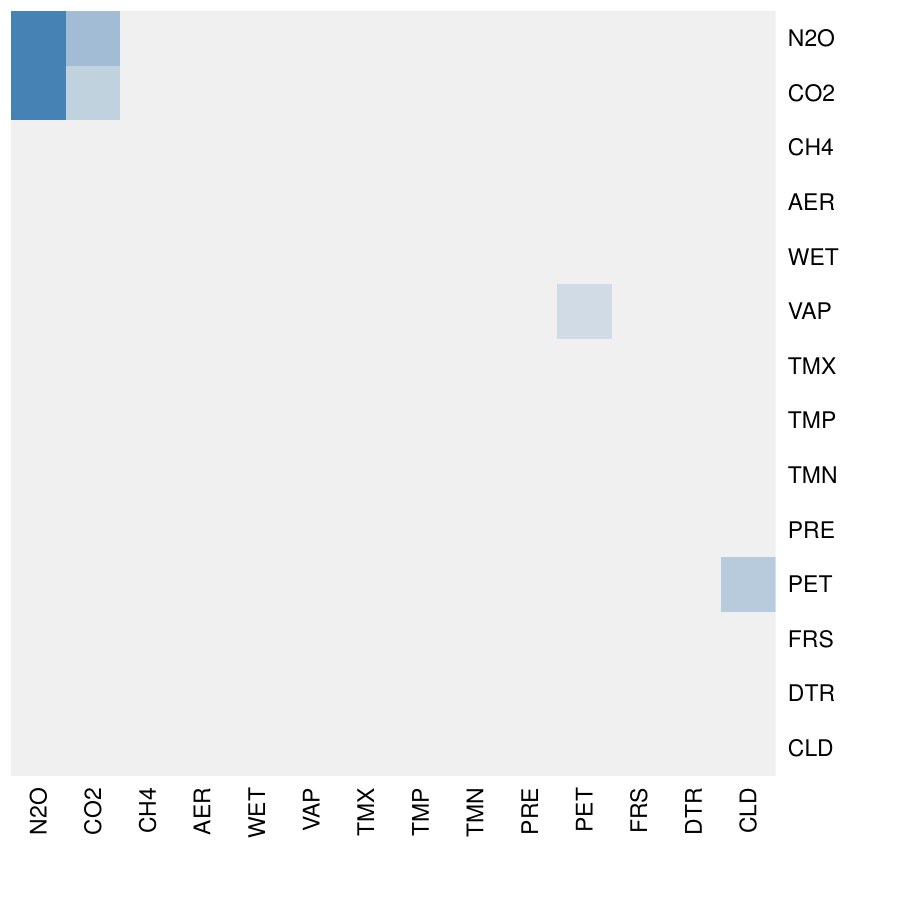}%
        \end{multicols}
    \vspace{-25pt}
    \caption{Estimated variable dependence networks at all locations and five distinct climate zones. Top left to bottom right: All locations; Humid continental (warm summer); Humid continental (hot summer); Semi-arid steppe; Humid subtropical; Mid-latitude desert.}
    \label{fig:heatmaps}
\end{figure*}

The estimated variable-dependence networks and heatmaps in Figures~\ref{fig:networks} and \ref{fig:heatmaps} reveal several interpretable patterns. Greenhouse-gas variables, including nitrous oxide ($\text{N}_2\text{O}$), carbon dioxide ($\text{CO}_2$), and methane ($\text{CH}_4$), exhibit strong mutual dependence across the selected locations. Temperature-related variables, including TMP, TMX, and TMN, are connected with rain days (WET), cloud cover (CLD), and diurnal temperature range (DTR). The estimated network also suggests that $\text{CO}_2$ is associated with TMX, DTR, potential evapotranspiration (PET), and precipitation (PRE), which is consistent with prior climate studies \citep{van2015causal}. The water-cycle-related variables VAP, PET, WET, and CLD also form a connected subnetwork.

The zone-specific networks display both common and region-specific patterns. The two humid continental zones have similar structures: temperature variables are connected with DTR and $\text{CO}_2$, while TMN and TMP are also linked to PET depending on the zone. In the humid continental zone with hot summer, temperature variables are additionally connected with CLD and WET. The semi-arid steppe zone shares several features with the humid continental zone with hot summer, but also exhibits distinctive links such as (PRE, WET), (PET, DTR), and (VAP, PRE), which reflect regional water-cycle dynamics. The humid subtropical zone continues to show the importance of temperature variables, which connect with PET, CLD, and WET. The mid-latitude desert zone has the sparsest dependence network, although the VAP--PET--CLD structure remains visible and $\text{CO}_2$ and $\text{N}_2\text{O}$ appear more prominent than $\text{CH}_4$.

\begin{figure}[!ht]
    \centering
        \includegraphics[scale=.36]{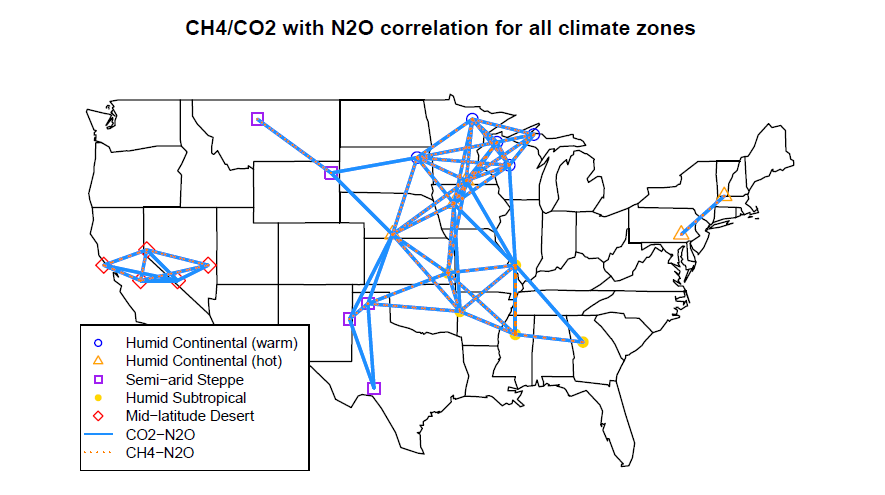}%
        \includegraphics[scale=.36]{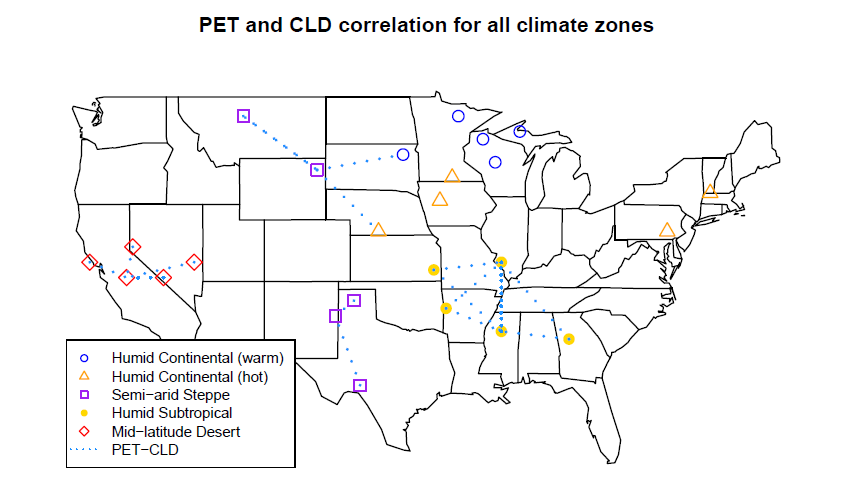}
    \vspace{-10pt}
    \caption{Left: Spatial dependencies of $\text{CH}_4/\text{CO}_2$ with $\text{N}_2\text{O}$ over all selected locations; Right: Spatial dependencies of PET and CLD for all climate zones.}
    \label{fig:4}
\end{figure}

Next, we present selected variable pairs to illustrate their estimated spatial structure. The left panel of Figure~\ref{fig:4} shows the spatial structures for $(\text{CO}_2,\text{N}_2\text{O})$ and $(\text{CH}_4,\text{N}_2\text{O})$, two greenhouse-gas relationships that are important in climate research \citep{tian2015global}. The two estimated spatial structures are similar and mainly follow short-distance connections. The $(\text{CO}_2,\text{N}_2\text{O})$ pair covers a slightly broader spatial range than the $(\text{CH}_4,\text{N}_2\text{O})$ pair, especially in the southern semi-arid steppe and mid-latitude desert regions. Compared with the pre-specified graph $\mathcal{J}_0$, both pairs recover a large fraction of the candidate spatial structure.

The right panel of Figure~\ref{fig:4} shows the estimated spatial structure for (PET, CLD), a variable pair known to be related \citep{naval2012potential}. Its estimated spatial structure is sparser than the pre-specified spatial structure, suggesting that only a subset of the candidate local spatial links is selected by the proposed estimator.

\subsection{Estimation Results for VAR(2) Model}
\begin{figure*}[!ht]
    \centering
        \begin{multicols}{3}
            \includegraphics[width=\linewidth]{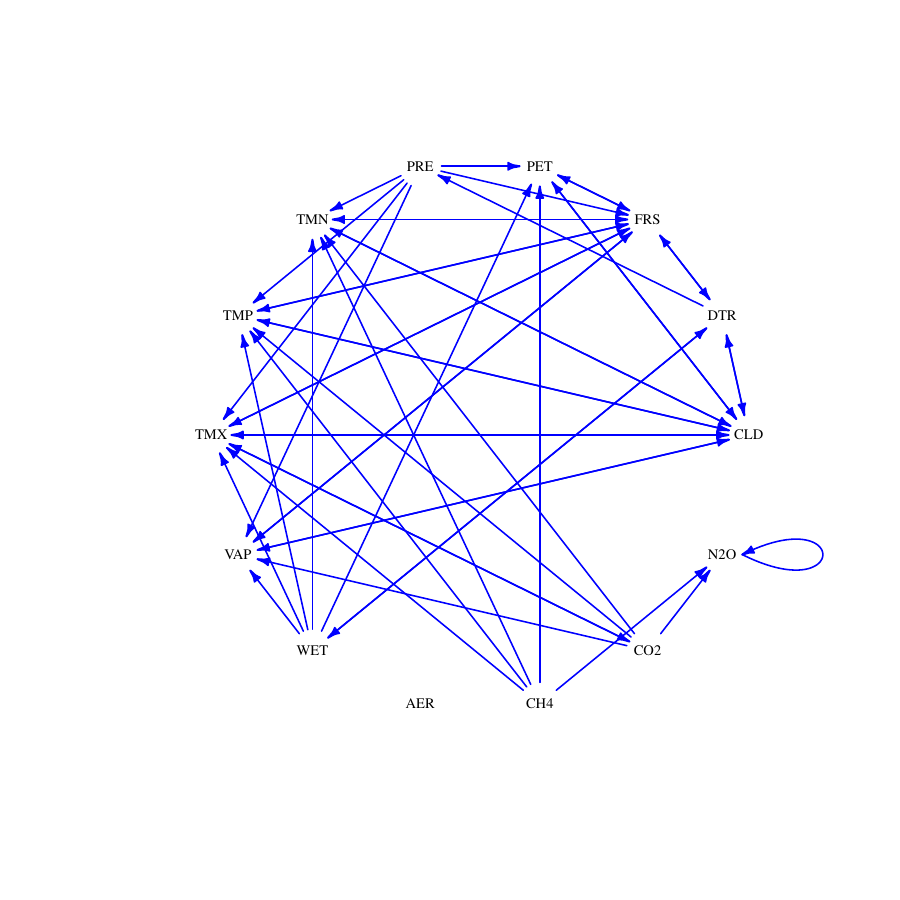}%
            \includegraphics[width=\linewidth]{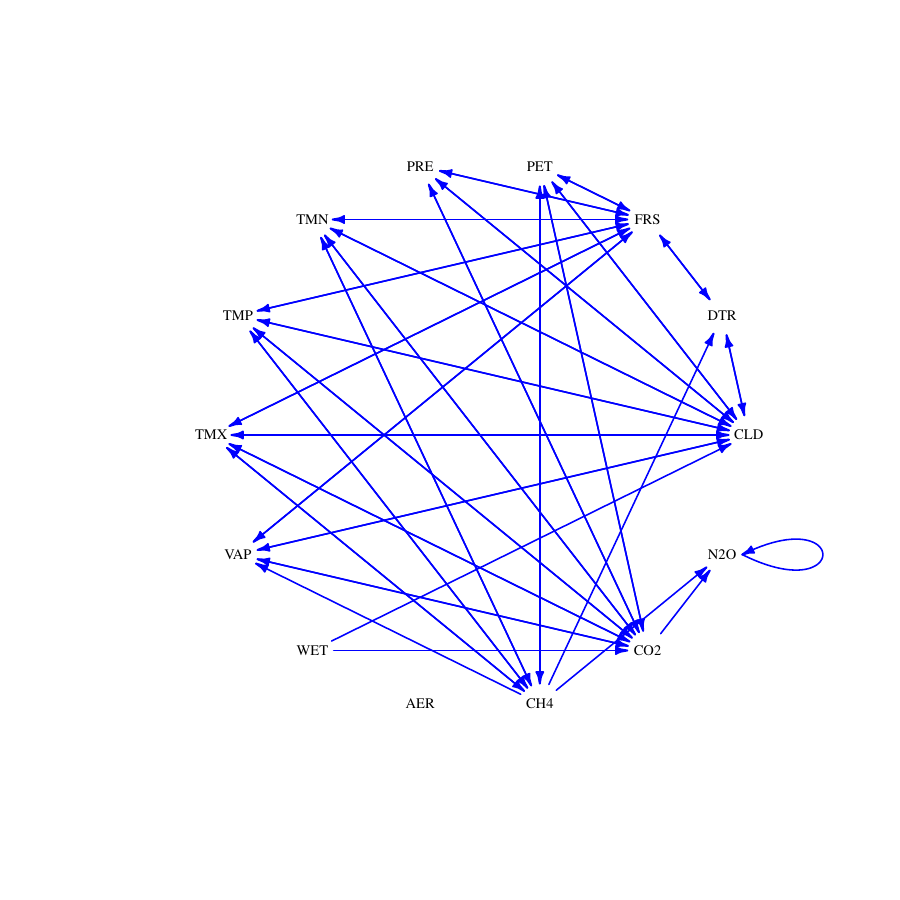}%
            \includegraphics[width=\linewidth]{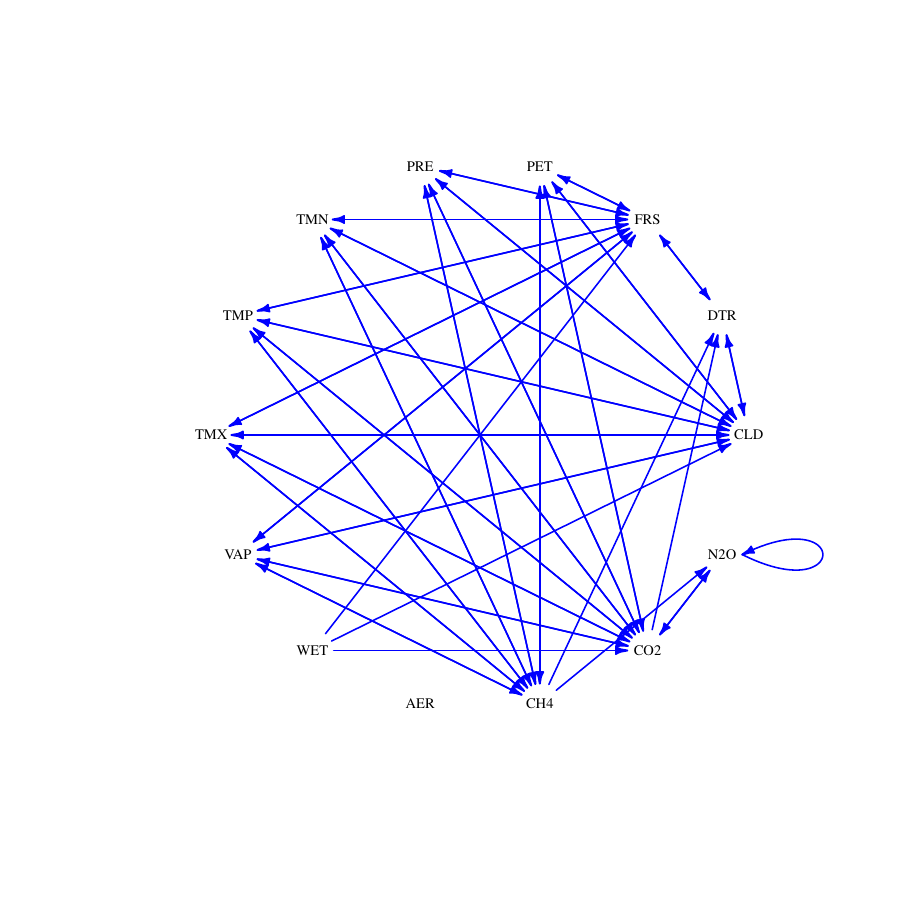}%
        \end{multicols}
        \begin{multicols}{3}
            \includegraphics[width=\linewidth]{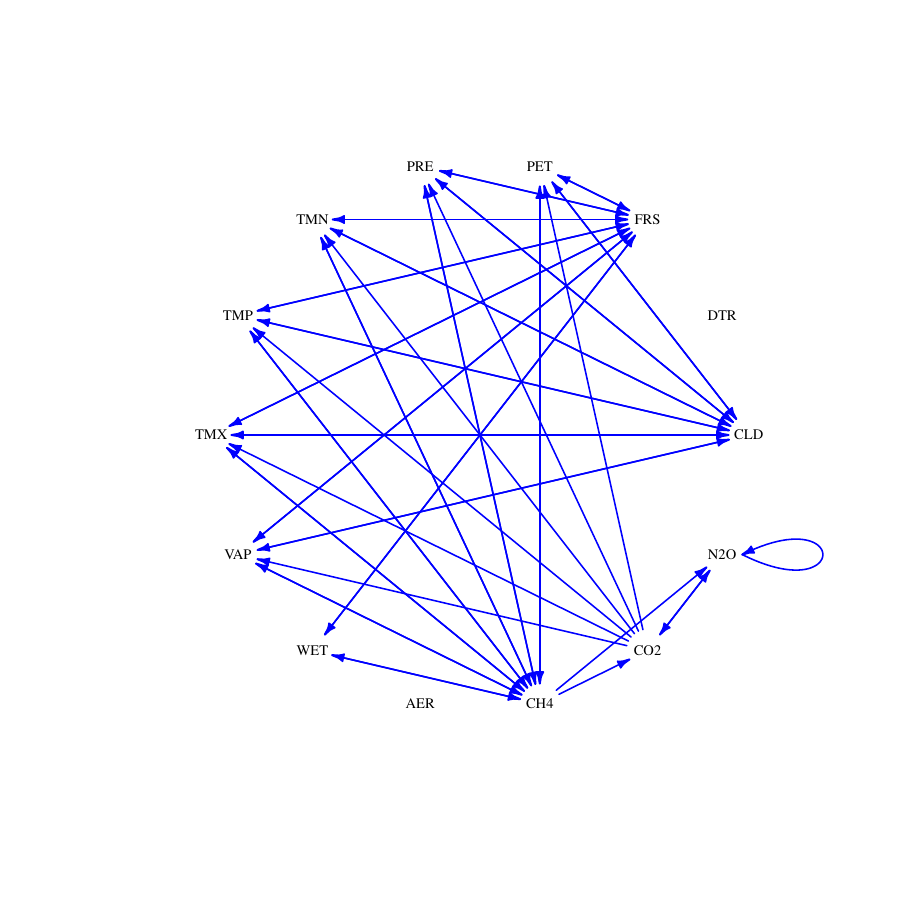}%
            \includegraphics[width=\linewidth]{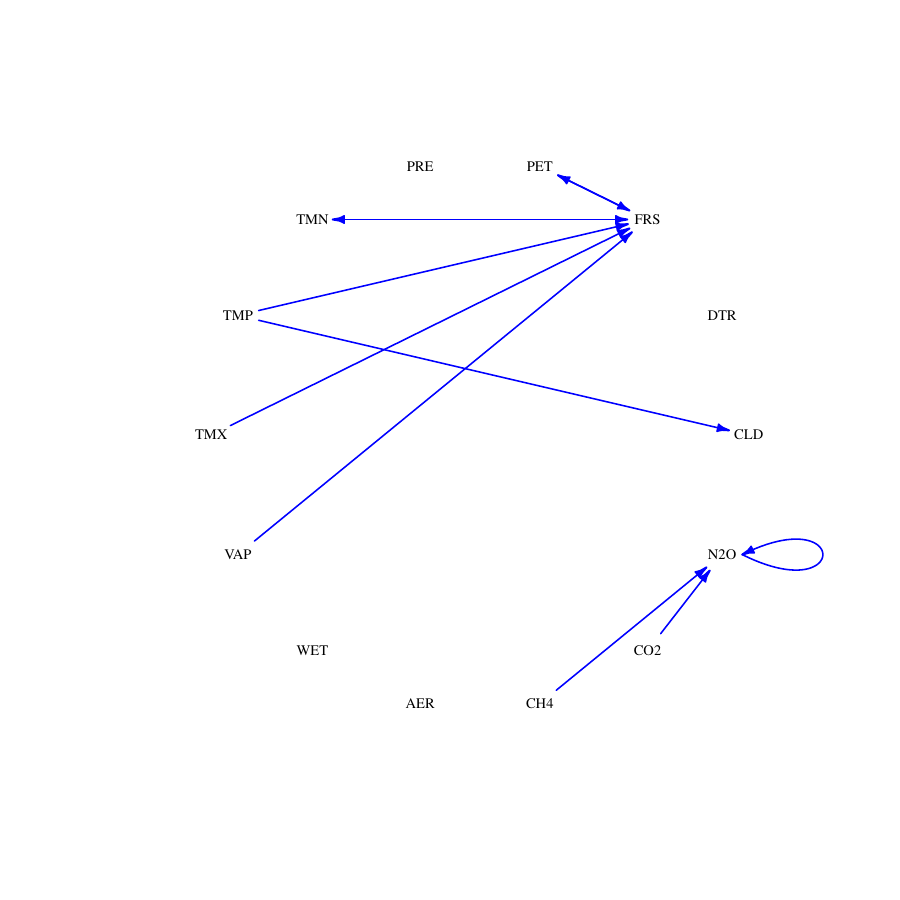}%
            \includegraphics[width=\linewidth]{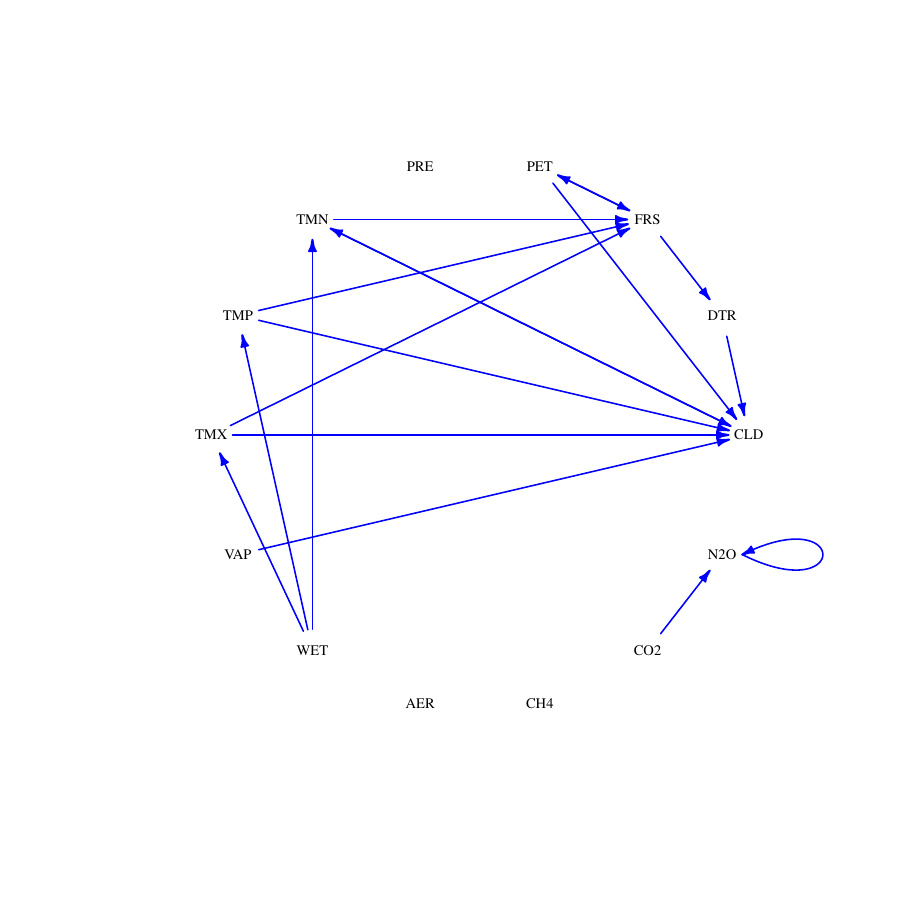}%
        \end{multicols}
    \vspace{-25pt}
    \caption{Estimated variable dependence networks at all locations and five distinct climate zones. Top left to bottom right: All locations; Humid continental (warm summer); Humid continental (hot summer); Semi-arid steppe; Humid subtropical; Mid-latitude desert.}
    \label{fig:network_lag1}
\end{figure*}

\begin{figure*}[!ht]
    \centering
        \begin{multicols}{3}
            \includegraphics[width=\linewidth]{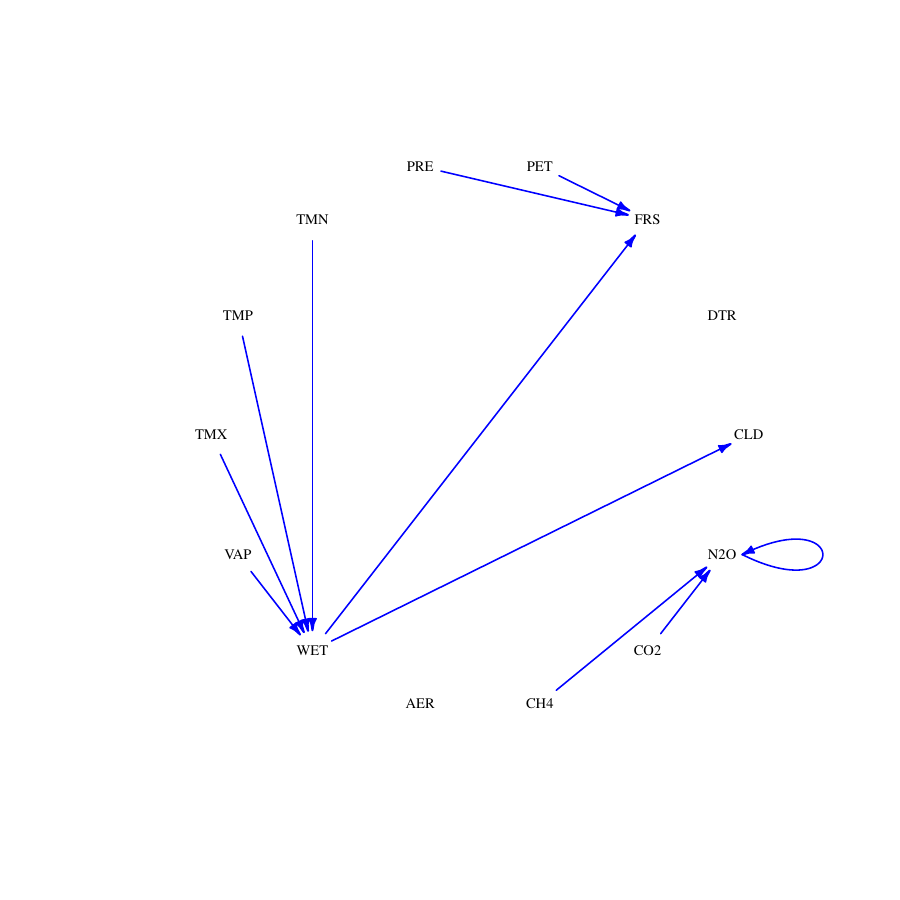}%
            \includegraphics[width=\linewidth]{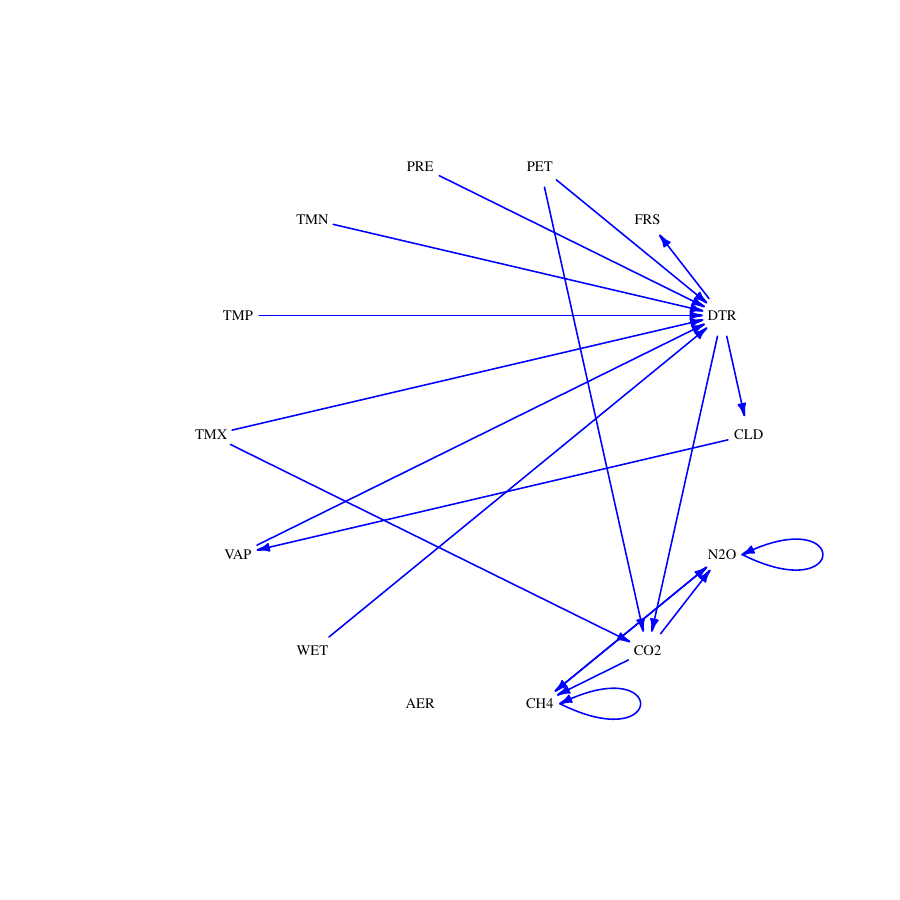}%
            \includegraphics[width=\linewidth]{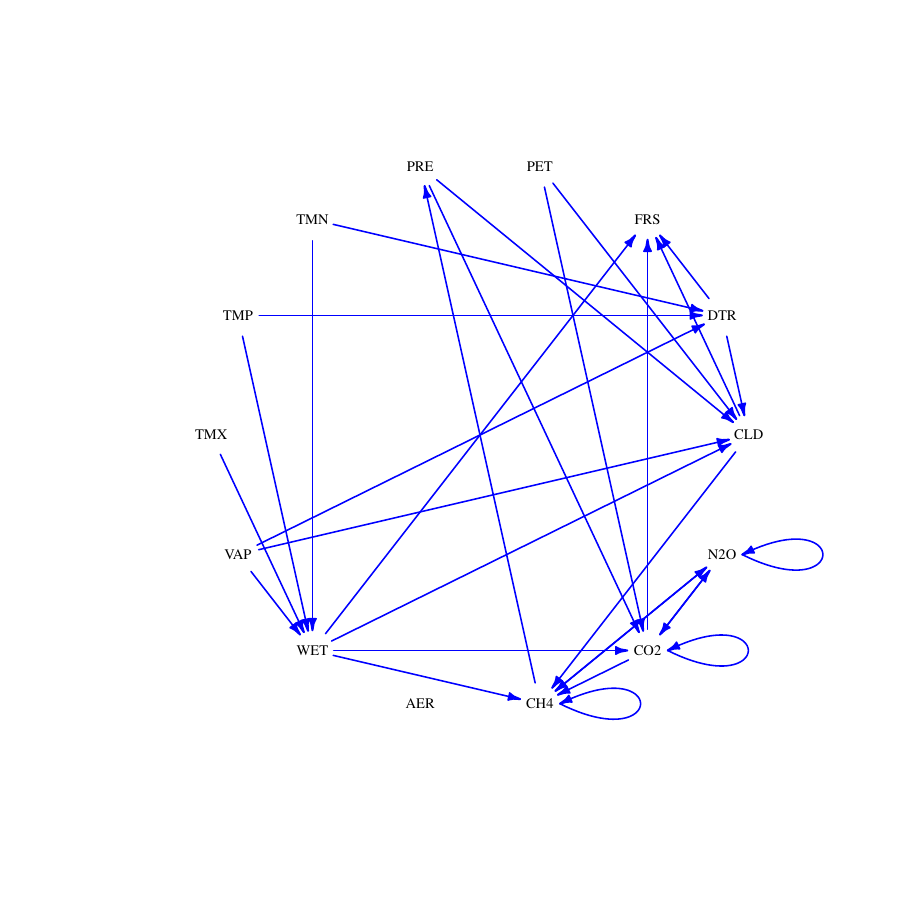}%
        \end{multicols}
        \begin{multicols}{3}
            \includegraphics[width=\linewidth]{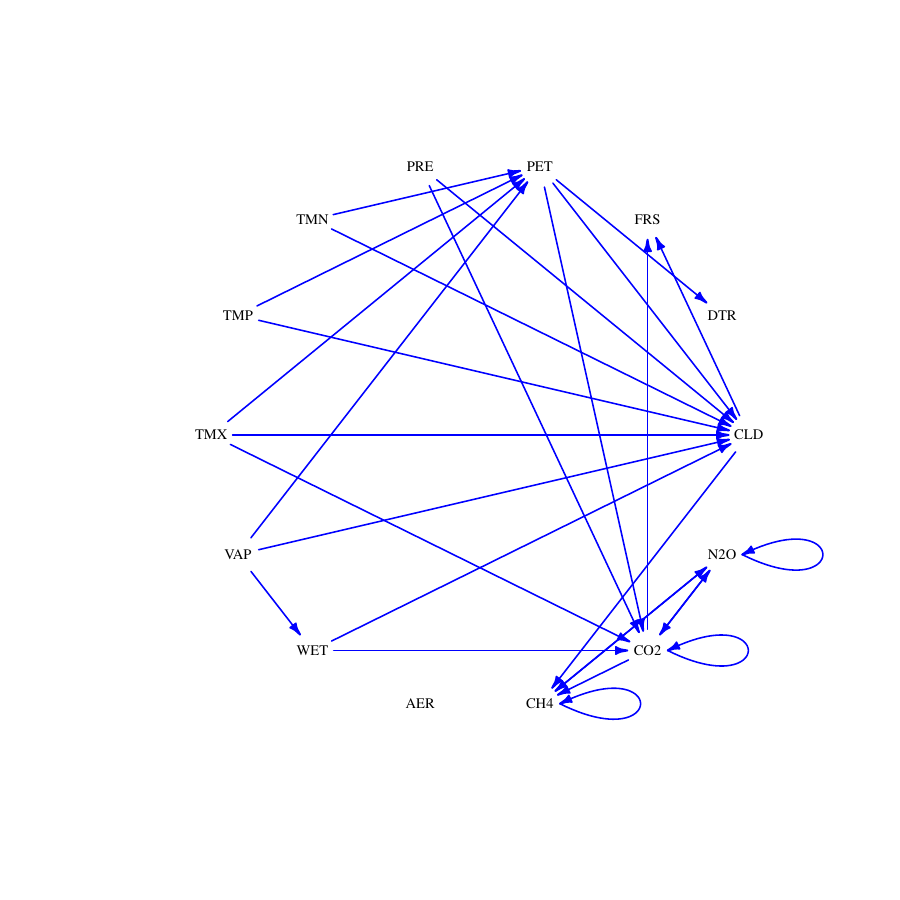}%
            \includegraphics[width=\linewidth]{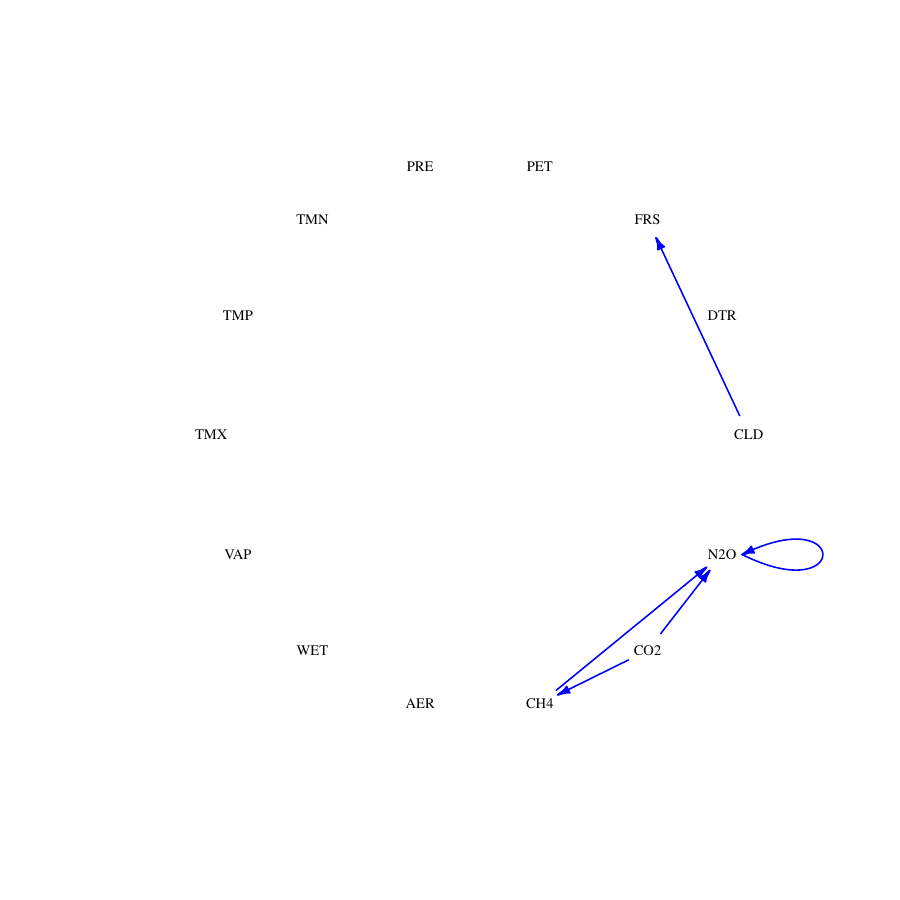}%
            \includegraphics[width=\linewidth]{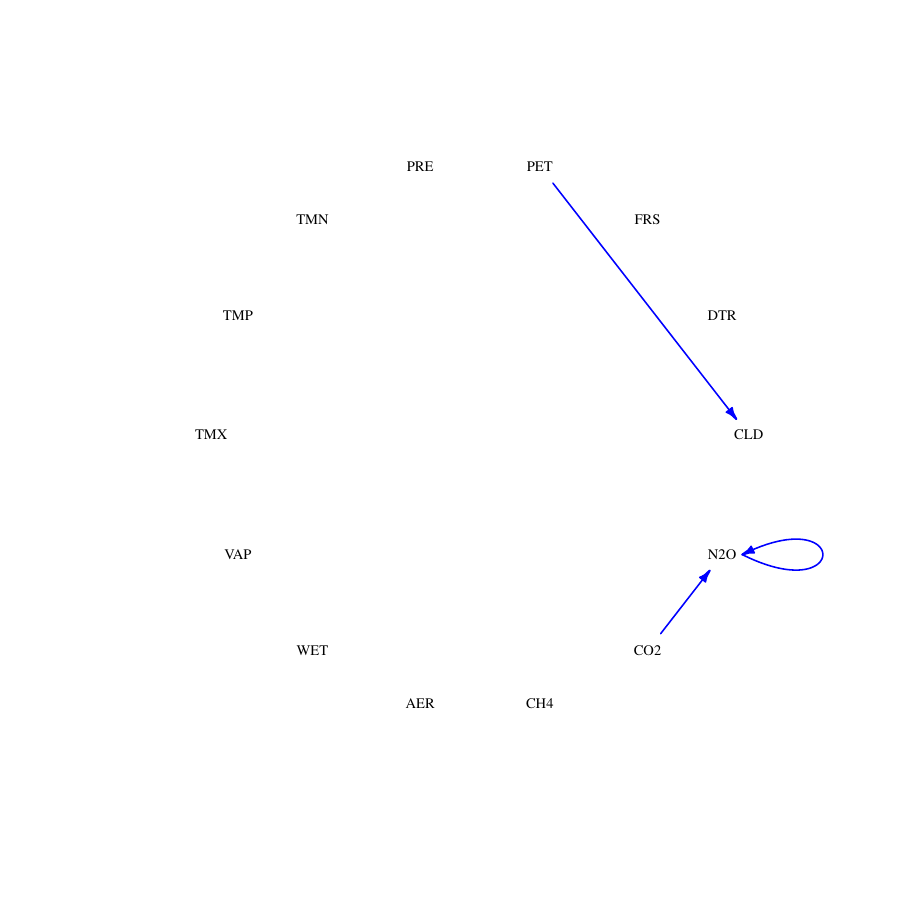}%
        \end{multicols}
    \vspace{-25pt}
    \caption{Estimated variable dependence networks at all locations and five distinct climate zones. Top left to bottom right: All locations; Humid continental (warm summer); Humid continental (hot summer); Semi-arid steppe; Humid subtropical; Mid-latitude desert.}
    \label{fig:network_lag2}
\end{figure*}

\begin{figure*}[!ht]
    \centering
        \begin{multicols}{3}
            \includegraphics[width=\linewidth]{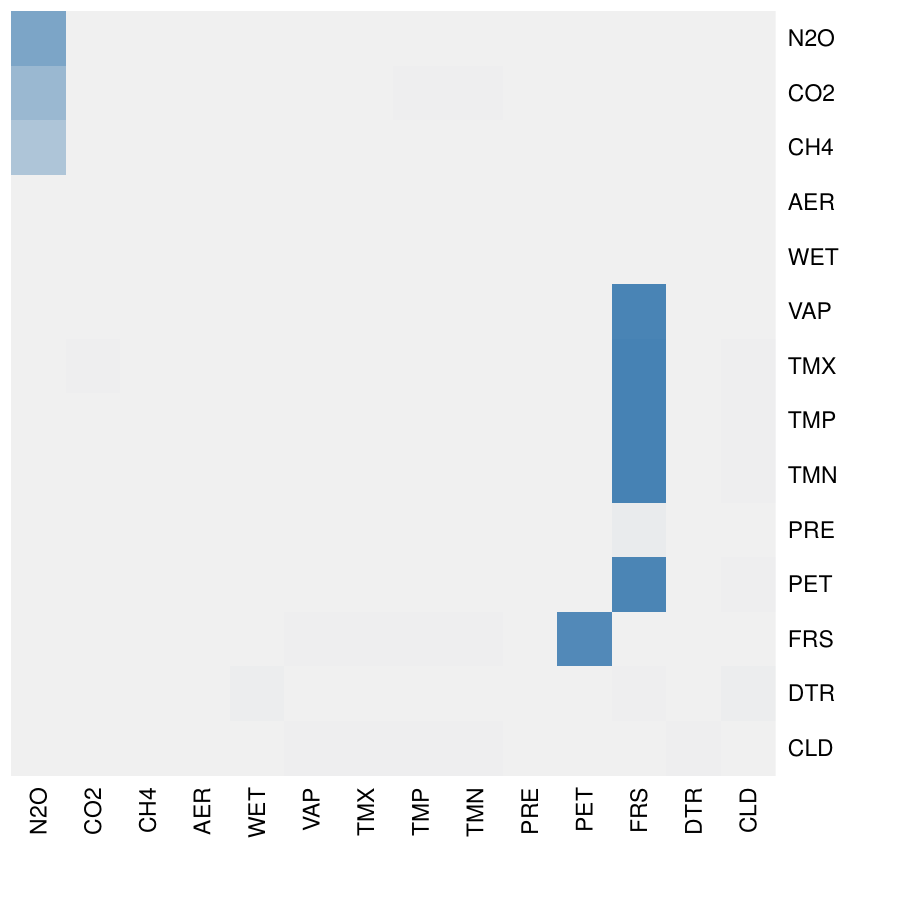}%
            \includegraphics[width=\linewidth]{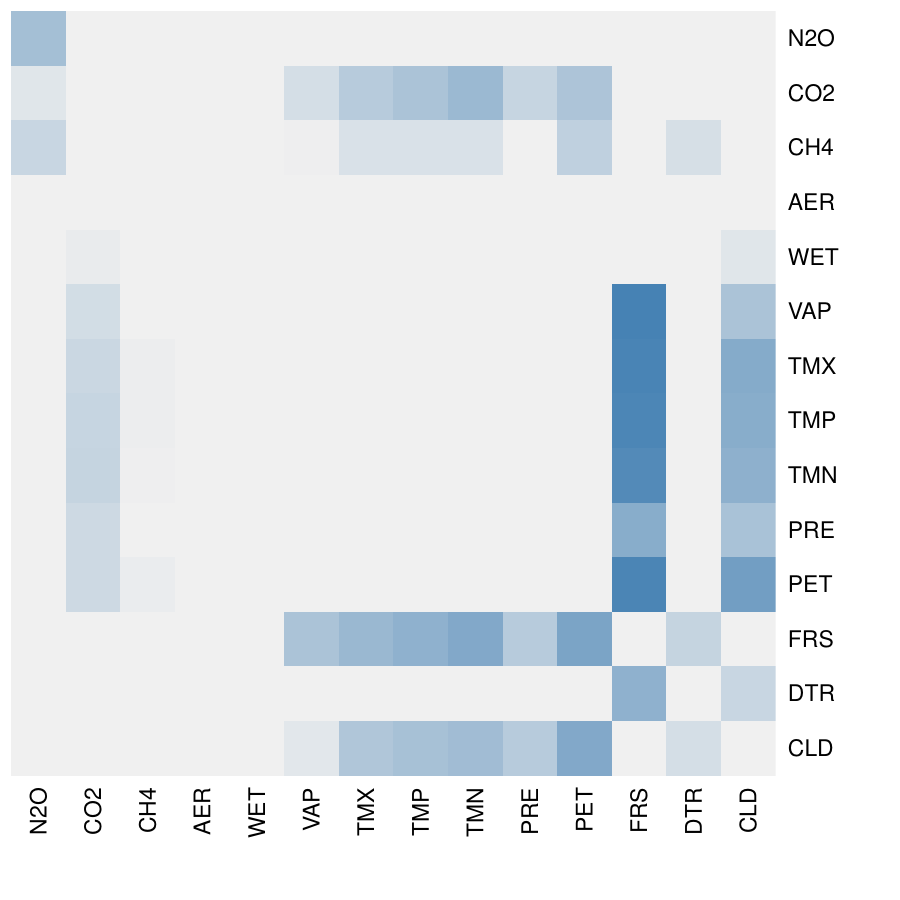}%
            \includegraphics[width=\linewidth]{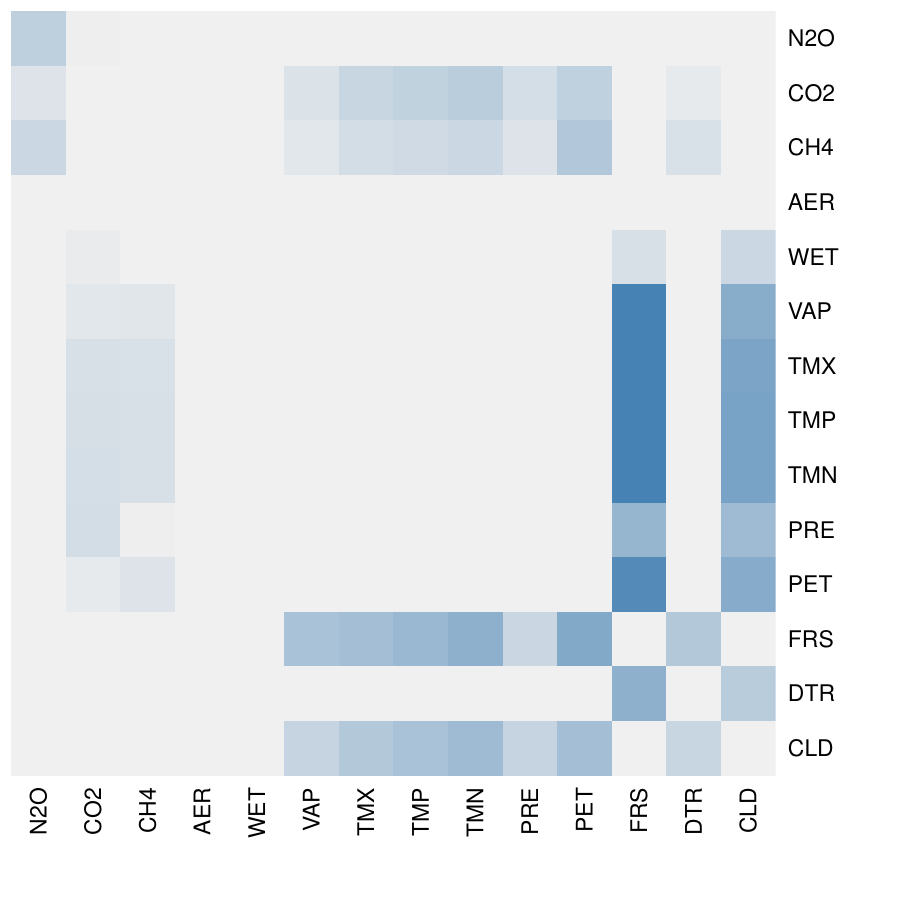}%
        \end{multicols}
        \begin{multicols}{3}
            \includegraphics[width=\linewidth]{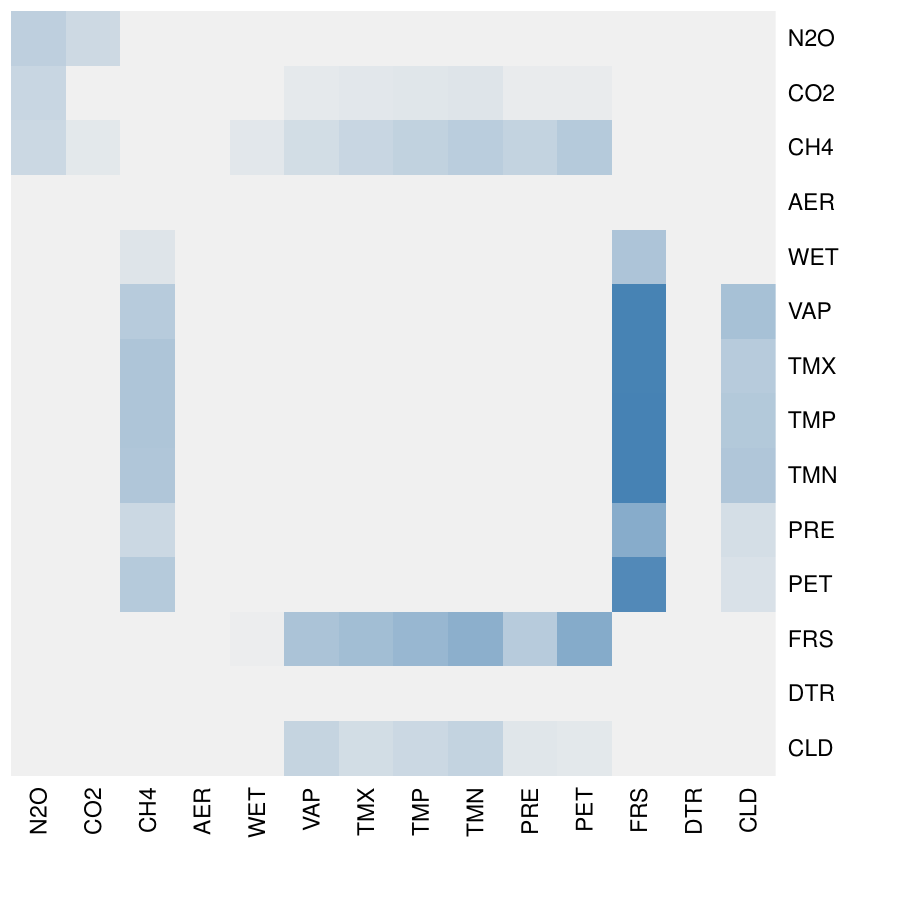}%
            \includegraphics[width=\linewidth]{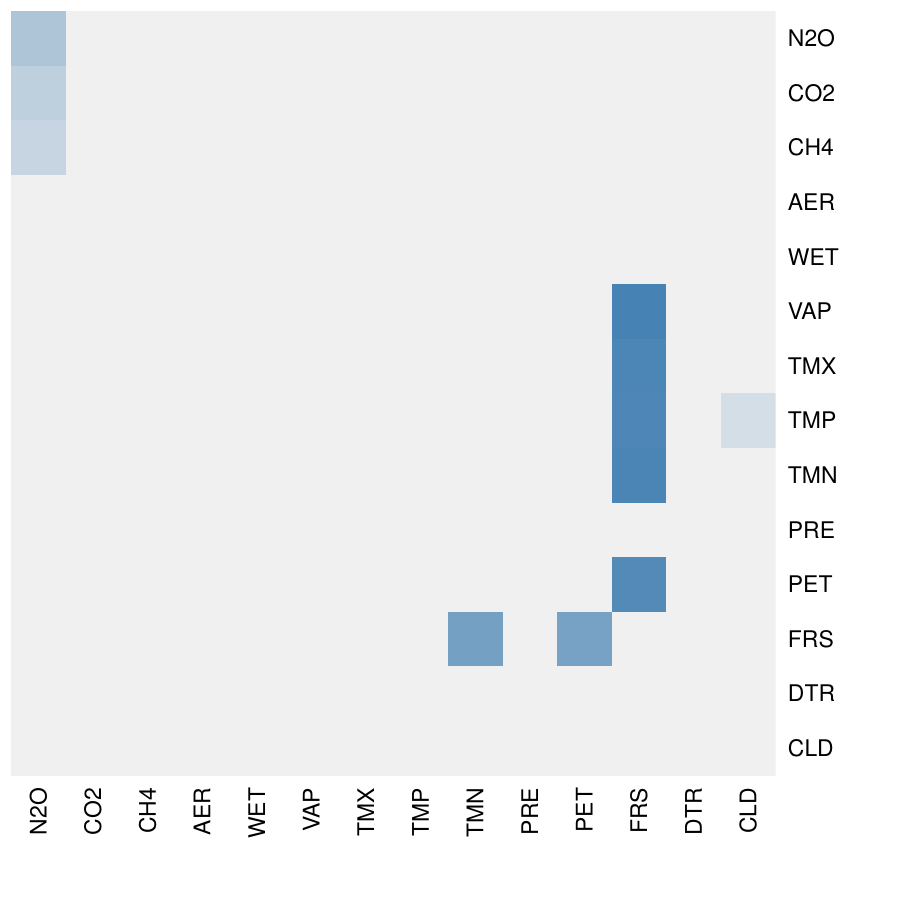}%
            \includegraphics[width=\linewidth]{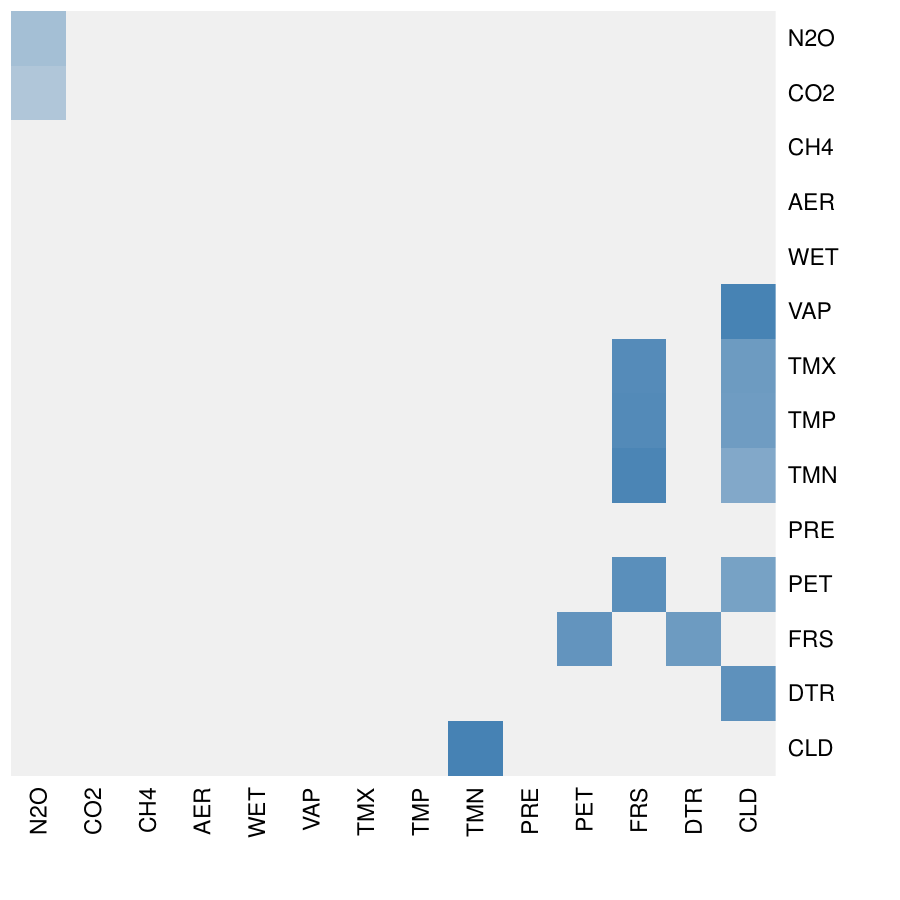}%
        \end{multicols}
    \vspace{-25pt}
    \caption{Estimated variable dependence networks at all locations and five distinct climate zones. Top left to bottom right: All locations; Humid continental (warm summer); Humid continental (hot summer); Semi-arid steppe; Humid subtropical; Mid-latitude desert.}
    \label{fig:heatmaps_lag1}
\end{figure*}

\begin{figure*}[!ht]
    \centering
        \begin{multicols}{3}
            \includegraphics[width=\linewidth]{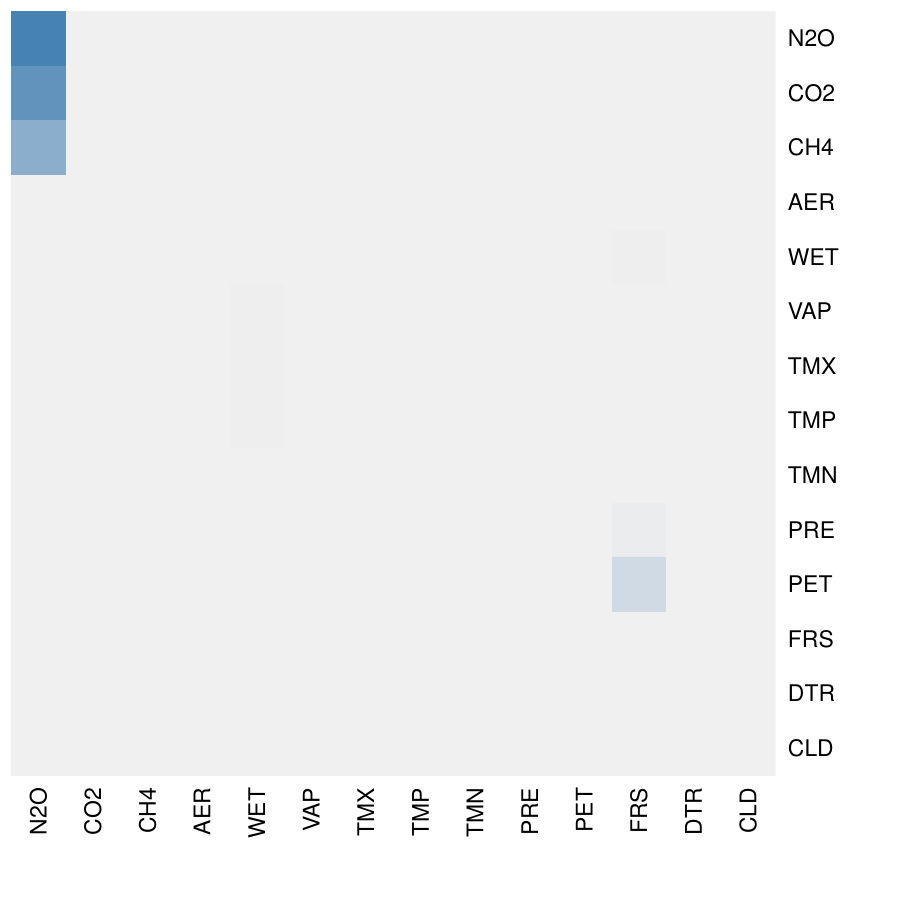}%
            \includegraphics[width=\linewidth]{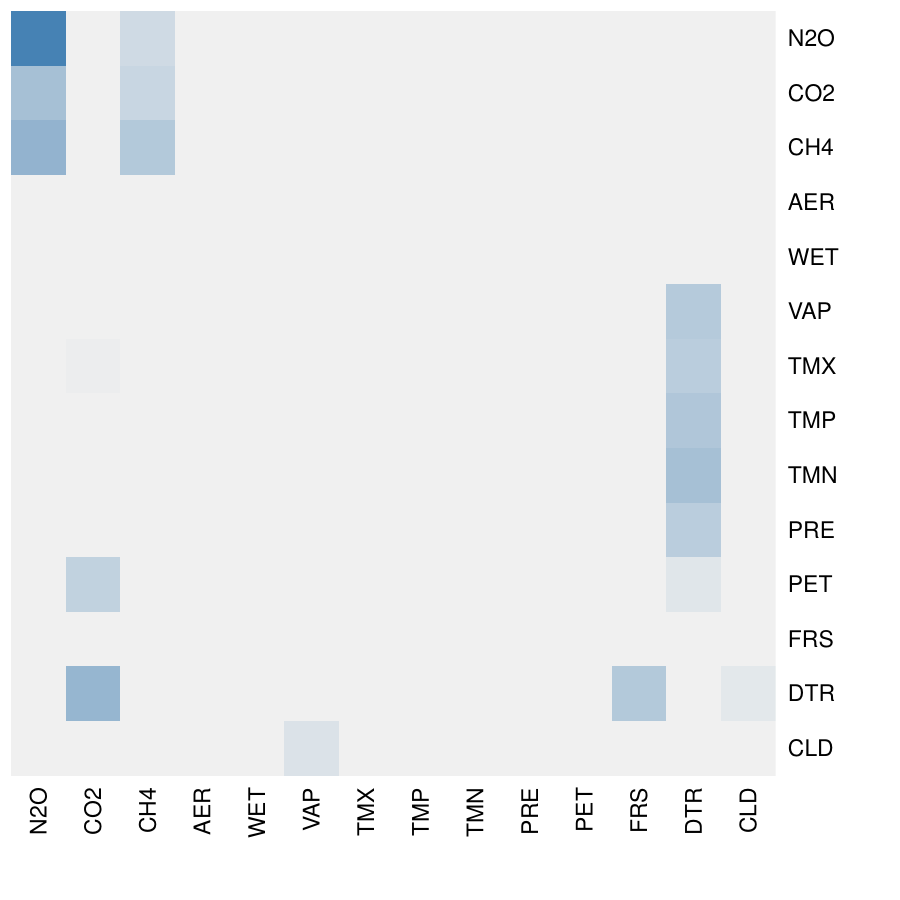}%
            \includegraphics[width=\linewidth]{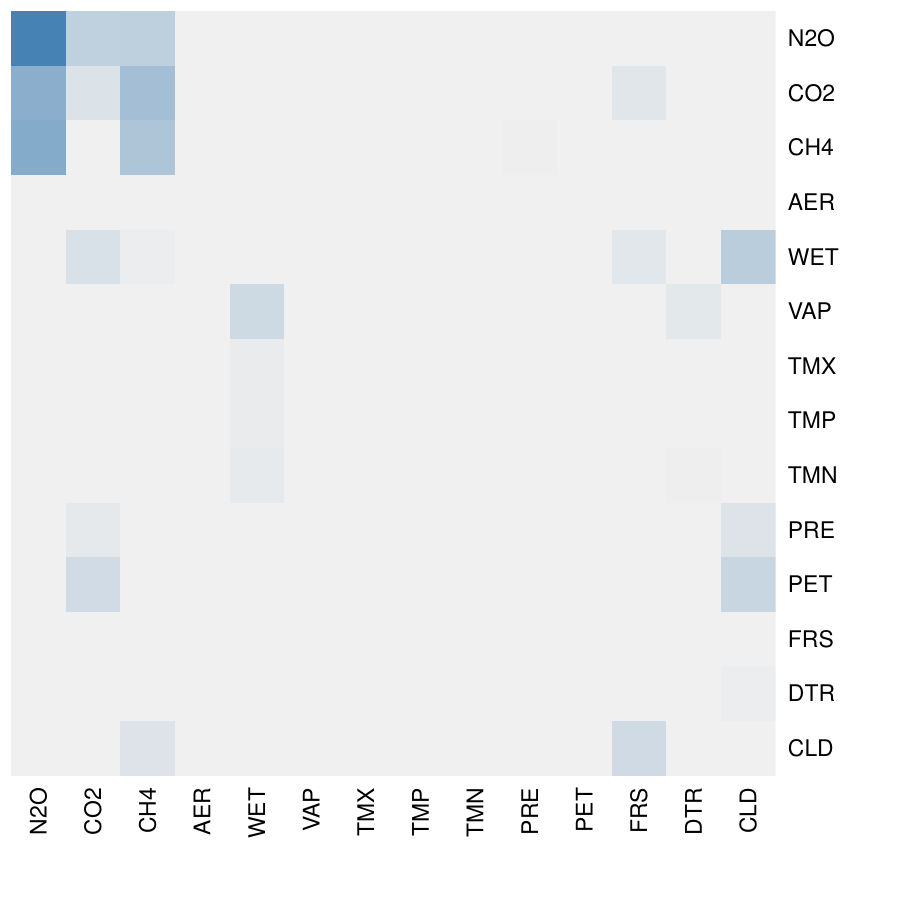}%
        \end{multicols}
        \begin{multicols}{3}
            \includegraphics[width=\linewidth]{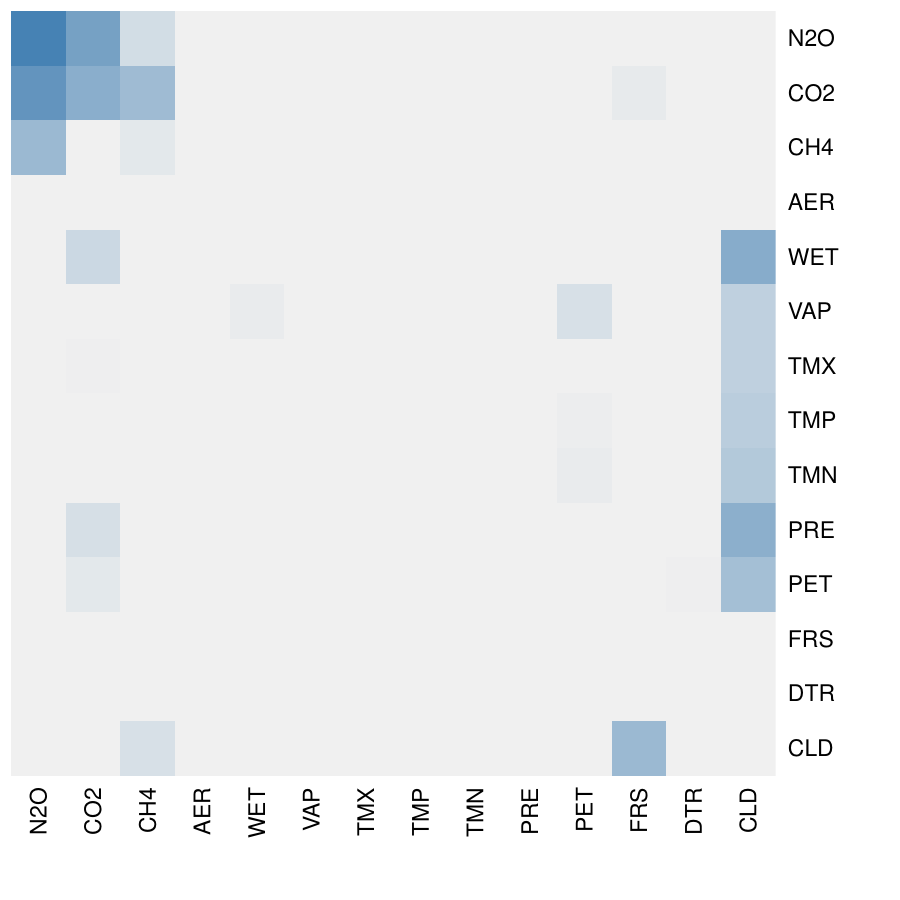}%
            \includegraphics[width=\linewidth]{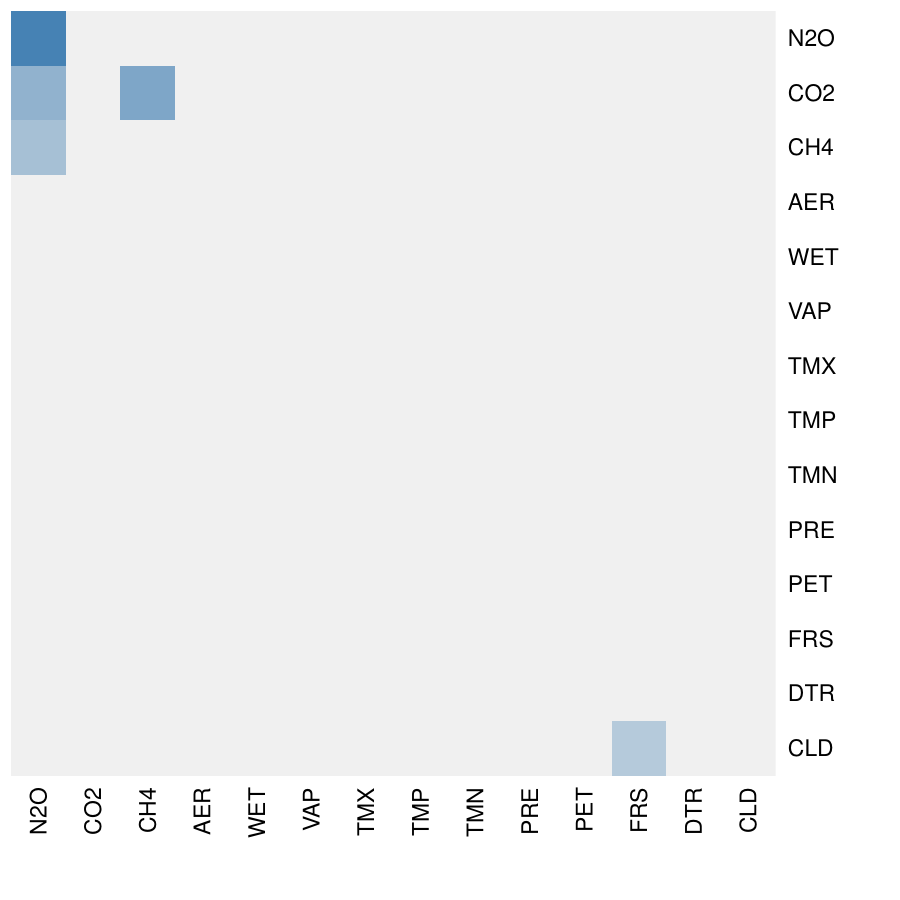}%
            \includegraphics[width=\linewidth]{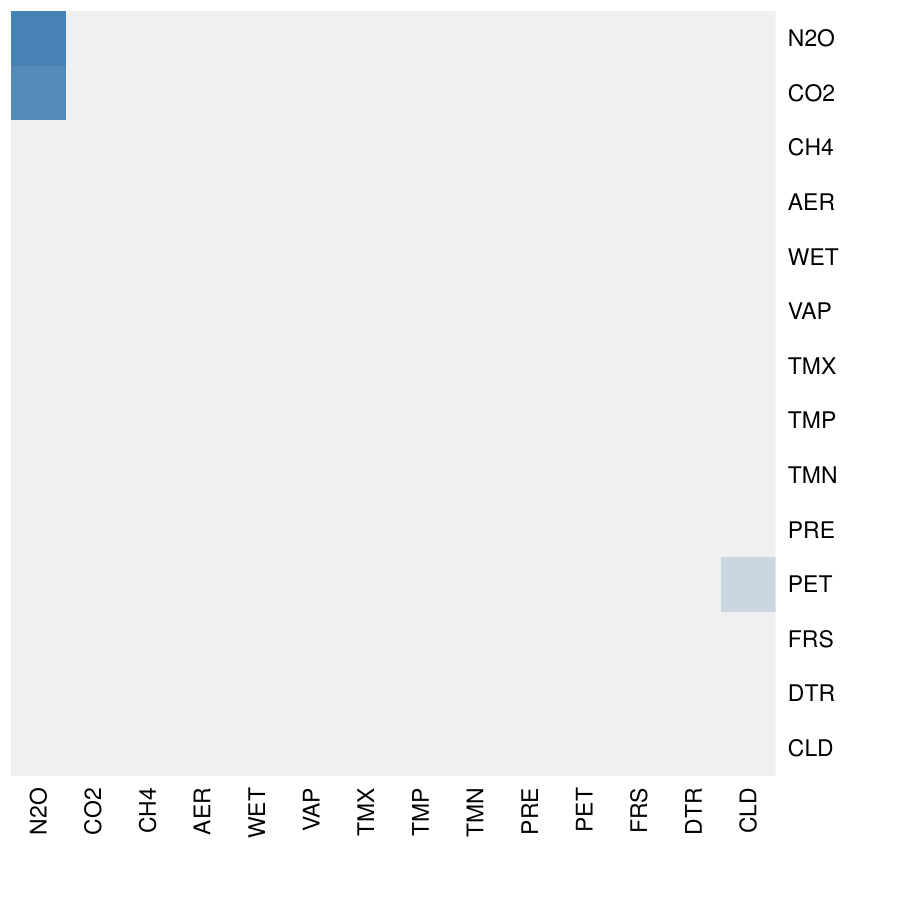}%
        \end{multicols}
    \vspace{-25pt}
    \caption{Estimated variable dependence networks at all locations and five distinct climate zones. Top left to bottom right: All locations; Humid continental (warm summer); Humid continental (hot summer); Semi-arid steppe; Humid subtropical; Mid-latitude desert.}
    \label{fig:heatmaps_lag2}
\end{figure*}

In this section, we provide the estimation results for applying VAR($d$) model with a time lag $d>1$ to the environmental data set. According to the aggregated data and scientific rationalities, we employ a VAR(2) model for estimating the variables dependence networks and further investigation. Figure~\ref{fig:network_lag1} and Figure~\ref{fig:network_lag2} illustrate the dependence networks. Figure~\ref{fig:heatmaps_lag1} and Figure~\ref{fig:heatmaps_lag2} show the corresponding heatmaps.

The VAR(2) estimates show that the lag-2 dependence networks are sparser and have different structure than the lag-1 networks. The estimated magnitudes are generally weaker at lag 2 than at lag 1, as shown in Figures~\ref{fig:network_lag1} and \ref{fig:network_lag2}. Temperature-related variables remain important at both lags and continue to connect with CLD and WET. Greenhouse-gas variables, however, display different lag-specific behavior: $\text{CO}_2$ and $\text{CH}_4$ do not show strong self-dependence at lag 1 but exhibit clearer lag-2 dependence. Water-cycle variables remain connected in the lag-1 network.

Methane ($\text{CH}_4$) also shows lag-specific dependence patterns. At lag 1, $\text{CH}_4$ is linked with temperature variables and PET; TMP, TMX, and TMN are also linked with DTR, CLD, and FRS in some regions. At lag 2, TMP, TMX, and TMN continue to influence WET, DTR, and CLD, respectively. The greenhouse-gas subnetwork is less dense than in the VAR(1) model: although $\text{CH}_4$ and $\text{CO}_2$ remain connected with $\text{N}_2\text{O}$, the direct $(\text{CH}_4,\text{CO}_2)$ pair is no longer selected.


\begin{figure}[!ht]
    \centering
    \includegraphics[scale=.55]{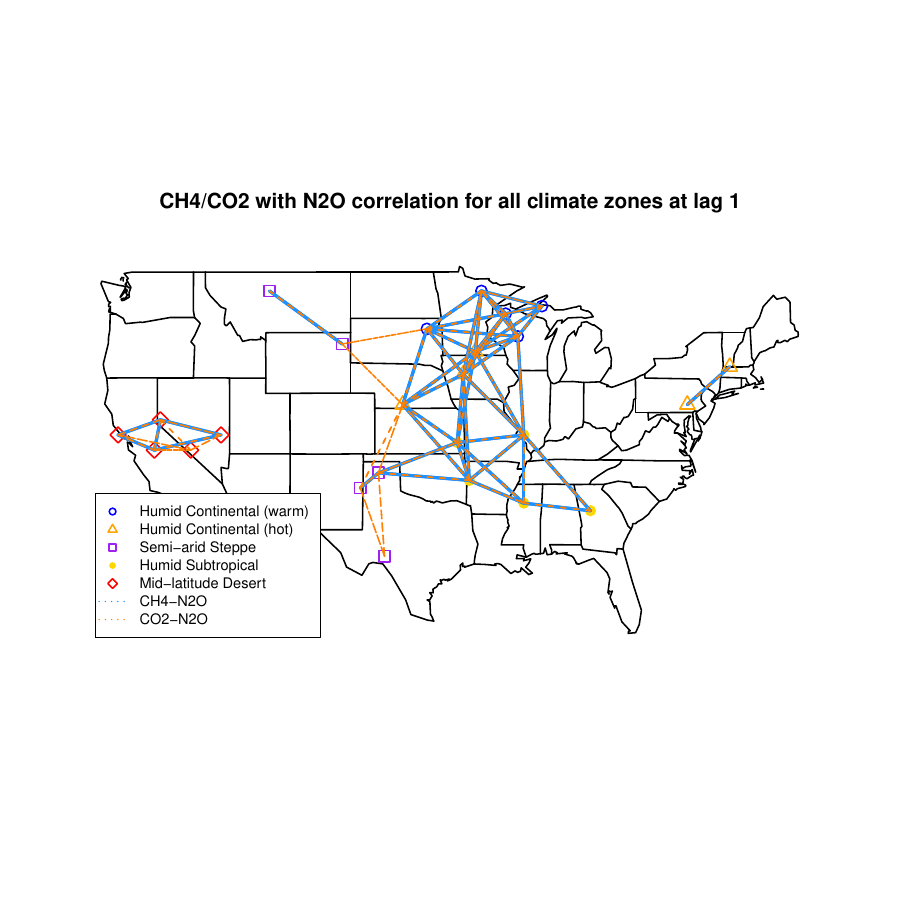}%
    \includegraphics[scale=.55]{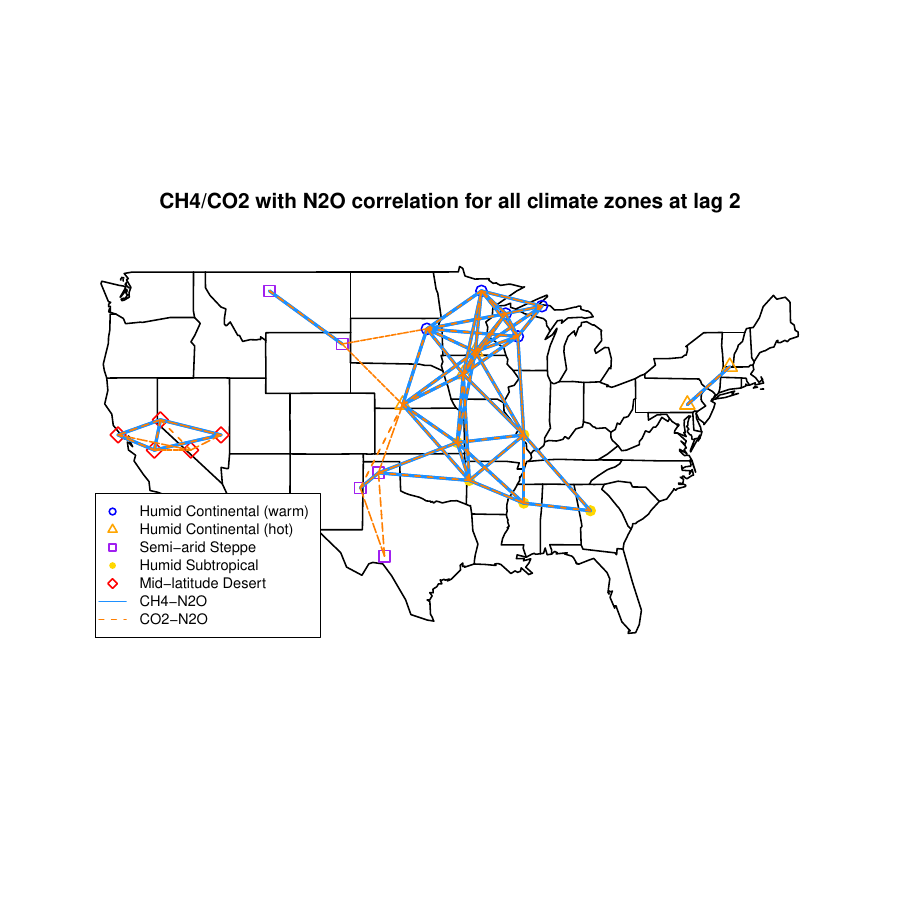}
    \vspace{-70pt}
    \caption{Estimated spatial structure for methane/carbon dioxide with nitrous oxide at different time lags.}
    \label{figure:8}
\end{figure}

\begin{figure}[!ht]
    \centering
    \includegraphics[scale=.55]{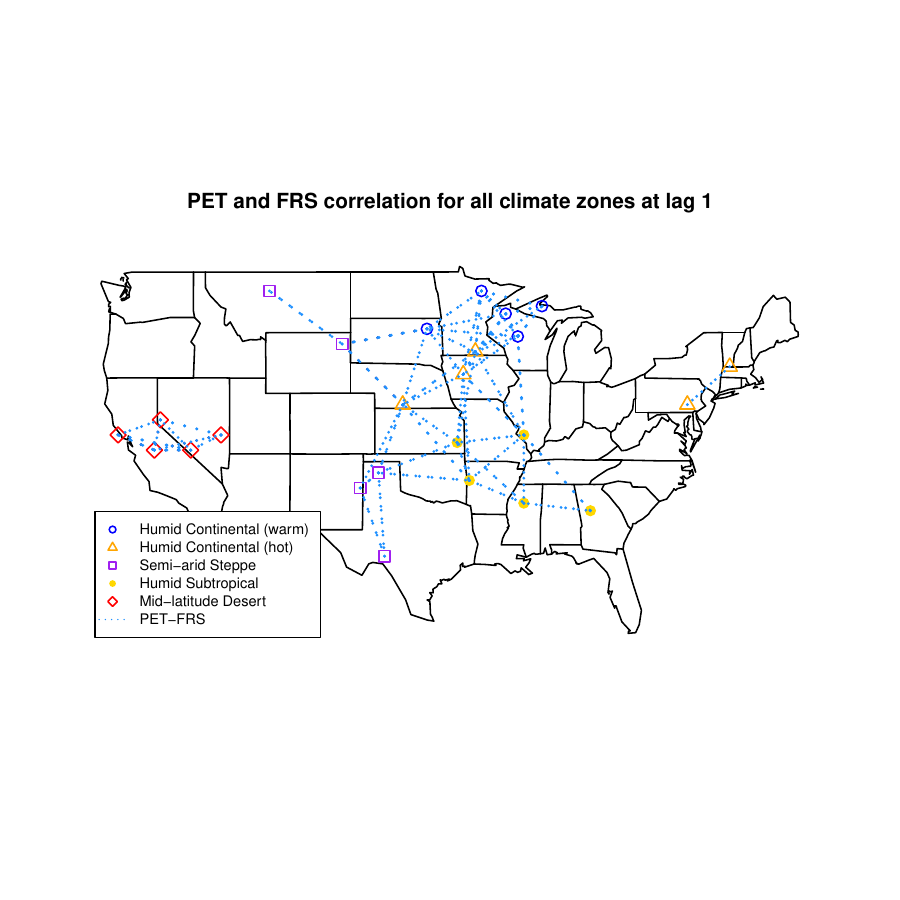}%
    \includegraphics[scale=.55]{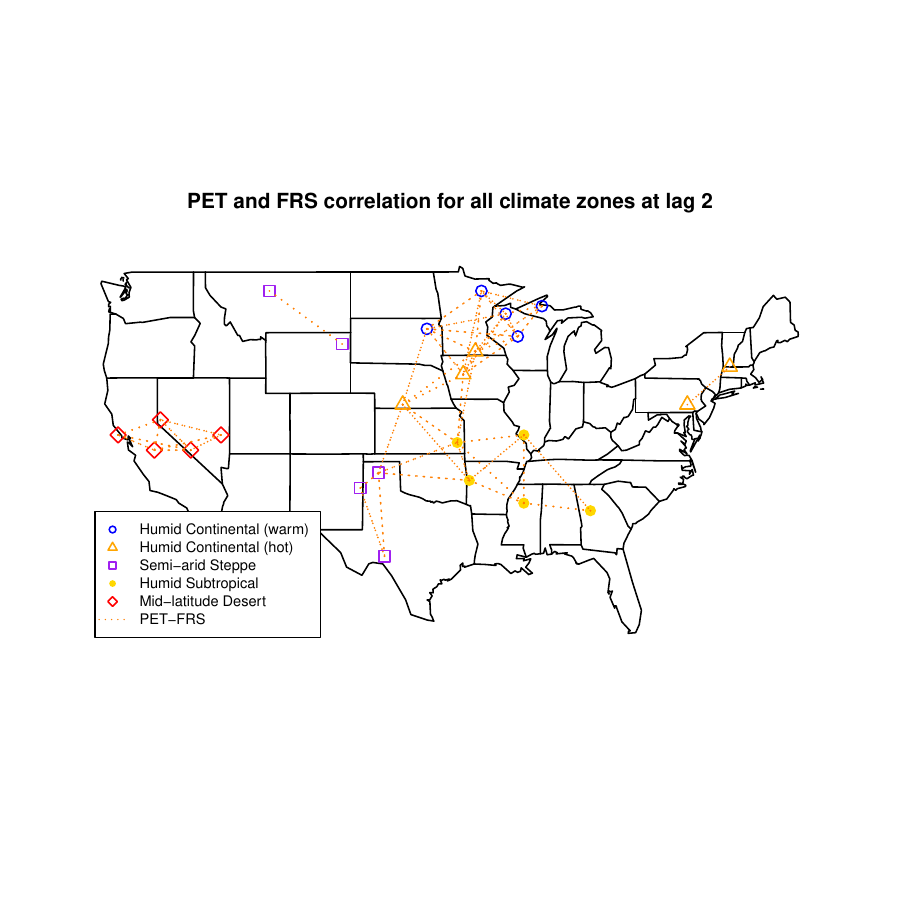}
    \vspace{-70pt}
    \caption{Estimated spatial structure for PET and FRS at different time lags.}
    \label{figure:9}
\end{figure}

Next, we separately estimate the spatial structure for different climate defined zones, and we focus on the change of spatial structure from lag 1 to lag 2 time points separately with fixed variable pairs, the results are illustrated in Figure~\ref{figure:8} and we also demonstrate the specific estimated spatial structure for (PET, FRS) variables pair in Figure~\ref{figure:9}.

\begin{figure}[!ht]
    \centering
    \includegraphics[scale=.75]{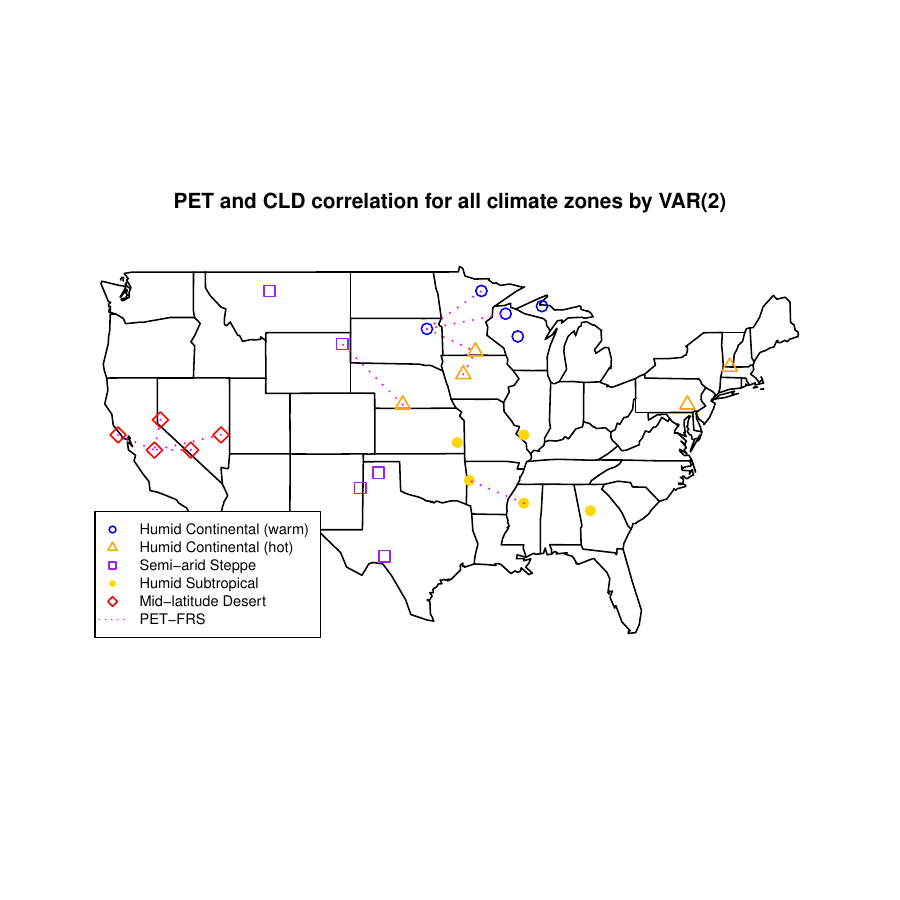}
    \vspace{-70pt}
    \caption{Spatial structure for PET and CLD by using VAR(2).}
    \label{figure:10}
\end{figure}

Then, let's compare the spatial structure for same variables pair. In the main context, we have presented the spatial structure for (PET, CLD) variables pair for VAR(1) model. Now, we illustrate the spatial structure for the same. Here, we only present the estimated spatial structure for time lag 1 in the VAR(2) model in Figure~\ref{figure:10}. It explicitly presents a sparser structure than VAR(1) model, specifically, the mid-latitude desert maintains the same structure while other climate zones extremely changed.

\section{Conclusion and Discussion}\label{sec:discuss}
This paper studied the estimation of high-dimensional multivariate VAR models with spatio-temporal structure. We proposed a structured weighted $\ell_1$-regularized estimator that decomposes each block transition matrix into a cross-variable dependence coefficient and a spatial transition matrix. This decomposition allows the model to incorporate prior spatial information while preserving flexibility in the dependence patterns among multiple variables.

The proposed estimator is computed through an alternating convex-search algorithm with ADMM updates. We established deterministic convergence to a first-order stationary point and derived high-probability error bounds for the cross-variable dependence matrix and the spatial transition matrices under standard stability and restricted-eigenvalue-type conditions. The simulation studies demonstrate that the proposed method improves support recovery and estimation accuracy relative to existing two-step $\ell_1$-regularized estimators, particularly in higher-dimensional settings. The climate-data application further illustrates that the method can recover interpretable variable-dependence networks and spatial interaction patterns.

The framework can be extended in several directions. First, other structured sparsity assumptions can be incorporated through alternative choices of the spatial graph $\mathcal{J}_0$ and the weight matrix $W$, including group sparsity or multiscale spatial penalties. Second, the VAR(1) formulation can be extended to VAR($d$) models with $d>1$ by estimating a separate structured transition matrix for each lag. Finally, the theoretical analysis can be refined to study model selection consistency and sharper convergence guarantees for the alternating algorithm under additional identifiability conditions.
\bibliography{Bibliography-MM-MC}
\end{document}